\documentclass[longauth,bibyear]{aa} % for the long lists of affiliations 
\usepackage{lscape}
\usepackage{float}
\usepackage{graphicx}
\usepackage{txfonts}
\begin{document} 

\title{An ALMA survey of submillimetre galaxies in the COSMOS field: The extent of the radio-emitting region revealed by 3~GHz imaging with the Very Large Array\thanks{Based on observations with the Karl G. Jansky Very Large Array (VLA) of the National Radio Astronomy Observatory (NRAO). The National Radio Astronomy Observatory is a facility of the National Science 
Foundation operated under cooperative agreement by Associated Universities, Inc.}}

   \author{O.~Miettinen\inst{1}, M.~Novak\inst{1}, V.~Smol\v{c}i\'{c}\inst{1}, I.~Delvecchio\inst{1}, M.~Aravena\inst{2}, D.~Brisbin\inst{2}, A.~Karim\inst{3}, E.~J.~Murphy\inst{4}, 
E.~Schinnerer\inst{5}, M.~Albrecht\inst{3}, H.~Aussel\inst{6}, F.~Bertoldi\inst{3}, P.~L.~Capak\inst{7,8}, C.~M.~Casey\inst{9}, F.~Civano\inst{10,11}, C.~C.~Hayward\inst{12,11}, 
N.~Herrera Ruiz\inst{13}, O.~Ilbert\inst{14}, C.~Jiang\inst{2,15,16}, C.~Laigle\inst{17}, O.~Le F{\`e}vre\inst{14}, B.~Magnelli\inst{3}, S.~Marchesi\inst{10,18,11}, 
H.~J.~McCracken\inst{17}, E.~Middelberg\inst{13}, A.~M.~Mu{\~n}oz Arancibia\inst{19}, F.~Navarrete\inst{3}, N.~D.~Padilla\inst{20,21}, D.~A.~Riechers\inst{22}, M.~Salvato\inst{23}, 
K.~S.~Scott\inst{24}, K.~Sheth\inst{25}, L.~A.~M.~Tasca\inst{14}, M.~Bondi\inst{26}, and G.~Zamorani\inst{27}  }

   \institute{Department of Physics, University of Zagreb, Bijeni\v{c}ka cesta 32, HR-10000 Zagreb, Croatia \\ \email{oskari@phy.hr} \and N\'ucleo de Astronom\'{\i}a, Facultad de Ingenier\'{\i}a, Universidad Diego Portales, Av. Ej\'ercito 441, Santiago, Chile \and Argelander-Institut f\"{u}r Astronomie, Universit\"{a}t Bonn, Auf dem H\"{u}gel 71, D-53121 Bonn, Germany \and National Radio Astronomy Observatory, 520 Edgemont Road, Charlottesville, VA 22903, USA \and Max-Planck-Institut f\"{u}r Astronomie, K\"{o}nigstuhl 17, 69117 Heidelberg, Germany \and AIM Unit\'e Mixte de Recherche CEA CNRS, Universit\'e Paris VII UMR n158, Paris, France \and Infrared Processing and Analysis Center, California Institute of Technology, MC 100-22, 770 South Wilson Ave., Pasadena, CA 91125, USA \and Spitzer Science Center, California Institute of Technology, Pasadena, CA 91125, USA \and Department of Astronomy, The University of Texas at Austin, 2515 Speedway Blvd Stop C1400, Austin, TX 78712, USA \and Yale Center for Astronomy and Astrophysics, 260 Whitney Avenue, New Haven, CT 06520, USA \and Harvard-Smithsonian Center for Astrophysics, 60 Garden Street, Cambridge, MA 02138, USA \and Center for Computational Astrophysics, Flatiron Institute, 162 Fifth Avenue, New York, NY 10010 \and Astronomisches Institut, Ruhr-Universit\"{a}t Bochum, Universit\"{a}tstrasse 150, 44801 Bochum, Germany \and Aix Marseille Universit\'e, CNRS, LAM (Laboratoire d'Astrophysique de Marseille), UMR 7326, 13388, Marseille, France \and CAS Key Laboratory for Research in Galaxies and Cosmology, Shanghai Astronomical Observatory, Nandan Road 80, Shanghai 200030, China \and Chinese Academy of Sciences South America Center for Astronomy, 7591245 Santiago, Chile \and Sorbonne Universit\'es, UPMC Universit\'e Paris 6 et CNRS, UMR 7095, Institut d'Astrophysique de Paris, 98 bis Boulevard Arago, 75014 Paris, France \and Dipartimento di Fisica e Astronomia, Universit\`{a} di Bologna, viale Berti Pichat 6/2, 40127 Bologna, Italy \and Instituto de F\'{\i}sica y Astronom\'{\i}a, Universidad de Valpara\'{\i}so, Av. Gran Breta\~{n}a 1111, Valpara\'{\i}so, Chile \and Instituto de Astrof\'{\i}sica, Pontificia Universidad Cat\'olica de Chile, Avda. Vicu\~{n}a Mackenna 4860, Santiago, Chile \and Centro de Astro-Ingenier\'{\i}a, Pontificia Universidad Cat\'olica de Chile, Avda. Vicu\~{n}a Mackenna 4860, 782-0436 Macul, Santiago, Chile \and Astronomy Department, Cornell University, 220 Space Sciences Building, Ithaca, NY 14853, USA \and Max-Planck-Institut f\"{u}r extraterrestrische Physik, Garching bei M\"{u}nchen, D-85741 Garching bei M\"{u}nchen, Germany \and North American ALMA Science Center, NRAO, 520 Edgemont Rd, Charlottesville, VA, USA 22903 \and NASA Headquarters, 300 E. St. SW, Washington, DC 20546, USA \and Osservatorio di Radioastronomia - INAF, Bologna, Via P. Gobetti 101, 40129 Bologna, Italy \and INAF - Osservatorio Astronomico di Bologna, via Ranzani 1, I-40127, Bologna, Italy  } 

   \date{Received ; accepted}

% \abstract{}{}{}{}{} 
% 5 {} token are mandatory
\authorrunning{Miettinen et al.}
\titlerunning{3~GHz radio sizes of the ALMA detected ASTE/AzTEC SMGs in COSMOS}

\abstract {The observed spatial scale of the radio continuum emission from star-forming galaxies can be used to investigate the spatial extent of active star formation, constrain the importance of cosmic-ray transport, and examine the effects of galaxy interactions.} {We determine the radio size distribution of a large sample of 152 submillimetre galaxies (SMGs) in the COSMOS field that were pre-selected at 1.1~mm, and later detected with the Atacama Large Millimetre/submillimetre Array (ALMA) in the observed-frame 1.3~mm dust continuum emission at a signal-to-noise (S/N) ratio of $\geq5$.} {We used the deep, subarcsecond-resolution ($1\sigma=2.3$~$\mu$Jy~beam$^{-1}$; $0\farcs75$) centimetre radio continuum observations taken by the Karl G.~Jansky Very Large Array (VLA)-COSMOS 3~GHz Large Project.} {One hundred and fifteen of the 152 target SMGs ($76\%\pm7\%$) were found to have a 3~GHz counterpart ($\geq4.2\sigma$), which renders the radio detection rate notably high. The median value of the deconvolved major axis full width at half maximum (FWHM) size at 3~GHz is derived to be $0\farcs59 \pm 0\farcs05$, or $4.6\pm0.4$~kpc in physical units, where the median redshift of the sources is $z=2.23\pm0.13$ (23\% are spectroscopic and 77\% are photometric values). The radio sizes are roughly log-normally distributed, and they show no evolutionary trend with redshift, or difference between different galaxy morphologies. We also derived the spectral indices between 1.4 and 3~GHz, and 3~GHz brightness temperatures for the sources, and the median values were found to be $\alpha_{\rm 1.4\, GHz}^{\rm 3\, GHz}=-0.67$ ($S_{\nu}\propto \nu^{\alpha}$) and $T_{\rm B}=12.6\pm2$~K. Three of the target SMGs, which are also detected with the Very Long Baseline Array (VLBA) at 1.4~GHz (AzTEC/C24b, 61, and 77a), show clearly higher brightness temperatures than the typical values, reaching $T_{\rm B}(3\,{\rm GHz})>10^{4.03}$~K for AzTEC/C61.} {The derived median radio spectral index agrees with a value expected for optically thin non-thermal synchrotron radiation, and the low median 3~GHz brightness temperature shows that the observed radio emission is predominantly powered by star formation and supernova activity. However, our results provide a strong indication of the presence of an active galactic nucleus in the VLBA and X-ray-detected SMG AzTEC/C61 (high $T_{\rm B}$ and an inverted radio spectrum). The median radio-emitting size we have derived is $\sim1.5-3$ times larger than the typical far-infrared dust-emitting sizes of SMGs, but similar to that of the SMGs' molecular gas component traced through mid-$J$ line emission of carbon monoxide. The physical conditions of SMGs probably render the diffusion of cosmic-ray electrons inefficient, and hence an unlikely process to lead to the observed extended radio sizes. Instead, our results point towards a scenario where SMGs are driven by galaxy interactions and mergers. Besides triggering vigorous starbursts, galaxy collisions can also pull out the magnetised fluids from the interacting disks, and give rise to a taffy-like synchrotron-emitting bridge. This provides an explanation for the spatially extended radio emission of SMGs, and can also cause a deviation from the well-known infrared-radio correlation owing to an excess radio emission. Nevertheless, further high-resolution observations are required to examine the other potential reasons for the very compact dust-emitting sizes of SMGs, such as the radial dust temperature and metallicity gradients.} 


   \maketitle
%
%________________________________________________________________

\section{Introduction}

Radio continuum imaging of extragalactic fields in the centimetre wavebands can be used as an efficient, dust-unbiased tool to search for 
star-forming galaxies and to probe their recent and ongoing star formation activity. The underlying physical reason for this is that the radio emission in question predominantly arises from 
two radiation mechanisms that are linked to the evolution of short-lived high-mass ($M\gtrsim8$~M$_{\sun}$) stars (see e.g. \cite{condon1992} for a review). 
First, the non-thermal synchrotron emission arises via radiative losses of ultrarelativistic electrons (with Lorentz factors $\gamma \gg 1$) that are gyrating in magnetic fields and accelerated in shock fronts of Type~II and Ib core-collapse supernova (SN; e.g. \cite{weiler1986}) ejecta and the shells of their expanding remnants (e.g. \cite{blandford1980}; \cite{bogdan1983}). Secondly, the thermal free-free radiation arises from the electron-ion Coulomb interactions in regions of ionised hydrogen atoms, that is \ion{H}{II} regions. Besides these two phenomena that are linked to the formation and explosive death of high-mass stars (\ion{H}{II} regions and SNe, respectively), the ultraviolet (UV)-optical photons emitted by the young, massive stars heat the dust grains mixed with the more abundant gas component of the galactic interstellar medium (ISM; e.g. \cite{mathis1983}). The energy absorbed by dust grains is then re–radiated as thermal emission in the infrared (IR; e.g. \cite{devereux1990}). On the basis of this connection, the spatial extent or distribution of the rest-frame (far-)IR dust continuum emission is expected to be fairly similar to that of radio continuum emission. This expectation is strongly supported by the remarkably tight correlation between the (far-)IR and radio continuum emissions observed in star-forming galaxies (e.g. \cite{vanderkruit1971}; \cite{dejong1985}; \cite{helou1985}; \cite{condon1991}; \cite{yun2001}; \cite{basu2015}). 

However, high angular resolution ($0\farcs16-0\farcs30$) (sub-)millimetre continuum imaging of submillimetre 
galaxies (SMGs) with the Atacama Large Millimetre/submillimetre Array (ALMA) have shown their rest-frame FIR-emitting sizes to be very compact with median diameters (full width at half maximum or FWHM) of only $\sim 1.4-3.1$~kpc (\cite{simpson2015}, hereafter S15; \cite{ikarashi2015}; \cite{hodge2016}; \cite{simpson2016}; see also \cite{bussmann2015}). When compared to the median major axis FWHM sizes of the radio emission from SMGs as measured through observations with the Very Large Array (VLA; \cite{thompson1980}; \cite{perley2011}) at 1.4~GHz (\cite{biggs2008}; see also \cite{chapman2004}) and 3~GHz (\cite{miettinen2015}, hereafter M15), the aforementioned FIR sizes appear to be $\sim1.4-4.4$ times smaller. As a physical cause of this potential spatial decoupling, S15 suggested that the diffusion of cosmic-ray (CR) electrons in the galactic magnetic field away from their sites of origin leads to a more extended size scale of the observed radio emission than that of FIR dust emission. However, this scenario is challenged by the rapid cooling (a few times $10^5$~yr or less) of CR electrons in strongly star-forming galaxies, which renders their diffusion scale length in a magneto-ionic medium very short, only $\lesssim100$~pc (see M15, and references therein). In contrast, less extreme, main-sequence star-forming galaxies with stellar masses of $M_{\star}\simeq 2\times10^{10}-1.6\times10^{11}$~M$_{\sun}$ and star formation rates of ${\rm SFR}\simeq 40-330$~M$_{\sun}$~yr$^{-1}$ at $z=1.3-3$ are found to have similar spatial extents of VLA radio and ALMA dust continuum emissions (average $\langle r_{\rm radio}/r_{\rm dust}\rangle=0.96\pm0.14$; \cite{rujopakarn2016}). On the other hand, using very deep ($1\sigma=572$~nJy~beam$^{-1}$), high-resolution ($0\farcs22$ FWHM) observations with the VLA at 10~GHz, Murphy et al. (2017) measured extremely compact radio sizes for their $z\sim0.3-3.2$ star-forming galaxies in the Great Observatories Origins Deep Survey-North (GOODS-N) field, the median major axis FWHM size being only 1.2~kpc, which is comparable to the median FIR-emitting size derived by Ikarashi et al. (2015) for their SMGs. Nevertheless, on the basis of the Rujopakarn et al. (2016) results, it seems that the dust-radio spatial decoupling could be a characteristic of more extreme SMGs.

A possible alternative explanation for a more extended radio emission is a population of relativistic CR electrons radiating in magnetic fields pulled out of the disks of gravitationally interacting galaxies (e.g. \cite{condon1993}; \cite{drzazga2011}; M15). Besides this effect, galaxy interactions and mergers, which are expected to be more frequent at high redshifts, can also lead to the vigorous SFRs observed in SMGs, which can sometimes reach values as high as thousands of solar masses per year (e.g. \cite{tacconi2008}; \cite{engel2010}). 

On the other hand, the spatial extent of the molecular gas component in SMGs, as probed through observations of the mid-$J$ transitions of the $^{12}$C$^{16}$O main isotopologue ($3\leq J_{\rm up}\leq7$, where $J_{\rm up}$ is the upper rotational energy level), is found to have a typical major axis FWHM of $\sim4$~kpc (\cite{tacconi2006}; \cite{engel2010}), and hence rather similar to the size scale of radio emission (M15). The mid-$J$ CO lines are probing the denser and warmer molecular gas component than the lower excitation $J_{\rm up}\leq2$ lines, and the full molecular gas reservoir is expected to occupy a larger galactic area. This was indeed demonstrated by the CO$(J=1-0)$ observations of $z=2.490-3.408$ SMGs with the VLA by Riechers et al. (2011a,b), where the spatial scale of the emission ($\sim6-15$~kpc in diameter) was found to be $\sim1.6-3$ times larger than that probed by higher $J$ ($3-2$, $4-3$, and $6-5$) CO transitions for the same SMGs by Tacconi et al. (2006) and Engel et al. (2010). A similar conclusion was reached by Ivison et al. (2011) for their sample of $z=2.202-2.486$ SMGs that were observed in both the $J=1-0$ and $J=3-2$ lines of CO (see also \cite{swinbank2011}; \cite{sharon2015}; \cite{spilker2015}). Besides CO, also the spatial scale of the $\lambda_{\rm rest}=158$~$\mu$m $[\ion{C}{II}]$ fine-structure line emission of SMGs is found to be more extended than the dust continuum emission (e.g. \cite{riechers2013}, 2014; \cite{oteo2016}).

If the molecular gas disk of an SMG traced by mid-$J$ CO line emission exhibits a lower dust temperature than the compact, starbursting central region, then the latter could outshine the more extended dust zone, and hence dominate the rest-frame FIR dust continuum size measurements. A radial dust temperature gradient could then lead to the aforementioned size mismatch between dust and radio emissions. Another candidate culprit for the highly centrally concentrated dust-emitting regions of SMGs, and hence for the dust-radio size discrepancy could be a strong radial metallicity gradient in the galactic disk. Because the dust and gas contents are linked to each other through metallicity (the dust-to-gas mass ratio increases with the gas-phase metallicity; e.g. \cite{draine2007}), a radial dust emission gradient can arise from a gradient in metallicity. Finally, it remains a possibility that the FIR-radio size difference is just illusory if the low-surface brightness outer parts of SMGs have been missed, and hence the corresponding FIR FWHM sizes have been correspondingly underestimated. 

To gain further insight into the characteristic radio emission sizes of SMGs, here we present a study of radio sizes of a large, well selected sample of SMGs in the deeply observed Cosmic Evolution Survey (COSMOS; \cite{scoville2007}) field using radio data from the Karl G.~Jansky VLA-COSMOS 3~GHz Large Project, which is a sensitive ($1\sigma$ noise of 2.3~$\mu$Jy~beam$^{-1}$), subarcsecond resolution ($0\farcs75$) survey (\cite{smolcic2017}). In Sect.~2, we describe our SMG sample, and the employed radio data. The analysis and results are presented in Sect.~3. The results are discussed in Sect.~4, and we summarise the results and recapitulate our conclusions in Sect.~5. 

The cosmology adopted in the present work corresponds to a spatially flat $\Lambda$CDM (Lambda cold dark matter) universe with the present-day dark energy density parameter $\Omega_{\Lambda}=0.7$, total (baryonic plus non-baryonic) matter density parameter $\Omega_{\rm m}=0.3$, and a Hubble constant of $H_0=70$~km~s$^{-1}$~Mpc$^{-1}$.

\section{Data}

\subsection{Source sample: The ASTE/AzTEC 1.1~mm selected sources followed up with ALMA at 1.3~mm}
 
The target SMGs of the present study were first uncovered by the $\lambda_{\rm obs}=1.1$~mm blank-field continuum survey over 
an area of 0.72~deg$^2$ or 37.5\% of the full 2~deg$^2$ COSMOS field carried out with the AzTEC bolometer array on the 10~m Atacama Submillimetre 
Telescope Experiment (ASTE; \cite{ezawa2004}) by Aretxaga et al. (2011). The angular resolution of these observations was $34\arcsec$ FWHM, and the $1\sigma$ rms noise level was 1.26~mJy~beam$^{-1}$. Of the 189 SMG candidates with signal-to-noise (S/N) ratios of at least 3.5 that Aretxaga et al. (2011) found (see their Table~1), the 129 brightest sources (S/N$_{\rm 1.1\, mm}\geq4.0$) were followed up with ALMA at $\lambda_{\rm obs}=1.3$~mm 
and $\sim1\farcs6 \times 0\farcs9$ resolution by M.~Aravena et al. (in prep.) (Cycle~2 ALMA project 2013.1.00118.S; PI: M.~Aravena). 
Altogether, 152 SMG candidates at an S/N$_{\rm 1.3\, mm}\geq5$ ($1\sigma_{\rm 1.3\, mm}\sim0.1$~mJy~beam$^{-1}$) were uncovered by our ALMA survey; this detection S/N$_{\rm 1.3\, mm}$ threshold yields a sample reliability of about 100\%, that is, with no contamination by spurious sources (M.~Aravena et al., in prep.). The number of target single-dish AzTEC sources that were resolved into more than one ALMA component (33) yields a multiplicity percentage of $26\% \pm 4\%$ (at the $\sim1\farcs6 \times 0\farcs9$ resolution of our ALMA data, and above the AzTEC flux density limit of $S_{\rm 1.1\, mm}\gtrsim3.5$~mJy), where the uncertainty represents the Poisson error on counting statistics. Under the assumption that the dust emissivity index is $\beta=1.5$, the 1.3~mm flux densities of our ALMA sources suggest that $S_{\rm 850\, \mu m}\gtrsim2$~mJy, and hence all of them fulfil a definition of SMGs as galaxies having $S_{\rm 850\, \mu m}\geq1$~mJy (\cite{coppin2015}; \cite{simpson2016}). The ALMA observations, which together with the source catalogue are described in detail by M.~Aravena et al. (in prep.), allowed us to accurately pinpoint the position of the actual SMGs giving rise to the millimetre continuum emission seen in the single-dish AzTEC map, and hence we could reliably identify the correct radio (and other wavelength) counterparts of the target SMGs (\cite{brisbin2017}).

An important aspect regarding the present work is that three of the identified SMGs, AzTEC/C24b, 61, and 77a, were detected with the Very Long Baseline Array (VLBA) at a high, $16.2\times7.3$ square milliarcsecond resolution at $\nu_{\rm obs}=1.4$~GHz with flux densities of $S_{\rm 1.4\, GHz}=134.2$~$\mu$Jy (C24b), $11.1$~mJy (C61), and $332.6$~$\mu$Jy (C77a) (N.~Herrera Ruiz et al., in prep.), which indicates the presence of either a radio-emitting active galactic nucleus (AGN) or a very compact nuclear starburst, or both (cf.~\cite{casey2009}). One of these three SMGs, AzTEC/C61, was also detected in the X-rays with \textit{Chandra} (the 1.8~Ms \textit{Chandra} COSMOS Survey (C-COSMOS; \cite{elvis2009}; \cite{civano2012}) and the \textit{Chandra} COSMOS Legacy Survey (\cite{civano2016})). The source is identified as CID-1787 in the \textit{Chandra} COSMOS Legacy Survey ($0\farcs25$ south-west from the ALMA position), and its flux density in the 0.5--2~keV (soft), 2--10~keV (hard), and 0.5--10~keV (full) bands is $S_{\rm X}=1.60 \times 10^{-15}$, $4.24 \times 10^{-15}$, and $6.54 \times 10^{-15}$~erg~cm$^{-2}$~s$^{-1}$, respectively (\cite{civano2016}). The absorption-corrected, rest-frame 2--10~keV luminosity of AzTEC/C61 is $L_{\rm 2-10\, keV}=1.7\times10^{44}$~erg~s$^{-1}$ (scaled from the photometric redshift of $z_{\rm phot}=1.56$ reported in the catalogue to $z_{\rm spec}=3.2671$; \cite{brisbin2017}), which is too high to arise only from stellar processes, and hence AzTEC/C61 very likely harbours an AGN.  

Seven additional, VLBA-non-detected SMGs in our sample, AzTEC/C11, 44b, 45, 56, 71b, 86, and 118, were also detected in the \textit{Chandra} X-ray imaging. The angular offsets of these X-ray sources (as reported in the Legacy Survey catalogue) from the ALMA positions range from $0\farcs15$ for AzTEC/C86 to $0\farcs95$ for AzTEC/C11. By using the source redshifts from Brisbin et al. (2017), we derived the absorption-corrected, rest-frame 2--10~keV luminosities of these sources to range from $L_{\rm 2-10\, keV}=7.3\times10^{42}$~erg~s$^{-1}$ for AzTEC/C71b to $<1.9\times10^{44}$~erg~s$^{-1}$ for AzTEC/C11 (C11 and C56 were not detected in the hard band) with a mean and its standard error ($\sigma / \sqrt{N}$, where $\sigma$ is the standard deviation and $N$ the sample size) of $ (4.0\pm1.3)\times10^{43}$~erg~s$^{-1}$ (based on the survival analysis technique described in Sect.~3.2).

\subsection{Very Large Array 3~GHz radio continuum data}

The radio observations used in the present paper were taken by the VLA-COSMOS 3~GHz Large Project (\cite{smolcic2017}). 
The project, data reduction, and imaging are fully described in Smol{\v c}i{\'c} et al. (2017), 
and a brief summary can be found in M15 (Sect.~2.2 therein). The final radio mosaic was restored with a circular synthesised beam size of $0\farcs75$ (FWHM), and the typical final $1\sigma$ root mean square (rms) noise level is $1\sigma=2.3$~$\mu$Jy~beam$^{-1}$. As we already discussed in M15, the effect of bandwidth smearing in the 3~GHz mosaic is negligible, and hence does not affect the radio size measurements (see \cite{smolcic2017} for more details). 

\section{Analysis and results}

\subsection{Identification of the 3~GHz counterparts}

The 3~GHz counterparts of our SMGs were identified using a two-step process. First, we identified all the 3~GHz sources in each of the 129 ALMA target field by eye inspection, and created a custom 3~GHz source catalogue. Secondly, our ALMA $\geq5\sigma$ source catalogue was cross-matched with the aforementioned 3~GHz source catalogue using a $1\arcsec$ matching radius. Altogether, 115 out of the 152 SMGs having a ${\rm S/N}_{\rm 1.3\, mm}\geq5$ were found to be associated with a 3~GHz source. This makes the percentage of the 3~GHz-detected SMGs in our sample to be $76\% \pm7\%$, where the uncertainty refers to the Poisson error. The other way round, the radio non-detection rate is about 24\%. The VLA 3~GHz contour maps of the 3~GHz detected SMGs are shown in Fig.~\ref{figure:maps}, selected sources are discussed in Appendix~B, and the basic 3~GHz radio source properties (e.g. peak position, peak surface brightness, and flux density) are listed in Table~\ref{table:results}. We note that 106 out of our 115 radio sources are common with the VLA-COSMOS 3~GHz Large Project catalogue ($0\farcs4$ matching radius), which is composed of $\geq5\sigma$ sources (\cite{smolcic2017}). The present 3~GHz flux densities are in good agreement with the catalogue values, the ratio between the two ranging from $S_{\rm 3\, GHz}^{\rm this\, work}/S_{\rm 3\, GHz}^{\rm cat.}=0.9$ to 2.7 with a median of 1.1.

\begin{figure}[!htb]
\centering
\resizebox{0.9\hsize}{!}{\includegraphics{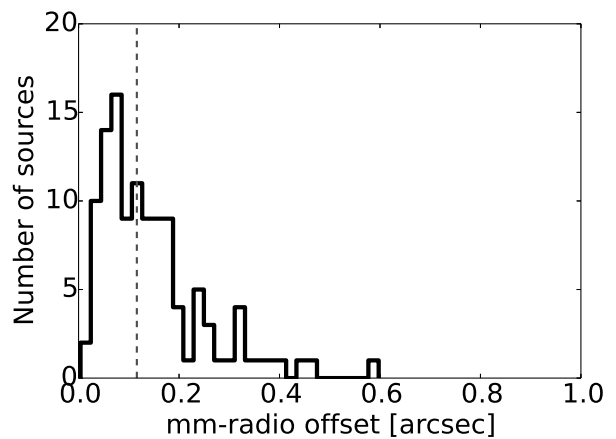}}
\caption{Distribution of the projected angular separations between the ALMA 1.3~mm and VLA 3~GHz peak positions. The bin size is $0\farcs02$. The $x$-axis extends to $1\arcsec$, which corresponds to the search radius used in the 3~GHz radio counterpart identification. The largest offset found is $0\farcs6$. The vertical dashed line marks the median separation of $0\farcs116$. }
\label{figure:offset}
\end{figure}

The projected separation between the ALMA 1.3~mm peak position and that of the VLA 3~GHz emission was found to range from 3.9 milliarcseconds to $0\farcs6$ with a mean (median) separation of $0\farcs144$ ($0\farcs116$) (see column~(11) in Table~\ref{table:results}, and Fig.~\ref{figure:offset}). The S/N ratios of the 3~GHz sources associated with our SMGs are in the range of ${\rm S/N}_{\rm 3\, GHz}=4.2-2\,712.7$, where AzTEC/C61, an SMG hosting a radio-loud AGN (Sects.~2.1 and 4.1), is the most significant detection. The detection S/N ratio with ALMA at 1.3~mm of these 3~GHz detected SMGs was found to be in the range of ${\rm S/N}_{\rm 1.3\, mm}=5.1-73.0$ with a mean (median) of $\langle {\rm S/N}_{\rm 1.3\, mm}\rangle=16.0$ (12.7), while those of the 3~GHz non-detected SMGs are ${\rm S/N}_{\rm 1.3\, mm}=5.1-25.1$ with a mean (median) of $\langle {\rm S/N}_{\rm 1.3\, mm}\rangle=9.0$ (7.3) (M.~Aravena et al., in prep.). Hence, the 3~GHz detected SMGs are brighter dust emitters on average than their radio non-detected counterparts as expected on the basis of the IR-radio correlation. 

\subsection{Redshift distribution of the 3~GHz detected and non-detected submillimetre galaxies}

\begin{figure}[!htb]
\centering
\resizebox{0.9\hsize}{!}{\includegraphics{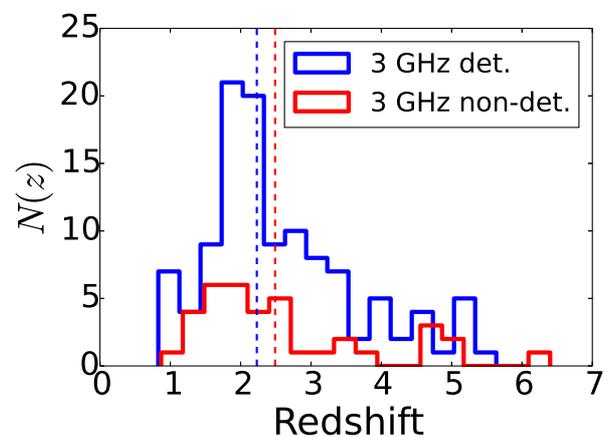}}
\caption{Redshift distributions of our 3~GHz detected ALMA SMGs (blue histogram) and SMGs not detected at 3~GHz (red histogram). 
The redshift bins have a width of $\Delta z = 0.3$. The lower redshift limits in the red histogram were placed in the bins corresponding to those values. The vertical dashed lines indicate the sample medians (blue: $\tilde{z}= 2.23\pm0.13$; red: $\tilde{z}= 2.49\pm0.27$).}
\label{figure:zdistributions}
\end{figure}

In Fig.~\ref{figure:zdistributions}, we show the redshift distributions of both the 3~GHz detected SMGs and the 37 SMGs that 
were not detected at 3~GHz. For details of the redshift determination, we refer to Brisbin et al. (2017). 
The sample of 3~GHz detections is composed of 66 photometric redshifts, 27 spectroscopic redshifts, ten redshifts constrained through synthetic redshift likelihood function by convolving the photo-$z$ likelihood function with a redshift likelihood function derived from the dust and radio indicators (the so-called synthetic redshifts in \cite{brisbin2017}), two redshifts derived through AGN template fitting, five redshifts based on the 3~GHz and submm flux density comparison, and five redshifts based on the peak wavelength of the FIR dust SED. The sample of 3~GHz non-detections contains 22 photometric redshifts, nine lower $z$ limits based on the upper limit to the 1.4~GHz flux density and the Carilli \& Yun (1999, 2000) method, three spectroscopic redshifts, one synthetic redshift, and two redshifts based on the peak position of the FIR dust spectral energy distribution (SED). 

The mean (median) redshift of the 3~GHz detected SMGs is $\langle z \rangle= 2.60\pm0.10$ ($\tilde{z}= 2.23\pm0.13$), while that of the 3~GHz non-detections is $\langle z \rangle= 2.99\pm0.22$ ($\tilde{z}= 2.49\pm0.27$), where the quoted $\pm$ uncertainties represent the standard errors of the mean and median values (the latter is estimated as $\sqrt{\pi/2}\simeq1.253$ times the standard error of the mean, which is strictly valid for a large sample and a normal distribution). To calculate the mean and median of the latter distribution, we applied survival analysis to take the lower redshift limits into account. We assumed that the right-censored data follow the same distribution as the uncensored values, and we used the Kaplan-Meier (K-M) method to construct a model of the input data. For this purpose, we used the Nondetects And Data Analysis for environmental data (NADA; \cite{helsel2005}) package for {\tt R}. As expected, the radio non-detected sources have a higher average redshift than the 3~GHz detections. A two-sided Kolmogorov–Smirnov (K-S) test between these two redshift distributions (and where the lower $z$ limits, or right-censored data were excluded) yields a K-S test statistic of $D_{\rm KS}=0.16$ (the maximum separation between the two cumulative distribution functions) and a K-S probability of $p_{\rm KS}= 0.61$ (a quantitative measure of the significance level of $D_{\rm KS}$) under the null hypothesis that the two samples are drawn from the same distribution. Owing to a fairly high $p_{\rm KS}$ value, we cannot reject the hypothesis that the redshift distributions are drawn from a common underlying parent distribution. However, the aforementioned $p_{\rm KS}$ value should be interpreted as an upper limit because the lower $z$ limits in the sample of 3~GHz non-detections were excluded in the K-S test (they are assumed to follow the distribution of uncensored data in our survival analysis). Indeed, on the basis of the positive $K$ correction at radio wavelengths, and negative $K$ correction in the (sub-)mm, one could expect a difference in the redshift distributions. As mentioned in Sect.~3.1, the 3~GHz non-detections are weaker ALMA sources on average compared to those that have a 3~GHz counterpart. We also note that the apparent excess of 3~GHz non-detections at $z\sim1-2$ in Fig.~\ref{figure:zdistributions} is the result of placing the nine lower redshift limits in the histogram bins corresponding to those values.

\subsection{Measurement of the size of the observed-frame 3~GHz radio emission}

Following the approach of M15, the 3~GHz radio sizes of the sources were measured using the NRAO Astronomical Image Processing System (AIPS) 
software package.\footnote{\tt http://www.aips.nrao.edu.} To determine the beam-deconvolved (intrinsic) sizes, we performed two-dimensional elliptical Gaussian fits to the image plane data using the AIPS task {\tt JMFIT}. The fitting was performed inside a box containing the source, and the fit was restricted to the pixel values of $\geq2.5\sigma$. The results are listed in Table~\ref{table:results}. 

The size measurement simulations by M15 (Appendix~D therein) suggest that the sizes provided by {\tt JMFIT} can be considered reliable because they were found to be consistent with the input sizes of the mock sources within the uncertainties assigned by the fitting task. Following M15 (and references therein), we considered a source to be resolved if its deconvolved FWHM size is larger than one-half the synthesised beam FWHM.\footnote{In the VLA-COSMOS 3~GHz Large Project catalogue, a source is taken to be resolved if it lies above the total flux density-to-peak surface brightness ratio given by $S_{\rm 3\, GHz}/I_{\rm 3\, GHz}=1+6\times ({\rm S/N})^{-1.44}$ (\cite{smolcic2017}). Seventy-seven of our sources satisfy this criterion, which is only a factor of 1.19 less than the number of sources (92) we consider resolved in the present study. For the remaining $92-77=15$ sources, our $S_{\rm 3\, GHz}/I_{\rm 3\, GHz}$ ratios are very close to the 3~GHz Large Project criterion, the median ratio between the two values being 0.9. Hence, the two definitions of resolved sources are in fairly good agreement despite the fact that the Large Project catalogue is based on a different source extraction and flux density measurement method ({\tt BLOBCAT}; \cite{hales2012}) than used in the present work (AIPS {\tt JMFIT}). However, {\tt BLOBCAT} does not provide parametric source sizes for a quantitative analysis, which is the main purpose of the present work.} The upper size limit for unresolved sources was set to one-half the synthesised beam FWHM ($<0\farcs38$). 

In the subsequent size analysis, we will use the deconvolved major axis FWHM as the diameter of the source 
(again following M15). Under the assumption of a simplified disk-like geometry, the major axis represents the physical extent of a disk galaxy, while the minor axis would be given by $\theta_{\rm min}=\theta_{\rm maj}\times \cos(i)$, where $i$ is the inclination angle defined so that for a disk viewed face-on ($i=0\degr$), $\theta_{\rm min}=\theta_{\rm maj}$. 
%We note that it is a fair assumption that an SMG has a disk-like configuration because such a flattened configuration is expected for gas in a merger-driven SMG where the highly dissipative dense gas falls in along trajectories parallel to the rotation axis of the system (e.g. \cite{barnes1996}).

To calculate the statistical parameters of the derived radio size distribution, we applied a similar K-M survival analysis as in Sect.~3.2 to take 
the upper size limits (left-censored data) into account. The mean, median, standard deviation, 95\% confidence interval for the mean, and the interquartile range (IQR) of the deconvolved major axis FWHMs are given in Table~\ref{table:stat}. For example, the median value of the angular deconvolved $\theta_{\rm maj}$ among the 115 SMGs detected at 3~GHz is $0\farcs59\pm0\farcs05$, where the quoted $\pm$ uncertainty represents the standard error of the median. The median major axis FWHM in linear units is $4.6\pm0.4$~kpc. 

The mean (median) value of the angular radio sizes of our eight X-ray detected SMGs is $0\farcs70\pm0\farcs09$ ($0\farcs54\pm0\farcs11$), while that of their linear sizes is $ 5.3\pm0.7$~kpc ($4.4\pm0.9$~kpc). Hence, these X-ray detected, potentially AGN-hosting SMGs do not stand out from the 3~GHz radio size distribution of our other SMGs, but rather have very similar sizes on average. We thus conclude that inclusion of the X-ray detected SMGs does not introduce any biases in the subsequent radio size analysis.

\begin{figure}[!htb]
\centering
\resizebox{0.9\hsize}{!}{\includegraphics{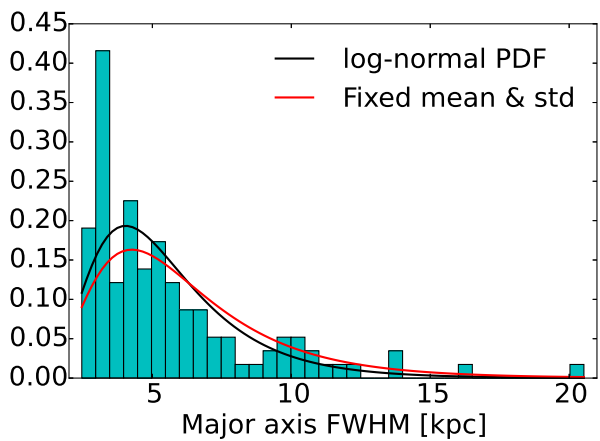}}
\caption{Distribution of the major axis FWHM sizes at 3~GHz shown as a normalised histogram. The upper size limits were placed in the bins corresponding to those values, and the bin width is 0.5~kpc. The solid black curve represents a best-fit PDF to a log-normal size distribution ($\mu=5.0$~kpc and $\sigma=2.9$~kpc). The PDF shown by the red solid curve was obtained by fixing the values of $\mu$ and $\sigma$ to those derived from a survival analysis ($\mu=5.5$~kpc and $\sigma=3.2$~kpc).}
\label{figure:lognormal}
\end{figure}

The distribution of the linear major axis FWHM sizes is shown in Fig.~\ref{figure:lognormal}. The data are presented as a normalised histogram, while the overlaid solid black curve represents a fit to a log-normal size distribution. The mean and standard deviation of the underlying normal distribution of this probability density function (PDF) are $\mu=5.0$~kpc and $\sigma= 2.9$~kpc. If the PDF is fit only to the uncensored data, these values are $\mu=5.6$~kpc and $\sigma=2.7$~kpc. For comparison, we also show a PDF with the values of $\mu$ and $\sigma$ tuned to those calculated using a survival analysis ($\mu=5.5$~kpc and $\sigma=3.2$~kpc). This comparison suggests that the observed radio sizes are fairly closely log-normally distributed. We note that the peak near $\sim3$~kpc in Fig.~\ref{figure:lognormal} results from the unresolved sources being placed in the bins corresponding to the the upper size limits, while those upper limits were taken into account in the calculation of the sample mean and median in our survival analysis.

\begin{figure}[!htb]
\centering
\resizebox{0.9\hsize}{!}{\includegraphics{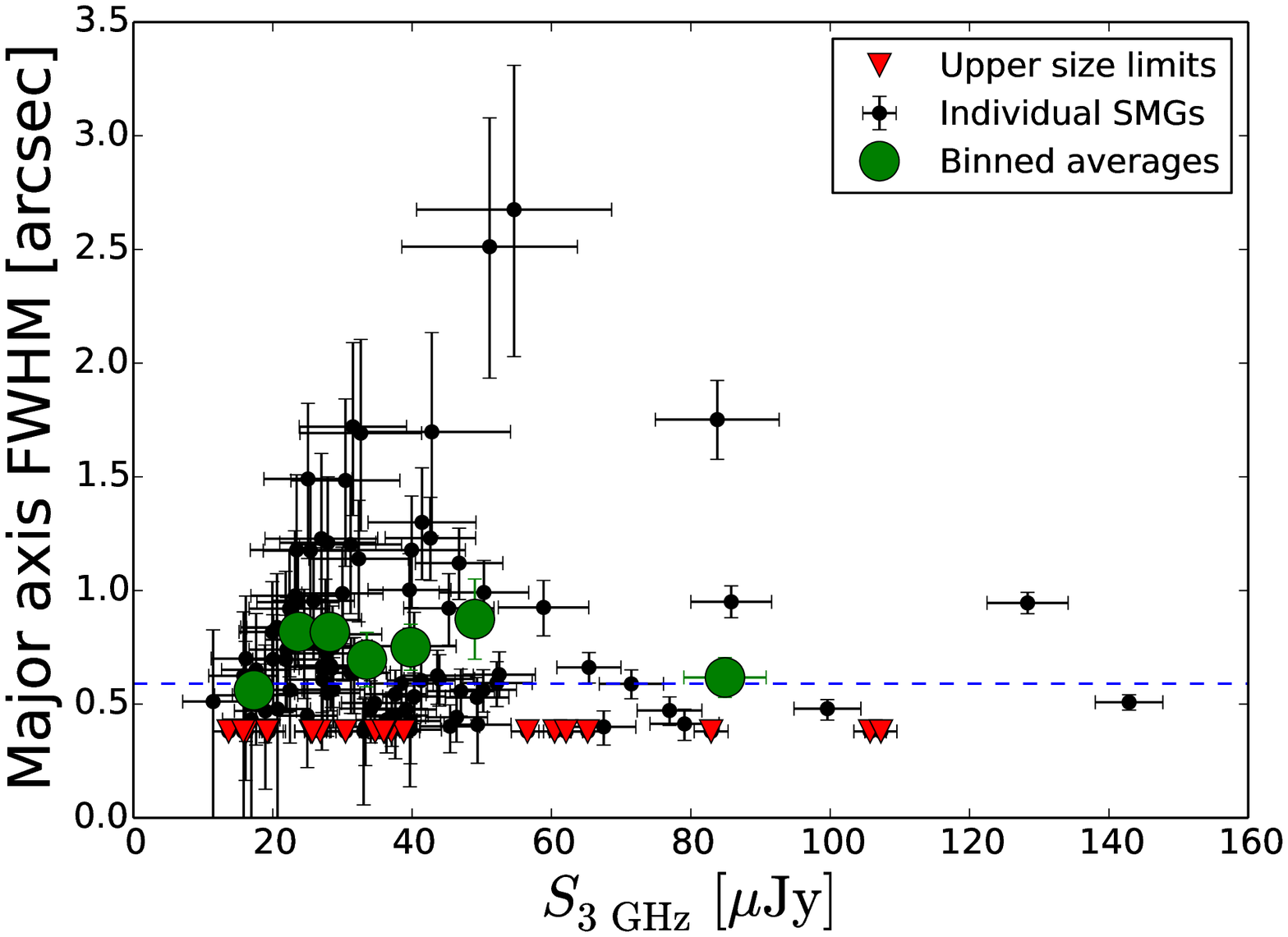}}
\resizebox{0.9\hsize}{!}{\includegraphics{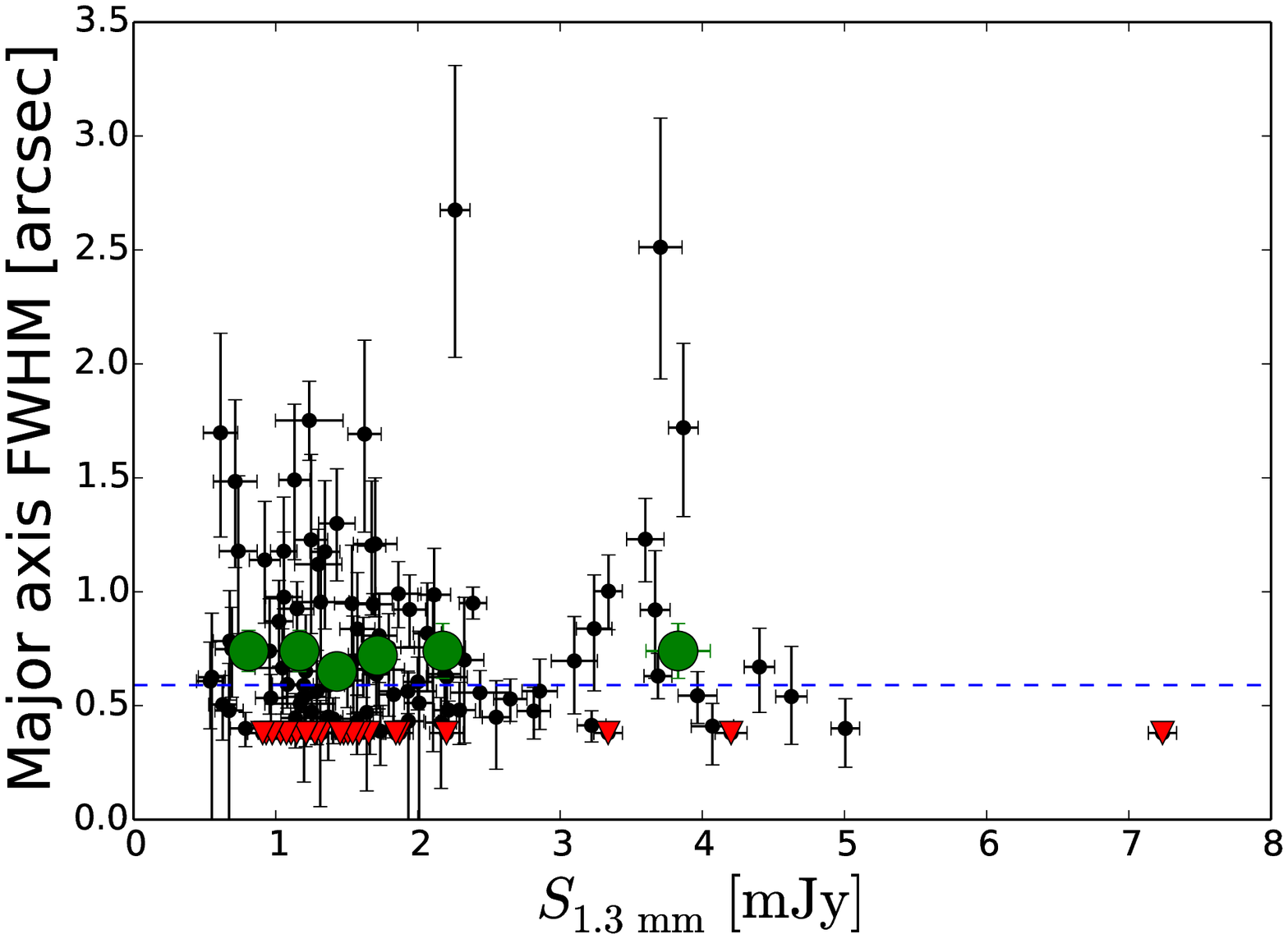}}
\caption{\textbf{Top:} Angular major axis FWHM size distribution of the 3~GHz emission as a function of the 3~GHz flux density. AzTEC/C61 and C77a, which have 3~GHz flux densities of $S_{\rm 3\, GHz}=11.5$~mJy and $S_{\rm 3\, GHz}=261$~$\mu$Jy, respectively, are omitted from the plot for legibility purposes. The horizontal dashed line marks the median major axis FWHM of $0\farcs59$. The upper size limits are indicated by red, downwards pointing triangles. The green filled circles represent the mean values of the binned data (each bin is equally populated by 16 SMGs, except the highest flux density bin, which contains 17 SMGs, and where C61 and C77a have not been taken into account), with the error bars showing the standard errors of the mean values. \textbf{Bottom:} Same as above but as a function of the ALMA 1.3~mm flux density. Again, the green filled circles represent the binned averages, where each bin contains 19 SMGs, except the highest flux density bin, which contains 20 SMGs (all the sources, including AzTEC/C61 and C77a, have been taken into account). }
\label{figure:fluxsize}
\end{figure}

In Fig.~\ref{figure:fluxsize}, we plot the deconvolved major axis FWHM sizes as a function of the VLA 3~GHz flux density (top panel) and ALMA 1.3~mm flux density (bottom panel). The binned data show that there is no correlation between the radio-emitting size and the radio or millimetre flux density, which agrees with the results presented by M15. To quantify this absence of correlation, we fit the binned data points using a linear regression line, and derived the relationships of the form $\theta_{\rm 3\, GHz}^{\rm maj} \propto (0.001 \pm 0.003)\times S_{\rm 3\, GHz}$ and $\theta_{\rm 3\, GHz}^{\rm maj} \propto (0.007 \pm 0.024)\times S_{\rm 1.3\, mm}$, where the uncertainty in the slope is based on the standard errors of the average major axis FWHM data points. %In the former case, the Pearson correlation coefficient of the binned data was found to be $r=-0.14$, while in the latter case it was derived to be $r=0.21$ ($r$ ranges from $-1$ to 1, and $r=0$ implies no (linear) correlation). 

\begin{figure}[!htb]
\centering
\resizebox{0.9\hsize}{!}{\includegraphics{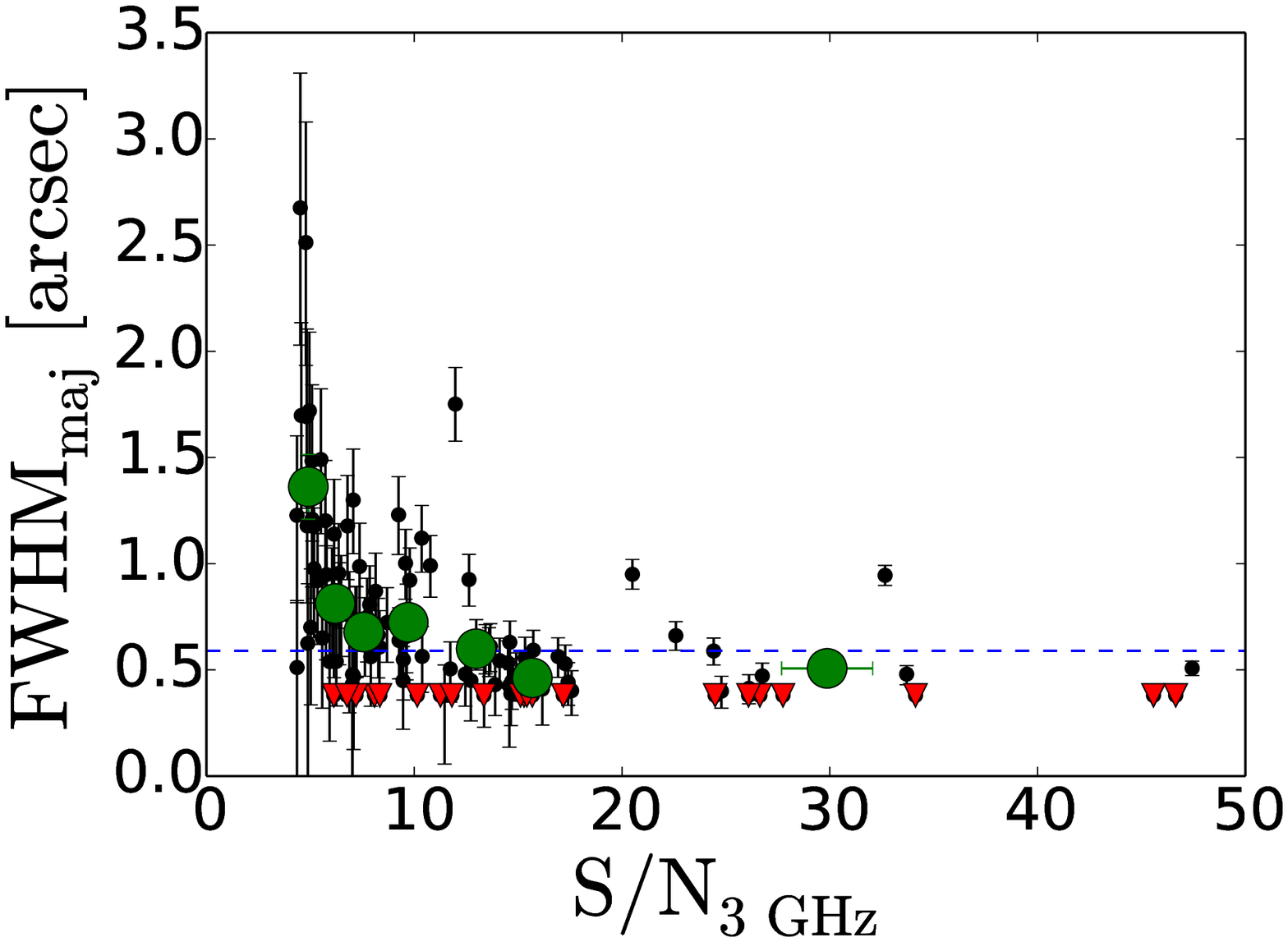}}
\caption{Angular major axis FWHM at 3~GHz plotted against the S/N ratio of the 3~GHz detection. AzTEC/C61 and C77a, which have S/N ratios of 2\,713 and 115, respectively, are omitted from the plot for legibility purposes. The horizontal dashed line marks the median major axis FWHM of $0\farcs59$. The upper size limits are indicated by red, downwards pointing triangles. The green filled circles represent the mean values of the binned data (each bin contains 16 SMGs, except the highest S/N bin, which contains 17 SMGs, and where C61 and C77a have not been taken into account), with the error bars showing the standard errors of the mean values.}
\label{figure:snr}
\end{figure}

In Fig.~\ref{figure:snr}, we plot the 3~GHz radio sizes as a function of the 3~GHz source S/N ratio. As expected, the sources with the lowest S/N ratios (the lowest S/N bin has an average value of ${\rm S/N}_{\rm 3\, GHz}=4.9$) appear larger on average than the more significant detections. However, the weakest sources also have large uncertainties in their size, and for example there are also three sources whose sizes ($0\farcs51$, $0\farcs62$, and $0\farcs70$) are smaller than the average size of the fourth highest S/N bin ($0\farcs72$, ${\rm S/N}_{\rm 3\, GHz}=9.7$). Hence, the observed S/N-size trend is not expected to significantly bias our statistical analysis. We note that nine out of 16 sources in the lowest S/N bin can be found in the VLA-COSMOS 3~GHz Large Project catalogue (comprised of $\geq 5\sigma$ detections; \cite{smolcic2017}), but only two of them (AzTEC/C2b and C10b) were classified as resolved on the basis of their $(S/I)_{\rm 3\, GHz}$ ratio (see footnote~2 above). Were we to cut our sample at an S/N ratio of $\geq5$, the mean and median major axis FWMH sizes would be $0\farcs67\pm0\farcs03$ and $0\farcs56\pm0\farcs04$, which are consistent with the full sample values within the quoted standard errors.

\begin{table}
\renewcommand{\footnoterule}{}
\caption{3~GHz major axis size (FWHM) distribution statistics.}
{\normalsize
\begin{minipage}{1\columnwidth}
\centering
\label{table:stat}
\begin{tabular}{c c}
\hline\hline 
Parameter & Value\tablefootmark{a} \\
\hline
Mean & $0\farcs72\pm0\farcs04$ ($5.5\pm0.3$~kpc) \\
Median & $0\farcs59\pm0\farcs05$ ($4.6\pm0.4$~kpc) \\
Standard deviation & $0\farcs42$ (3.2~kpc) \\ 
95\% confidence interval\tablefootmark{b} & $0\farcs64-0\farcs80$ (5.0--6.1~kpc)\\ 
IQR\tablefootmark{c} & $0\farcs41-0\farcs92$ (3.2--6.7~kpc)\\
\hline 
\end{tabular} 
\tablefoot{\tablefoottext{a}{The sample size of the deconvolved major axis FWHM sizes is 115. The linear size values in kpc are given in parentheses.}\tablefoottext{b}{A two-sided 95\% confidence interval for the mean value computed using the K-M method.}\tablefoottext{c}{The interquartile range or the values that fall between the 25th and 75th percentiles (the first and third quartiles, respectively).}}
\end{minipage} 
}
\end{table}

\subsection{Spectral index between the observed frequencies of 1.4~GHz and 3~GHz, and 3~GHz brightness temperature} 

Besides the radio-emitting sizes, we also derived the two-point radio spectral indices between the observed-frame
1.4~GHz and 3~GHz frequencies ($\alpha_{\rm 1.4\, GHz}^{\rm 3\, GHz}$, where we use the convention $S_{\nu} \propto \nu^{\alpha}$), and the observed-frame 3~GHz brightness temperatures ($T_{\rm B}$) for our sources to constrain the nature of the energy source (high-mass star formation and SN activity versus AGN) of the observed radio emission. The 1.4~GHz flux densities were taken from the final catalogue of the VLA-COSMOS Deep project (\cite{schinnerer2010}), which was merged with the revised and updated version of the VLA-COSMOS Large project catalogue (\cite{schinnerer2007}; \cite{bondi2008}). The source properties listed in this Joint 1.4~GHz catalogue were derived using the $2\farcs5$ resolution Deep project data. This angular resolution is 3.3 times coarser than that of the new 3~GHz radio mosaic. The four exceptions were AzTEC/C2a, C5, C22a, and C22b, for which we adopted the 1.4~GHz flux densities from M15 (the SMGs AzTEC~8, 1, 11-S, and 11-N therein, respectively) because either the source was not included in the 1.4~GHz catalogue or the catalogue value was inconsistent with the image flux density (for details, see M15; Table~4 and references therein). A 1.4~GHz counterpart was found for 54 or 47\% of our 3~GHz detected SMGs (see also \cite{brisbin2017}). We note that two of the 3~GHz non-detected SMGs, AzTEC/C97a and C100a, were found to have a significant 1.4~GHz counterpart in the Joint catalogue ($6.5\sigma$ and $5.8\sigma$, respectively). A $3\sigma$ upper limit (typically $<36$~$\mu$Jy~beam$^{-1}$) to the 1.4~GHz flux density was placed for non-detections (61 sources). We did not correct for the aforementioned mismatch between the 1.4~GHz and 3~GHz angular resolutions because our compact radio sources are not resolved in the $2\farcs5$ resolution 1.4~GHz data, which is shown by the (near) equality of the 1.4~GHz catalogue peak surface brightnesses and flux densities. As a sanity check, we cross-matched our 3~GHz source catalogue with a source catalogue created with {\tt BLOBCAT} from the 3~GHz mosaic of about $2\farcs5$ angular resolution (3\,722 sources at a ${\rm S/N}\geq 5$). We found 47 common sources within a $1\arcsec$ matching radius. We then cross-matched these 47 source positions with the aforementioned 1.4~GHz Joint catalogue, and found 38 matches within $1\arcsec$. The spectral indices derived for these sources at a common angular resolution were found to be in good agreement on average with those based on the $0\farcs75$ resolution 3~GHz data; the mean (median) ratio between the spectral indices derived from different resolution data and those based on the matched-resolution data was found to be 1.03 (0.91). 

\begin{figure}[!htb]
\centering
\resizebox{0.9\hsize}{!}{\includegraphics{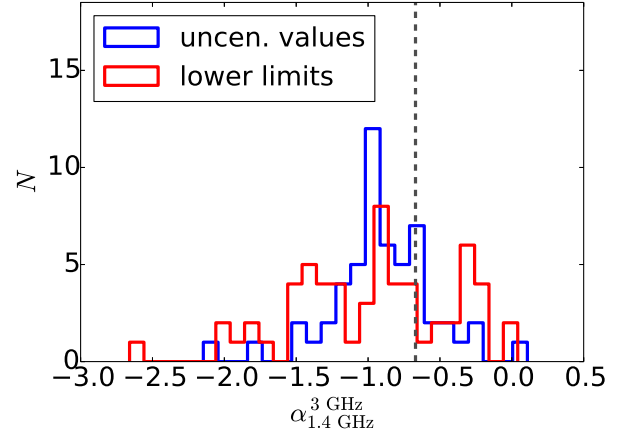}}
\resizebox{0.9\hsize}{!}{\includegraphics{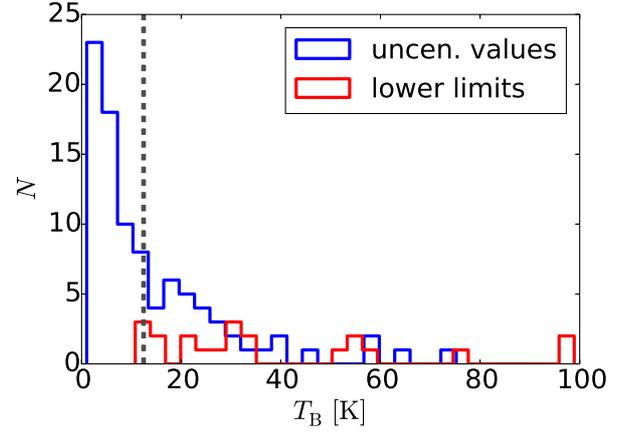}}
\caption{\textbf{Top:} Distribution of the 1.4--3~GHz radio spectral indices. The distributions of the uncensored values and the lower limits (placed in the bins corresponding to those values) are shown separately by the blue and red histograms, respectively. In both cases, 
the histogram bin size is 0.1. The sample median, shown by the vertical dashed line, is $\alpha_{\rm 1.4\, GHz}^{\rm 3\, GHz}=-0.67$ (same for the full sample and a sample from which the three VLBA-detected sources were removed; Table~\ref{table:stat2}). \textbf{Bottom:} Distribution of the $\nu_{\rm obs}=3$~GHz brightness temperatures, where the VLBA-detected sources with high $T_{\rm B}$ values are omitted for legibility purposes. Again, the uncensored values and the lower limits are plotted separately. The bin size is 3~K. The median values, $T_{\rm B}=12.6$~K for the full sample and $T_{\rm B}=12.3$~K for a sample from which the VLBA-detected sources were removed, are marked by the vertical dashed lines (the two lines are nearly indistinguishable owing to a minute, 0.3~K difference in the medians).}
\label{figure:histograms}
\end{figure}

The derived $\alpha_{\rm 1.4\, GHz}^{\rm 3\, GHz}$ indices are listed in Col.~(12) in Table~\ref{table:results}, where the quoted uncertainties were propagated from those of the flux densities. The distribution of the spectral indices is shown in the top panel in Fig.~\ref{figure:histograms}, and the corresponding statistical parameters are given in Table~\ref{table:stat2}. 

The observed 3~GHz flux densities were converted into a brightness temperature defined via the Rayleigh-Jeans approximation, that is 

\begin{equation} 
\label{eq:TB}
T_{\rm B}=\frac{c^2}{2k_{\rm B}\nu^2}\frac{S_{\nu}}{\Omega}=1.22\times \left(\frac{\nu}{{\rm GHz}} \right)^{-2}\left(\frac{S_{\nu}}{\mu{\rm Jy}} \right)\left(\frac{\theta_{\rm maj}}{\arcsec} \right)^{-2}\,{\rm K}\,,
\end{equation} 
where $c$ is the speed of light, $k_{\rm B}$ is the Boltzmann constant, and the solid angle subtended by the Gaussian source was derived from 
$\Omega=\pi \theta_{\rm maj}^2/(4 \ln 2)=1.133\times \theta_{\rm maj}^2$. The derived values of $T_{\rm B}$ are listed in Col.~(13) 
in Table~\ref{table:results}. The uncertainties in $T_{\rm B}$ were derived from those associated with $S_{\rm 3\, GHz}$ and the average of the $\pm$ error of the 3~GHz major axis FWHM size. The distribution of the $T_{\rm B}$ values is shown in the bottom panel in Fig.~\ref{figure:histograms}, while the statistical parameters are given in Table~\ref{table:stat2}. To convert the observed-frame $T_{\rm B}$ to the source rest frame, a value obtained from Eq.~(\ref{eq:TB}) should be multiplied by the inverse of the cosmological scale factor, $a^{-1}(t)=(1+z)$. The spectral indices and brightness temprature properties of our SMGs are discussed further in Sect.~4.1.

\begin{table}
\renewcommand{\footnoterule}{}
\caption{1.4--3~GHz radio spectral index and 3~GHz brightness temperature statistics.}
{\scriptsize
\begin{minipage}{1\columnwidth}
\centering
\label{table:stat2}
\begin{tabular}{c c c}
\hline\hline 
Parameter & $\alpha_{\rm 1.4\, GHz}^{\rm 3\, GHz}$ & $T_{\rm B}$ [K]\\
\hline
Mean & \ldots\tablefootmark{a} & $23.2\pm1.6$ ($20.4\pm1.5$)\\
Median & $-0.67$ ($-0.67$)\tablefootmark{b} & $12.6\pm2.0$ ($12.3\pm1.9$)\\
Standard deviation & 0.51 (0.31) & 16.8 (15.7)\\ 
95\% confidence interval & \ldots\tablefootmark{a} & 20.1--26.3 (17.5--23.3)\\ 
IQR & $-0.94$\tablefootmark{c} [$-0.94-(-0.26)$] & 5.4--31.0 (5.1--28.3)\\
\hline 
\end{tabular} 
\tablefoot{For each parameter, we give a full sample value (based on 115 sources), while the values given in parentheses refer to a sample from which the three VLBA-detections were omitted.\tablefoottext{a}{The value could not be determined owing to a large number of right-censored data points (61, i.e. 53\% of the whole sample).}\tablefoottext{b}{The standard error of the median, which is proportional to the standard error of the mean (see Sect.~3.2), could not be determined owing to a large number of censored data points.}\tablefoottext{c}{Only the 25th percentile could be determined, while the third quartile could not be calculated owing to a large amount of right-censored data.}}
\end{minipage} 
}
\end{table}

\begin{figure}[!htb]
\centering
\resizebox{0.9\hsize}{!}{\includegraphics{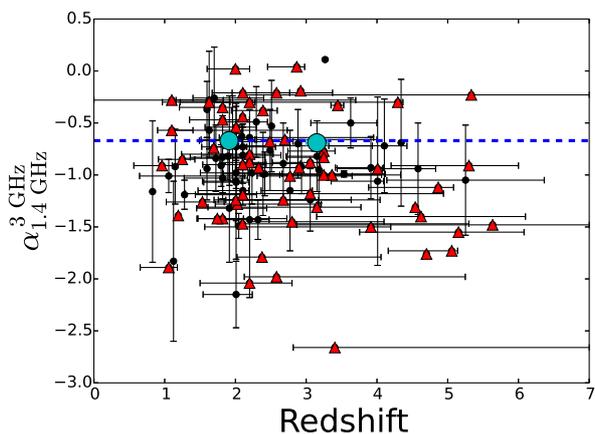}}
\caption{Radio spectral index between the observed-frame frequencies of 1.4~GHz and 3~GHz plotted as a function of redshift. The red up-pointing triangles show the lower $\alpha_{\rm 1.4\, GHz}^{\rm 3\, GHz}$ limits. The cyan filled circles represent the median values of the binned data with the error bars showing the corresponding standard errors. The lower redshift bin contains 57 SMGs, while the higher redshift bin contains 58 SMGs. The large number of censored $\alpha_{\rm 1.4\, GHz}^{\rm 3\, GHz}$ values allowed us to only halve the sample to apply survival analysis. The horizontal dashed line marks the median spectral index of $\alpha_{\rm 1.4\, GHz}^{\rm 3\, GHz}=-0.67$, which was calculated using a survival analysis technique.}
\label{figure:alphavsredshift}
\end{figure}

\begin{figure}[!htb]
\centering
\resizebox{0.9\hsize}{!}{\includegraphics{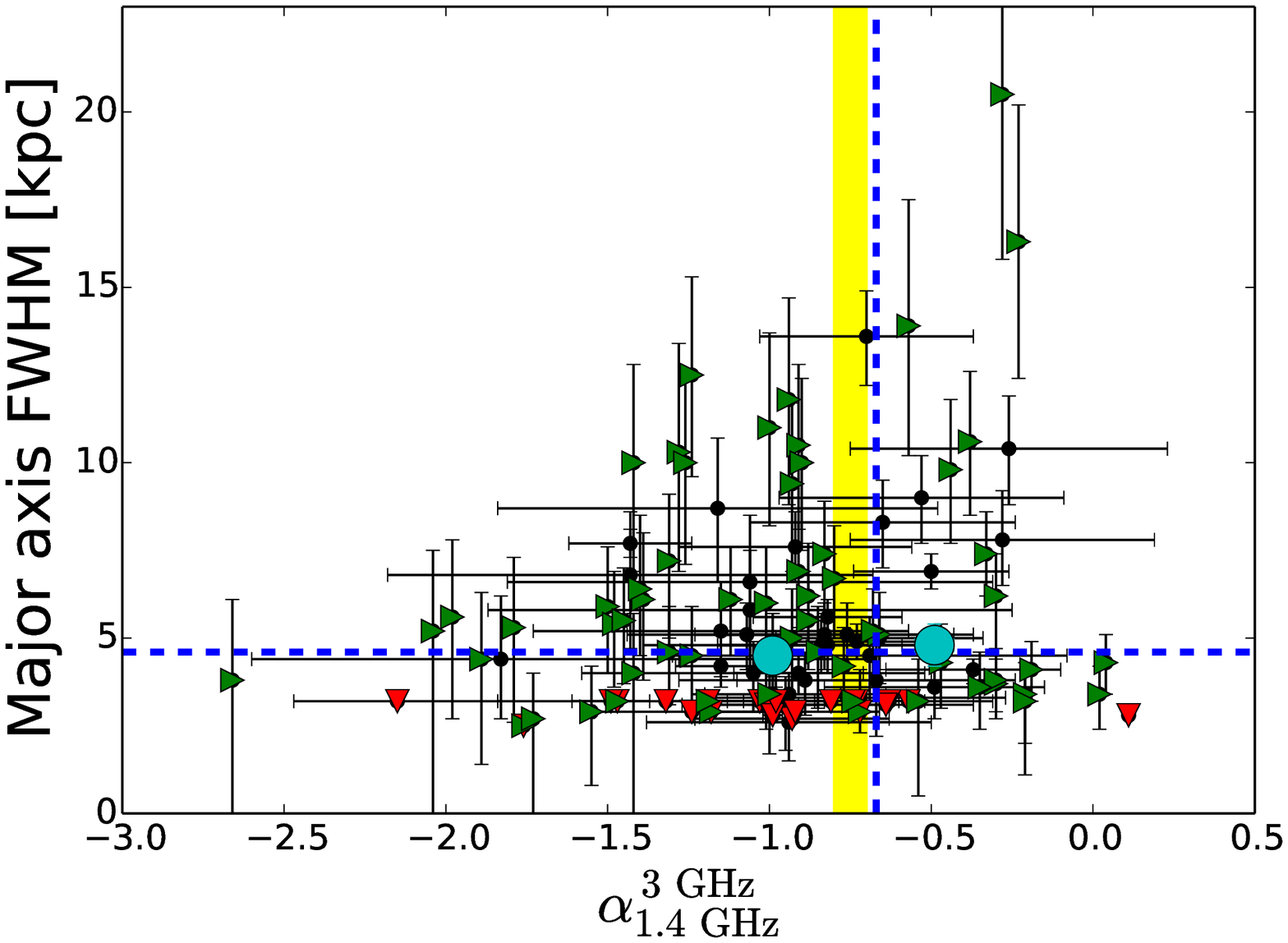}}
\resizebox{0.9\hsize}{!}{\includegraphics{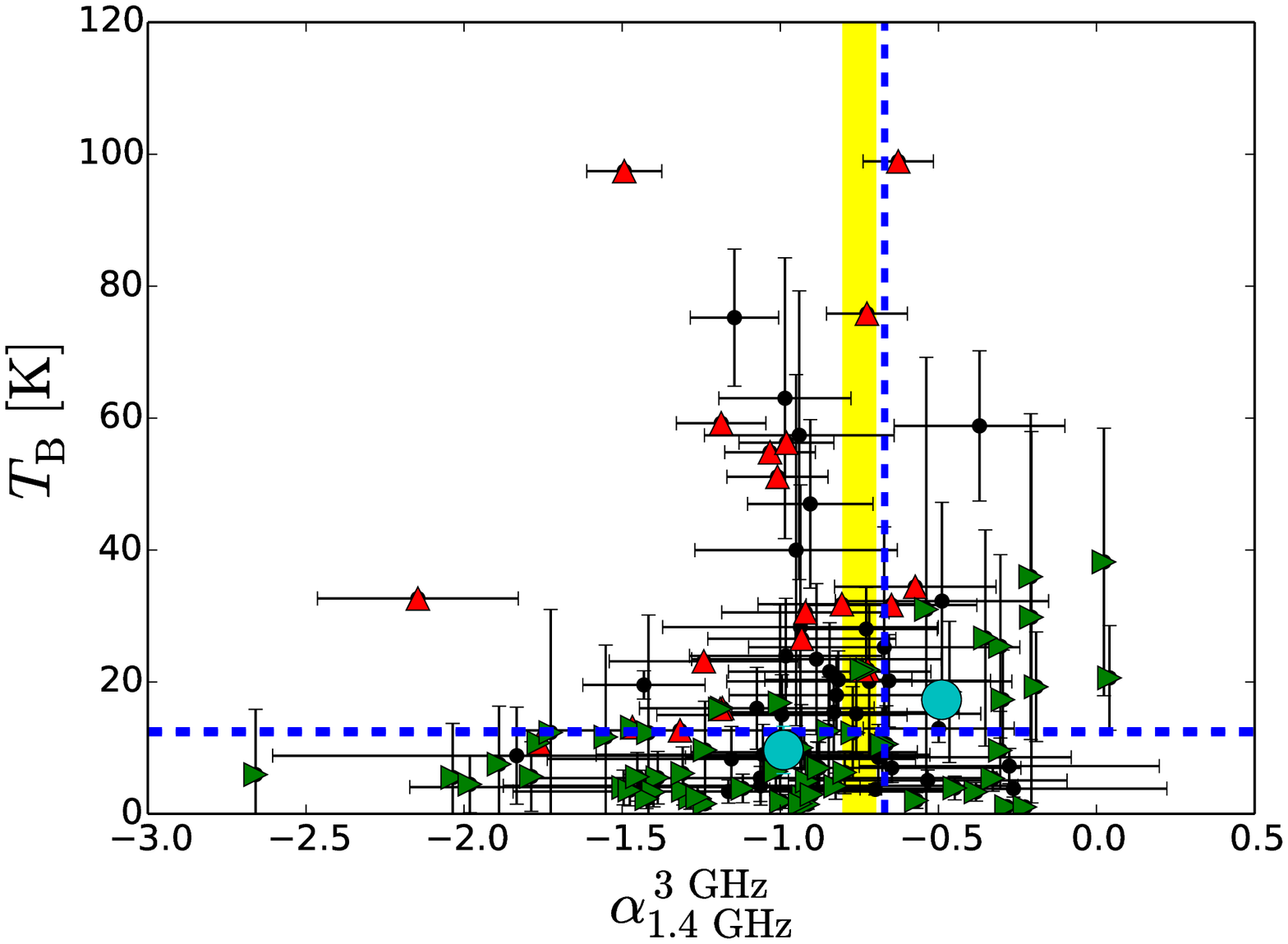}}
\caption{\textbf{Top:} Linear major axis FWHM size at 3~GHz plotted against the radio spectral index between the observed frequencies of 1.4~GHz and 3~GHz. The green right-pointing triangles indicate the lower $\alpha_{\rm 1.4\, GHz}^{\rm 3\, GHz}$ limits, while the red down-pointing triangles show the upper size limits. The cyan filled circles represent the median values of the binned data with the error bars showing the standard errors of the median values. The steeper spectral index bin contains 57 SMGs, while the flatter one contains 58 SMGs. The large number of censored data points allowed us to only halve the sample to apply survival analysis. The horizontal dashed line marks the median major axis FWHM size of our SMGs (4.6~kpc), and the vertical dashed line marks the sample median spectral index of $\alpha_{\rm 1.4\, GHz}^{\rm 3\, GHz}=-0.67$. For reference, the yellow shaded band shows the radio spectral index range of $\alpha_{\rm synch}\in[-0.8,\, -0.7]$, which is typical of the non-thermal synchrotron radio emission from star-forming galaxies. \textbf{Bottom:} 3~GHz brightness temperature as a function of $\alpha_{\rm 1.4\, GHz}^{\rm 3\, GHz}$. The symbols are as in the top panel, except that the red triangles pointing up indicate the lower $T_{\rm B}$ limits. The two partly overlapping horizontal dashed lines mark the median $T_{\rm B}$ values for the full sample (12.6~K) and for a sample from which the VLBA detected sources were removed (12.3~K). The VLBA detected high-$T_{\rm B}$ sources are not shown for the sake of clarity, but they are included in the binned data.}
\label{figure:alpha}
\end{figure}

\section{Discussion}

\subsection{Radio spectral indices between 1.4~GHz and 3~GHz, and 3~GHz brightness temperatures of the ASTE/AzTEC 1.1~mm selected submillimetre galaxies}

\subsubsection{Spectral indices}

At the redshifts of our sources, which range from $z_{\rm spec}=0.829$ for AzTEC/C71b to $z_{\rm FIR}=5.63$ for AzTEC/C106, the 3~GHz observations are probing the rest-frame frequencies of $\nu_{\rm rest}=5.5- 19.9$~GHz ($\lambda_{\rm rest}= 1.5-5.5$~cm). These rest-frame frequencies are expected to be dominated by non-thermal synchrotron emission with the contribution from thermal p$^+$e$^-$ free-free emission becoming increasingly important at higher frequencies, and starting to dominate at $\nu_{\rm rest}\simeq30$~GHz (e.g. \cite{condon1992}; \cite{murphy2012b}; \cite{marvil2015}; \cite{miettinen2017}).

The median 1.4--3~GHz spectral index we derived for our SMG sample, $\alpha_{\rm 1.4\, GHz}^{\rm 3\, GHz}=-0.67$, is consistent with a value expected for an optically thin ($\tau_{\nu} \ll 1$) non-thermal synchrotron radiation from an ensemble of CR electrons whose energy distribution has a power-law form of $N(E_{\rm e})\propto E_{\rm e}^p$ with $p = 2\alpha-1 \simeq -2.4$ (see e.g. \cite{reynolds1992}; \cite{deeg1993}; \cite{achterberg2001}; and e.g. \cite{leroy2011} and \cite{marvil2015} for observational data). This is a reassuring result given that more than half (53\%) of our $\alpha_{\rm 1.4\, GHz}^{\rm 3\, GHz}$ values are lower limits, which hampers our survival analysis to estimate the sample median. 

The spectral indices for our VLBA-detected SMGs are $\alpha_{\rm 1.4\, GHz}^{\rm 3\, GHz}=-1.15\pm0.14$ (C24b), $0.11\pm0.01$ (C61), and $-0.99\pm0.03$ (C77a). The spectral index for AzTEC/C24b and C77a is steeper than the classic synchrotron spectral index, which suggests that CR electrons in these sources have cooled through different energy-loss mechanisms (see Appendix~D). On the other hand, AzTEC/C61 exhibits a mildly inverted radio spectrum, which is yet another indication of the presence of an AGN; a rising radio spectrum is a characteristic of self-absorbed, black hole accretion-driven synchrotron jet component (e.g. \cite{blandford1979}; \cite{falcke1996}; \cite{nagar2000}).

\subsubsection{Brightness temperatures}

The median observed-frame 3~GHz brightness temperature we derived is $T_{\rm B}=12.6\pm2.0$~K for the full source sample, and $T_{\rm B}=12.3\pm1.9$~K for the sample from which the VLBA-detections were removed. However, the VLBA-detections stand out as having much higher brightness temperatures than the typical values found here, that is $T_{\rm B}=75.2\pm10.4$~K (AzTEC/C24b), $>10\,832$~K (C61), and $>243.9$~K (C77a). Because of the intermediate angular resolution of our 3~GHz observations, that is $0\farcs75$, the $T_{\rm B}$ values we derived should be interpreted as beam-averaged quantities, and the presence of a faint radio AGN in a low-$T_{\rm B}$ source cannot be excluded. However, it is likely that the observed 3~GHz radio emission from our SMGs is predominantly powered by SN activity (hence linked to star formation) with the exception of the VLBA and X-ray detected SMG AzTEC/C61, where the very high observed (rest-frame) brightness temperature of $>10^{4.03}$~K ($>10^{4.66}$~K) together with a positive spectral index between the rest-frame frequencies of 6.0~GHz and 12.8~GHz suggest an AGN-powered radio emission (e.g. \cite{condon1991}; \cite{murphyetal2013}; \cite{barcos2015}). %We note, however, that the absence of a high brightness temperature radio core does not rule out the presence of an AGN.

\subsubsection{Searching for correlations}

In Fig.~\ref{figure:alphavsredshift}, we plot the 1.4--3~GHz radio spectral indices as a function of redshift. The cyan filled circles shown in the plot represent the binned version of the data. Because of the large percentage of lower spectral index limits (53\%), we could split the full sample into only two parts, and compute the corresponding median values using a method of survival analysis described in Sect.~3.2. In the lower redshift bin ($z=1.91\pm0.06$), the median spectral index is $\alpha_{\rm 1.4\, GHz}^{\rm 3\, GHz}=-0.67\pm 0.06$, which equals the full sample median (and the one purified from VLBA detections), and in the higher redshift bin ($z=3.15\pm 0.15$) the value is very similar, namely $\alpha_{\rm 1.4\, GHz}^{\rm 3\, GHz}= -0.69\pm0.08$. The latter value remains practically unchanged if the flat spectral index source AzTEC/C61 is excluded from the analysis ($\alpha_{\rm 1.4\, GHz}^{\rm 3\, GHz}= -0.70\pm0.03$). 
Hence, we see no evolution in the median spectral index as a function of redshift. In part, this absence of evolution in the spectral index might reflect the fact that the redshift interval between the two binned data points is fairly narrow, only $\Delta z = 3.15- 1.91=1.24$, where no significant evolution is expected at these redshifts. We note, however, that because the rest-frame frequency probed is higher at higher redshifts, the contribution of the thermal free-free emission to the observed radio emission is expected to be higher (causing flattening of the spectral index). On the other hand, this effect could be competing with a boosted CR cooling at high redshifts (steepening of the spectral index; see Appendix~D.2), which can lead to an apparent flat trend between the radio spectral index and redshift.

Local luminous ($L_{\rm IR}>10^{11}$~L$_{\sun}$ in the rest-frame $8-1\,000$~$\mu$m) and ultraluminous ($L_{\rm IR}>10^{12}$~L$_{\sun}$) infrared galaxies, or (U)LIRGs, exhibit a trend of more compact radio emitters having a flatter radio spectral index (\cite{condon1991}; \cite{murphyetal2013}). The physical interpretation of this trend in these studies was that an increased photon absorption by free electrons, that is free-free absorption, becomes more important in more compact sources. For a given mass of ionised gas, this can be understood as an increasing free-free optical thickness, $\tau_{\nu}^{\rm ff}\propto n_{\rm e}^2$, where $n_{\rm e}$ is the number density of free thermal electrons. 

To see if our SMGs exhibit the aforementioned trend, we plot the 3~GHz linear major axis FWHM sizes as a function of $\alpha_{\rm 1.4\, GHz}^{\rm 3\, GHz}$ in the top panel in Fig.~\ref{figure:alpha}. No correlation is visible in the plot, and the binned censored data suggest that, on average, the radio size remains constant over the observed spectral index range. 

Moreover, in the bottom panel in Fig.~\ref{figure:alpha} we plot the 3~GHz $T_{\rm B}$ values as a function of $\alpha_{\rm 1.4\, GHz}^{\rm 3\, GHz}$. Because $T_{\rm B} \propto \Omega^{-1}$ (see Eq.~(\ref{eq:TB})), one would expect to see flatter spectral indices for sources with higher brightness temperature as has been seen among local (U)LIRGs in accordance with their radio size-spectral index anticorrelation (\cite{murphyetal2013}). Although this is not clearly discernible in our data, the binned data exhibit a hint that the flattest $\alpha_{\rm 1.4\, GHz}^{\rm 3\, GHz}$ values are found among sources with elevated $T_{\rm B}$. We note that this is not driven by the high-$T_{\rm B}$ and inverted-spectrum VLBA source AzTEC/C61; the median $T_{\rm B}$ value of the flatter $\alpha_{\rm 1.4\, GHz}^{\rm 3\, GHz}$ bin is $17.3\pm2.2$~K whether AzTEC/C61 is included or not. The steeper $\alpha_{\rm 1.4\, GHz}^{\rm 3\, GHz}$ bin has a median $T_{\rm B}$ value of $9.7\pm3.6$~K, and hence close to the flatter $\alpha_{\rm 1.4\, GHz}^{\rm 3\, GHz}$ bin's $T_{\rm B}$ value within the $1\sigma$ ranges. This statistically non-significant (or only marginally significant) behaviour is consistent with the finding that we do not see an anticorrelation between the radio size and $\alpha_{\rm 1.4\, GHz}^{\rm 3\, GHz}$ in the upper panel in Fig.~\ref{figure:alpha}. Although the calculation of the median bins suffers from the large amount of censored data, the lack of clear correlations in Fig.~\ref{figure:alpha} is consistent with the M15 study of 1.1~mm selected AzTEC SMGs in COSMOS.

\subsection{Radio-emitting sizes of submillimetre galaxies, and comparison with the spatial extent of dust, molecular gas, and stellar emission}

The median 3~GHz radio-emitting size (deconvolved major axis FWHM) we derived is $4.6 \pm 0.4$~kpc. This is consistent with the median 3~GHz radio size of $4.1\pm 0.8$~kpc derived by M15 for a much smaller (by a factor of 7.7), but partly overlapping sample of AzTEC SMGs in COSMOS (the quoted median size was scaled to the present cosmology and the revised redshifts from \cite{brisbin2017}). 

In Fig.~\ref{figure:sizes}, we show the derived 3~GHz radio size distribution of our 1.1~mm selected, 
ALMA 1.3~mm detected SMGs (cf.~Fig.~6 in M15). For comparison, we plot the 1.4~GHz radio size distribution of the SMGs from Biggs \& Ivison (2008), 1.1~mm and 870~$\mu$m dust emission sizes from Ikarashi et al. (2015), Hodge et al. (2016), and Simpson et al. (2016), the CO-emitting sizes from Tacconi et al. (2006), and the stellar emission sizes from Chen et al. (2015). As illustrated in Fig.~\ref{figure:sizes}, the present SMG sample size is significantly larger than those in the aforementioned previous works (by factors of 1.5 to 16). In the following subsections, we compare the derived radio sizes with those determined at other wavelengths and for other types of emissions in the previous studies.

\subsubsection{Comparison with the 1.4~GHz radio sizes from Biggs \& Ivison (2008)}

As described in more detail in M15, Biggs \& Ivison (2008) based their radio size study on 1.4~GHz observations taken with the VLA and the Multi-Element Radio Linked Interferometer Network (MERLIN). The angular resolution of their combined VLA plus MERLIN data set was about $0\farcs52 \times 0\farcs48$. The median 1.4~GHz radio size for the Biggs \& Ivison (2008) SMG sample is $6.1\pm1.1$~kpc, which is $1.3\pm0.3$ times larger than our 3~GHz median size. This discrepancy is not statistically significant owing to the standard errors of the sample medians. The Biggs \& Ivison (2008) SMGs lie at spectroscopic redshifts of $z_{\rm spec}=1.147-2.689$, and have single-dish 1.1~mm flux densities of $S_{\rm 1.1~mm}=1.4^{+0.6}_{-0.7}-6.0^{+1.4}_{-1.4}$~mJy (computed by assuming $\beta=1.5$; see M15). The deboosted ASTE/AzTEC 1.1~mm flux densities of our target SMGs lie in the range of $S_{\rm 1.1~mm}=3.5^{+1.1}_{-1.1}-13.0^{+1.1}_{-1.0}$~mJy (\cite{aretxaga2011}), and the majority of them (109/129 or 84.5\%) fall in the flux density range studied by Biggs \& Ivison (2008). Moreover, because the average radio size does not appear to depend on the millimetre flux density (Fig.~\ref{figure:fluxsize}, bottom panel), the aforementioned size comparison is justified in terms of source brightness. If we limit our SMG sample to the redshift range examined by Biggs \& Ivison (2008), we end up with a subsample of 65 SMGs with a median 3~GHz radio size of $4.6\pm0.4$~kpc, which is identical to that of our full sample.

A two-sided K-S test between our uncensored and Biggs \& Ivison (2008) radio size distributions in a common redshift interval yields a K-S test statistic of $D_{\rm KS}=0.25$ and a K-S probability of $p_{\rm KS}=0.71$ (cf.~Sect.~3.2). Hence, it seems possible that the two size distributions share a common underlying parent distribution.

\subsubsection{Comparison with the rest-frame far-infrared dust-emitting sizes revealed by ALMA}

The distribution of the ALMA 1.1~mm sizes of SMGs derived by Ikarashi et al. (2015) has a median FWHM value of only $1.4\pm0.3$~kpc, which is $3.3\pm0.8$ times smaller than our median radio size. As we already described in M15, Ikarashi et al. (2015) studied sources with ALMA 1.1~mm flux densities in the range of $S_{\rm 1.1~mm}=(1.23\pm0.07)-(3.45\pm0.10)$~mJy. Assuming that $\beta=1.5$, this corresponds to a 1.3~mm flux density range of $S_{\rm 1.3~mm}=(0.69\pm0.04)-(1.92\pm0.06)$~mJy, while that of our 3~GHz detected SMGs is $S_{\rm 1.3~mm}=(0.54\pm0.10)-(7.24\pm0.10)$~mJy. Seventy, or 61\%, of our 3~GHz detected SMGs have a ALMA millimetre flux density in the range examined by Ikarashi et al. (2015), and hence most of the sources are comparably bright. This, together with the finding that the average radio size does not depend on the ALMA millimetre flux density (Fig.~\ref{figure:fluxsize}, bottom panel), makes the aforementioned size comparison meaningful. Of the 13 SMGs studied by Ikarashi et al. (2015), only three have a photometric redshift available, while for the remaining ten sources the authors derived a lower redshift limit by assuming a uniform redshift probability at $z=3-6$ (see their Table~1). For this reason, the $\lambda_{\rm obs}=1.1$~mm physical sizes of these sources should be taken as approximate values. Ikarashi et al. (2017) revisited the 1.1~mm sizes of two of the SMGs from Ikarashi et al. (2015) by employing additional ALMA observations that covered baselines up to $1\,500$~k$\lambda$. The sizes measured in the $uv$-plane were found to be consistent with their previous measurements. Moreover, for the source that was consistent with a single symmetric Gaussian model, the size measured in the image plane was found to be consistent with that measured from the visibility data. 

For comparison, Gonz{\'a}lez-L{\'o}pez et al. (2017) detected 12 dusty star-forming galaxies at a ${\rm S/N}\geq5$ with ALMA at 1.1~mm towards three strong-lensing galaxy clusters from the Frontier Fields Survey. The demagnified (intrinsic) flux densities of these ALMA sources determined through two-dimensional elliptical Gaussian model fits in the $uv$-plane were found to be about $S_{\rm 1.1~mm}^{\rm demag}=0.14-1.63$~mJy ($S_{\rm 1.3~mm}^{\rm demag}=0.08-0.91$~mJy for $\beta=1.5$), and hence they are generally fainter than our SMGs (only 11 of our 3~GHz detected SMGs lie in the aforementioned flux density interval). From the intrinsic effective radii and the axial ratios reported by the authors in their Table~5, we calculated the angular major axis sizes of $0\farcs04-0\farcs41$ with a median of $0\farcs23\pm0\farcs05$. This is very similar to the median angular FWHM size at 1.1~mm from Ikarashi et al. (2015), that is $0\farcs22\pm0\farcs03$.

Simpson et al. (2015) based their study of the dust-emitting sizes of SCUBA-2 (Submillimetre Common User Bolometer Array 2) 850~$\mu$m selected SMGs in the Ultra Deep Survey (UDS) field on high-resolution ($0\farcs35 \times 0\farcs25$) 870~$\mu$m ALMA observations. Their target SMGs have single-dish 1.1~mm flux densities of $S_{\rm 1.1~mm}=3.2-6.5$~mJy (assuming $\beta=1.5$; see M15). The vast majority of our target SMGs (113 or 88\% out of the 129 single-dish AzTEC detections) lie within this flux density range, and hence our full, flux-limited sample is well-suited for a comparison with the S15 SMGs. For a subsample of their most significant ALMA detections (23 of the AS2UDS SMGs with ${\rm S/N}_{\rm 870\,\mu m}>10$), S15 derived a median deconvolved major axis FWHM of $0\farcs30\pm0\farcs04$. Simpson et al. (2016) derived photometric redshifts for 18 of these sources, and using a survival analysis (to take the upper size limit for UDS392.0 into account) we derived a median deconvolved major axis FWHM of $2.7\pm0.4$~kpc. This is $1.9\pm0.5$ times larger than the Ikarashi et al. (2015) sample median size, but still $1.7\pm0.3$ times smaller than our median radio size. 

Hodge et al. (2016) investigated the dust-emitting sizes of the ALMA 870~$\mu$m detected SMGs in the Extended \textit{Chandra} Deep Field South (ECDFS), that is the so-called ALESS SMGs (\cite{hodge2013}; \cite{karim2013}). The authors employed very high resolution ($0\farcs17 \times 0\farcs15$) ALMA Band~7 ($\lambda_{\rm obs}=870$~$\mu$m) follow-up observations. For the 16 sources that they detected at a ${\rm S/N}_{\rm 870\,\mu m}>10$, the median major axis FWHM size derived through two-dimensional Gaussian fitting in the image plane is $3.1\pm0.3$~kpc (scaled to our cosmology). This is $1.5\pm0.2$ times smaller than our median radio size, but very similar to the aforementioned median FWHM size of the 870~$\mu$m detected AS2UDS SMGs (the ratio between the two is $1.1\pm0.2$). Hodge et al. (2016) also measured the source sizes in the $uv$-plane, and on the basis of this they concluded that the image plane sizes can be considered robust (also no evidence of emission being resolved out was found). For comparison, Lindroos et al. (2016) stacked the $u-v$ visibility data of the ALESS SMGs, and derived an average FWHM size of $0\farcs40 \pm 0\farcs06$ through a Gaussian model fit. At the median redshift of the ALESS SMGs, $z=2.3$ (\cite{simpson2014}), this corresponds to $3.3\pm0.5$~kpc, which is in excellent agreement with the Hodge et al. (2016) results, regardless of the fact that Lindroos et al. (2016) used the original ALESS data of $1\farcs6 \times 1\farcs2$ angular resolution, which is sufficient to only marginally resolve the largest sources.

To allow us to make a fairer comparison with the ALESS sample from Hodge et al. (2016), we limit their sample to those SMGs that have 870~$\mu$m flux densities corresponding to our AzTEC 1.1~mm flux density range in the parent sample (AzTEC/C1--C129; $3.5~{\rm mJy}\leq S_{\rm 1.1\, mm} \leq13.0$~mJy). Under the assumption that $\beta=1.5$, this flux density range is $8.0~{\rm mJy}\leq S_{\rm 870\, \mu m} \leq29.5$~mJy. The original Large APEX BOlometer CAmera (LABOCA) 870~$\mu$m detected SMGs, or the so-called LESS SMGs (\cite{weiss2009}) that fall in this flux density range are LESS~1--16, 22, 23, 30, and 35, where LESS~1, 2, 3, 7, 15, 22, 23, and 35 were resolved into multiple components with ALMA (\cite{hodge2013}; \cite{karim2013}). Hodge et al. (2016) derived a dust-emitting size for seven of these sources, and the median size for these is $2.9\pm0.6$~kpc, which is consistent with their full sample and $1.6\pm0.4$ times smaller than our median radio size. Moreover, the observed 1.4~GHz flux densities of the Hodge et al. (2016) target sources, which range from $S_{\rm 1.4\, GHz}<24$~$\mu$Jy to 90~$\mu$Jy (see \cite{thomson2014}; Table~3 therein), fall within the 1.4~GHz flux density range of our 3~GHz detections ($S_{\rm 1.4\, GHz}<36$~$\mu$Jy--10.59~mJy, or up to 554.5~$\mu$Jy if AzTEC/C61 is not considered). Also from this standpoint, together with the finding that the average size at 3~GHz does not depend on the 3~GHz flux density (Fig.~\ref{figure:fluxsize}, top panel), a direct size comparison with the ALESS sources appears to be justified. Finally, we point out that four of the ALESS SMGs studied by Hodge et al. (2016) are associated with an X-ray source (ALESS~17.1, 45.1, 67.1, and 73.1; \cite{wang2013}), and their dust-emitting sizes were found to be similar to the other sources, which is similar to our conclusion concerning the radio sizes of AGN-host SMGs and their pure star-forming counterparts (Sect.~3.3). 

\subsubsection{Comparison with the CO-emitting sizes from Tacconi et al. (2006)}

As can be seen in Fig.~\ref{figure:sizes}, the CO-emitting sizes of SMGs 
derived through observations of the $J=3-2$ and $J=7-6$ rotational lines by Tacconi et al. (2006) 
have a median value ($4.0\pm1.0$~kpc) that is fairly similar to that of our radio 
sizes; the ratio between the two is $0.9\pm0.2$, that is consistent with unity. As we described in M15, the single-dish 1.1~mm flux densities of the Tacconi et al. (2006) SMGs, again assuming that $\beta=1.5$, are $S_{\rm 1.1~mm}\simeq3.3-4.3$~mJy. Half (64/129 or 49.6\%) of our target AzTEC detections lie within this flux density range. However, as explained in Sect.~1, the CO lines employed by Tacconi et al. (2006) are probing the dense, warm phase of the molecular gas component, while lower excitation $J_{\rm up}\leq2$ transitions have revealed more extended, diffuse, and less actively star-forming molecular gas reservoirs in SMGs. Another indication of different molecular gas phases in SMGs is provided by the finding that the $J_{\rm up}\geq2$ CO lines are subthermally excited (see \cite{bothwell2013}). That the spatial scales of radio emission and warm CO gas appear to be similar on average might be an indication that the dense gas, which is shielded from the interstellar UV radiation field, is heated by the hadronic component of CRs (protons) via ionisation (e.g. \cite{goldsmith1978}), while the dust heating by radiation from young stars is restricted to the more compact central parts of the galaxy.  

\subsubsection{Comparison with the size scale of the rest-frame ultraviolet stellar emission}

To investigate how the 3~GHz radio sizes of our SMGs compare with their stellar emission sizes, we retrieved the 
\textit{Hubble}/Advanced Camera for Surveys (ACS) $I$-band sizes from the publicly available COSMOS morphological catalogues\footnote{The 
catalogues are available through the NASA/IPAC Infrared Science Archive (IRSA) at {\tt http://irsa.ipac.caltech.edu}.}. We made use of all the five available catalogues, namely the Cassata morphological catalogue, which contains 232\,022 galaxies down to $I_{\rm AB}({\rm ACS})=25$~mag (\cite{cassata2007}); the Tasca morphological catalogue, which contains 237\,914 galaxies with $I_{\rm AB}({\rm ACS})<24.5$~mag (\cite{tasca2009}); the COSMOS/ACS morphological catalogue from the Columbia University group (Zamojski morphology catalogue; \cite{zamojski2007}), which contains 40\,666 galaxies with $I_{\rm AB}({\rm ACS})<23$~mag; the Z{\"u}rich structure \& morphology catalogue v1.0, which contains 131\,532 galaxies (measurements down to $I_{\rm AB}=24$~mag (\cite{scarlata2007}), and to $I_{\rm AB}=22.5$~mag (\cite{sargent2007})); and the COSMOS 2005 morphology catalogue, which contains 157\,610 galaxies, and where the faintest galaxy has a peak surface brightness of $I_{\rm AB}=23.9$~mag. 

We used a search radius of $1\arcsec$ with respect to the ALMA 1.3~mm source positions, and found 28, 29, 4, 15, and 18 matches with the Cassata, Tasca, Zamojski, 
Z{\"u}rich, and COSMOS catalogues, respectively. The adopted ACS/$I$-band sizes were those corresponding to the half-light radii, and the catalogue values in pixels were 
converted to arcseconds ($0\farcs03$~pixel$^{-1}$ or $0\farcs05$~pixel$^{-1}$ depending on the catalogue). Following Zahid et al. (2015), from the Z{\"u}rich catalogue, 
we took the semi-major axis length of an ellipse encompassing 50\% of total light ($a_{50}$), and circularised it to obtain the half-light radius using the formula 

\begin{equation} 
R_{1/2}=a_{50} \times \sqrt{\frac{b}{a}}\,,
\end{equation}
where $b/a$ is the minor-to-major axis ratio. On average, the circularised half-light radii were found to be 1.43 times smaller than the values of $a_{50}$.

The catalogue-based angular half-light radii are tabulated in Table~\ref{table:optical}. In the case a given source was found in more than one of the aforementioned catalogues, we adopted the average value of the reported half-light radii. At the redshifts of the 33 ACS/$I$-band detected sources listed in Table~\ref{table:optical}, $z= 1.06-4.68$, the observed mean wavelength of $\lambda_{\rm obs}=814$~nm corresponds to the rest-frame wavelengths of $\lambda_{\rm rest}\simeq 143-395$~nm, that is far-UV to near-UV radiation. For six additional SMGs, AzTEC/C2a, C4, C5, C10b, C17, and C42, we could obtain the observed-frame near-IR sizes derived by Toft et al. (2014; see also M15). For AzTEC/C42, the authors used the \textit{Hubble}/Wide Field Camera 3 (WFC3) $H_{160}$-band (mean $\lambda_{\rm obs}=1.54$~$\mu$m) observations, which probe the rest-frame near-UV ($\lambda_{\rm rest}\simeq 333$~nm) at the source redshift. For the other five sources they employed the stacked $Y$, $J$, $H$, and $K_{\rm s}$-band images from the UltraVISTA survey (mean observed central wavelength of $\lambda_{\rm obs}=1.52$~$\mu$m), which are probing the rest-frame mid-UV to near-UV radiation at the source redshifts of $z=2.9-5.3$.

In Fig.~\ref{figure:optradio}, we plot the rest-frame UV sizes (parameterised as ${\rm FWHM}=2\times R_{1/2}$) against the 3~GHz radio size (major axis FWHM). As illustrated by this plot, we see both the cases of radio emission being more extended ($16/36=44.4\%$) or more compact ($20/36=55.6\%$) than the spatial scale of stellar emission. One caveat to this size comparison is that the assumed relationship 
of ${\rm FWHM}=C(n)\times R_{1/2}$ with $C(n)=2$ is strictly valid only for a circular Gaussian profile with a S\'ersic index of $n=0.5$, while the numerical factor is $C(n)=0.83$ for an exponential profile ($n=1$) and only $C(n)=1.33\times10^{-4}$ for the de Vaucouleurs $n=4$ profile (e.g. \cite{voigt2010}). Because the S\'ersic index values of the rest-frame UV size measurements reported in Table~\ref{table:optical} are generally not available (only for AzTEC/C52, 59, and 66 the values $n=6.4^{+0.9}_{-0.7}$, $n=0.4^{+0.2}_{-0.2}$, and $n=1.0^{+0.2}_{-0.3}$ were reported in the Z{\"u}rich catalogue), we restricted our calculation to the simplified assumption that $C(n)=2$. 

\begin{figure}[!htb]
\centering
\resizebox{\hsize}{!}{\includegraphics{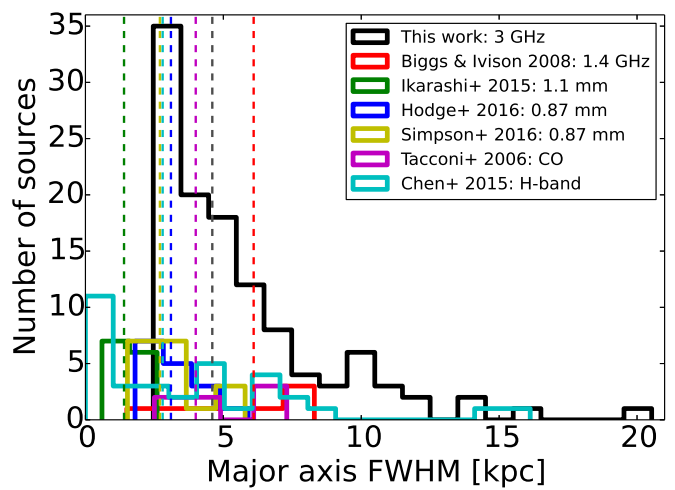}}
\caption{Distributions of the SMG sizes (major axis FWHM) measured in radio, dust, CO, and stellar emissions. The black histogram shows the sizes of our COSMOS ASTE/AzTEC SMGs as seen at $\nu_{\rm obs}=3$~GHz. The red, green, blue, yellow, magenta, and cyan histograms show the 1.4~GHz sizes from Biggs \& Ivison (2008), 1.1~mm (rest-frame FIR) sizes from Ikarashi et al. (2015), 870~$\mu$m sizes from Hodge et al. (2016) and Simpson et al. (2016), CO-emitting sizes from Tacconi et al. (2006), and the stellar emission sizes from Chen et al. (2015). The bin size is 1~kpc. The upper size limits were placed in the bins corresponding to those values, which is causing the apparent peak in our radio size distribution. The vertical dashed lines show the corresponding median sizes (4.6~kpc for our SMGs, 6.1~kpc for the Biggs \& Ivison (2008) SMGs, 1.4~kpc for the Ikarashi et al. (2015) SMGs, 3.1~kpc and 2.7~kpc for the Hodge et al. (2016) and Simpson et al. (2016) SMGs, respectively, 4.0~kpc for the CO sizes from Tacconi et al. (2006), and 2.8~kpc for the stellar sizes from Chen et al. (2015); survival analysis was used to take the upper size limits into account when calculating the median sizes). See text for details.}
\label{figure:sizes}
\end{figure}

\begin{figure}[!htb]
\centering
\resizebox{0.9\hsize}{!}{\includegraphics{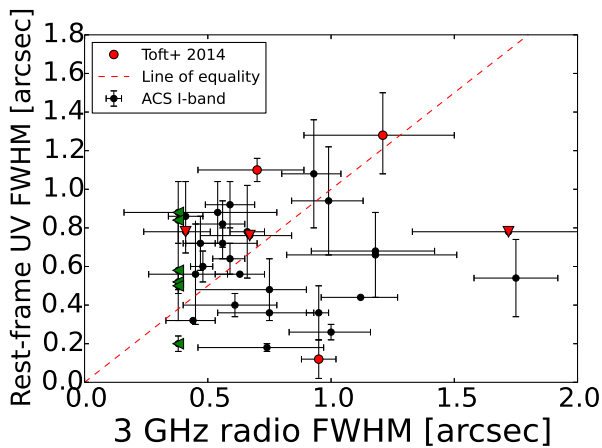}}
\caption{Rest-frame UV size (FWHM, which was assumed to be two times the half-light radius) plotted against the 3~GHz major axis FWHM. The UV sizes were derived from the morphology catalogues (see text for details) except those indicated by red data points, which were derived by Toft et al. (2014). The downwards pointing triangles indicate upper limits to the UV sizes, while the left-pointing triangles indicate the upper radio size limits. The red dashed line is the one-to-one relationship.}
\label{figure:optradio}
\end{figure}

In Fig.~\ref{figure:sizes}, we also plot the rest-frame UV-optical stellar emission sizes of the so-called MAIN ALESS SMGs derived by Chen et al. (2015) using the \textit{Hubble}/WFC3 $H_{160}$-band observations. Instead of all the 75 $H_{160}$-band sources listed by Chen et al. (2015) in their Table~1, we only considered the 35 resolved main $H_{160}$-band components because they are most clearly associated with the ALESS SMGs (the H1 components in \cite{chen2015}; see their Fig.~10). The FWHM sizes plotted in Fig.~\ref{figure:sizes} were calculated by using the effective, semi-major axis lengths (half-light radii) and the S\'ersic indices reported by Chen et al. (2015; Table~1 therein), and following the method outlined in Voight \& Bridle (2010). We found a wide range of the $C(n)={\rm FWHM}/R_{1/2}$ values for these sources, ranging from $1.2 \times 10^{-4}$ to 2.9 with a mean (median) of 0.98 (0.83), where the median S\'ersic index was derived to be $n=1$. The median stellar emission FWHM extent we derived is $2.8\pm0.9$~kpc, which is $1.6\pm0.5$ times smaller than our median radio size. We note that for one of the analysed ALESS sources, ALESS~039.1, the S\'ersic index was too small ($n=0.1$) to yield a physical FWHM size solutions (${\rm FWHM}\propto k^{-n}$, where $k=1.9992\times n - 0.3271$; \cite{voigt2010}), and hence the effective sample size considered here is 34. Among the ALESS SMGs that were detected towards those LESS sources that are equally bright to our AzTEC 1.1~mm sources (see above), Chen et al. (2015) derived stellar sizes for seven main $H_{160}$-band sources. For these sources, the median $H_{160}$-band FWHM size is $1.7\pm1.0$~kpc, which is $2.7\pm1.6$ times smaller than our median radio size. Consistent with our aforementioned source-by-source radio-UV size comparison, the stellar emission sizes of the ALESS SMGs span a wide range of FWHM values, from only 0.2~pc to 16.1~kpc (1.1~pc to 5.0~kpc for the flux-limited sample), and can be either larger or smaller than the radio-emitting region.  

Although the median $H_{160}$-band FWHM size of the Chen et al. (2015) sample is comparable to the typical size of the dust-emitting region, in some cases the size scale of stellar emission appears to be more extended than the dust-emitting region, similar to that found for the radio emission. The latter situation can be explained by a pre-existing, older stellar component, which extends beyond the (central) region of active, obscured star formation. For example, a merger remnant system is expected to host a central starburst region, surrounded by a more extended distribution of stars (e.g. \cite{wuyts2010}). However, measurements of the spatial scale of stellar emission of SMGs can be hampered by a strong, and possibly differential dust extinction in the rest-frame UV regime (e.g. \cite{swinbank2010}), which is also sensitive to the viewing angle and source geometry. In addition, the stellar population effects can influence the apparent stellar emission size scale (e.g. older stars are redder than hot, young stars; e.g. \cite{searle1973}), and depending on the source redshift the observed wavelength can fall either blueward or redward of the Balmer break at $3\,646~\AA$ or the $4\,000~\AA$ break. If a galaxy exhibits a colour gradient, its apparent size depends on the wavelength of the observations, and for example a galaxy with a redder central region than the surrounding disk would appear more compact at longer (redder) observed wavelengths (e.g. \cite{wellons2015}).

\begin{table*}[!htb]
\renewcommand{\footnoterule}{}
\caption{Radii in the rest-frame UV for a subset of 39 target SMGs.}
{\small
\begin{minipage}{2\columnwidth}
\centering
\label{table:optical}
\begin{tabular}{c c c c c c c c c}
\hline\hline 
ID & CC\tablefootmark{a} & TC\tablefootmark{b} & ZC\tablefootmark{c} & ZSMC\tablefootmark{d} & C2005\tablefootmark{e} & T14\tablefootmark{f} & $r_{\rm UV}$\tablefootmark{g} & $r_{\rm radio}/r_{\rm UV}$\tablefootmark{h} \\
   & $R_{1/2}$ [\arcsec] & $R_{1/2}$ [\arcsec] & $R_{1/2}$ [\arcsec] & $R_{1/2}$ [\arcsec] & $R_{1/2}$ [\arcsec] & $R_{1/2}$ [\arcsec] & [\arcsec] & \\
\hline
C2a & \ldots & \ldots & \ldots & \ldots & \ldots & $<0.39$ & $<0.39$ & $>0.31$ \\[1ex]
C4 & \ldots & \ldots & \ldots & \ldots & \ldots & $<0.39$ & $<0.39$ & $>1.71$ \\[1ex]
C5 & \ldots & \ldots & \ldots & \ldots & \ldots & $<0.38$ & $<0.38$ & $>0.62$ \\[1ex]
C9b & 0.37 & 0.17 & \ldots & \ldots & \ldots & \ldots & $0.27\pm0.10$ & $3.19^{+2.46}_{-1.05}$ \\[1ex]
C10b & \ldots & \ldots & \ldots & \ldots & \ldots & $0.64^{+0.11}_{-0.10}$ & $0.64^{+0.11}_{-0.10}$ & $0.95^{+0.44}_{-0.36}$ \\[1ex]
C12 & 0.40 & 0.32 & \ldots & \ldots & \ldots & \ldots & $0.36\pm0.04$ & $0.78^{+0.31}_{-0.28}$ \\[1ex]
C17 & \ldots & \ldots & \ldots & \ldots & \ldots & $0.55\pm0.03$ & $0.55\pm0.03$ & $0.64^{+0.22}_{-0.24}$ \\[1ex]
C18 & 0.28 & \ldots & \ldots & \ldots & \ldots & \ldots & 0.28 & $1.13^{+0.53}_{-0.54}$ \\[1ex]
C19 & 0.65 & 0.34 & 0.34 & 0.36 & 0.35 & \ldots & $0.41\pm0.06$ & $0.68^{+0.25}_{-0.20}$ \\[1ex]
C22a & 0.43 & 0.26 & \ldots & 0.25 & 0.28 & \ldots & $0.30\pm0.04$ & $0.80^{+0.20}_{-0.17}$ \\[1ex]
C23 & 0.28 & 0.36 & \ldots & \ldots & \ldots & \ldots & $0.32\pm0.04$ & $0.92^{+0.24}_{-0.20}$ \\[1ex]
C25 & \ldots & 0.22 & \ldots & \ldots & \ldots & \ldots & 0.22 & $2.55^{+0.34}_{-0.37}$ \\[1ex]
C28a & 0.25 & 0.11 & \ldots & \ldots & \ldots & \ldots & $0.18\pm0.07$ & $2.64^{+1.36}_{-0.84}$ \\[1ex]
C33a & \ldots & \ldots & \ldots & \ldots & 0.16 & \ldots & 0.16 & $1.38^{+0.28}_{-0.35}$ \\[1ex]
C36 & 0.67 & 0.22 & \ldots & 0.46 & 0.38 & \ldots & $0.43\pm0.09$ & $0.48^{+0.23}_{-0.15}$ \\[1ex]
C42 & \ldots & \ldots & \ldots & \ldots & \ldots & $0.06\pm0.05$ & $0.06\pm0.05$ & $7.92^{+43.08}_{-3.92}$ \\[1ex]
C44b & 0.11 & 0.08 & \ldots & \ldots\tablefootmark{i} & 0.08 & \ldots & $0.09\pm0.01$ & $4.11^{+1.95}_{-1.81}$ \\[1ex]
C45 & 0.27 & 0.14 & \ldots & 0.17 & 0.21 & \ldots & $0.20\pm0.03$ & $1.53^{+0.76}_{-0.66}$ \\[1ex]
C47 & 0.54 & 0.11 & \ldots & \ldots & 0.22 & \ldots & $0.29\pm0.13$ & $<1.19$ \\[1ex]
C48a & 0.40 & 0.15 & \ldots & \ldots & \ldots & \ldots & $0.28\pm0.13$ & $0.80^{+1.13}_{-0.48}$ \\[1ex]
C51b & 0.67 & 0.36 & \ldots & 0.29 & 0.44 & \ldots & $0.44\pm0.08$ & \ldots\tablefootmark{j} \\[1ex]
C52 & 0.98 & 0.16 & 0.57 & 0.60 & 0.39 & \ldots & $0.54\pm0.14$ & $0.86^{+0.44}_{-0.27}$ \\[1ex]
C56 & 0.15 & 0.10 & \ldots & \ldots & \ldots & \ldots & $0.13\pm0.02$ & $3.85^{+1.42}_{-1.08}$ \\[1ex]
C59 & 0.67 & 0.28 & \ldots & 0.42 & 0.39 & \ldots & $0.44\pm0.08$ & $<0.53$ \\[1ex]
C65 & 0.56 & 0.30 & \ldots & 0.28 & 0.31 & \ldots & $0.36\pm0.07$ & $0.65^{+0.26}_{-0.17}$  \\[1ex]
C66 & 0.57 & 0.32 & \ldots & 0.42 & 0.54 & \ldots & $0.46\pm0.06$ & $0.64^{+0.22}_{-0.17}$ \\[1ex]
C67 & 0.27 & 0.28 & \ldots & 0.31 & 0.16 & \ldots & $0.26\pm0.03$ & $<0.83$ \\[1ex]
C84b & 0.72 & 0.32 & \ldots & 0.69 & 0.15 & \ldots & $0.47\pm0.14$ & $1.05^{+0.66}_{-0.36}$ \\[1ex]
C86 & 0.13 & 0.08 & \ldots & \ldots\tablefootmark{i} & 0.08 & \ldots & $0.10\pm0.02$ & $<2.38$ \\[1ex]
C90c & \ldots & 0.18 & \ldots & \ldots & \ldots & \ldots & 0.18 & $2.08^{+0.50}_{-0.58}$ \\[1ex]
C95 & \ldots & \ldots & \ldots & \ldots & 0.34 & \ldots & 0.34 & $1.74^{+0.35}_{-0.39}$ \\[1ex]
C97a & 0.48 & 0.19 & \ldots & 0.31 & 0.32 & \ldots & $0.33\pm0.06$ & \ldots\tablefootmark{j} \\[1ex]
C101b & 0.44 & 0.22 & \ldots & \ldots & \ldots & \ldots & $0.33\pm0.11$ & $1.79^{+1.64}_{-0.86}$ \\[1ex]
C105 & 0.32 & 0.16 & \ldots & \ldots & \ldots & \ldots & $0.24\pm0.08$ & $1.56^{+1.25}_{-0.86}$ \\[1ex]
C112 & 0.51 & 0.27 & \ldots & \ldots & \ldots & \ldots & $0.39\pm0.12$ & $0.85^{+0.50}_{-0.27}$ \\[1ex]
C113 & \ldots & \ldots & 0.42 & \ldots & \ldots & \ldots & 0.42 & $<0.45$ \\[1ex]
C122a & 0.69 & 0.23 & 0.45 & 0.37 & 0.48 & \ldots & $0.44\pm0.08$ & $0.61^{+0.47}_{-0.46}$ \\[1ex]
C126 & 0.37 & 0.35 & \ldots & \ldots & \ldots & \ldots & $0.36\pm0.01$ & \ldots\tablefootmark{j} \\[1ex]
C127 & 0.23 & 0.26 & \ldots & \ldots & \ldots & \ldots & $0.25\pm0.01$ & $<0.79$ \\[1ex]
\hline 
\end{tabular} 
\tablefoot{Unless otherwise stated, an ellipsis means that the source was not found in the catalogue (within a search radius of $1\arcsec$).\tablefoottext{a}{Half-light radius from the Cassata's morphological catalogue (\cite{cassata2007}).}\tablefoottext{b}{Half-light radius from the Tasca's morphological catalogue v1.0 (\cite{tasca2009}).}\tablefoottext{c}{Half-light radius from the COSMOS/ACS morphological catalogue from the Columbia University group v1.0 (\cite{zamojski2007}).}\tablefoottext{d}{Circularised half-light radius calculated from the semi-major axis length of an ellipse encompassing 50\% of total light from the Z{\"u}rich structure \& morphology catalogue v1.0 (\cite{scarlata2007}; \cite{sargent2007}).}\tablefoottext{e}{Half-light radius from the COSMOS 2005 Morphology Catalogue.}\tablefoottext{f}{Effective radius from Toft et al. (2014).}\tablefoottext{g}{The mean of the quoted rest-frame UV radii, where the quoted uncertainty represents the standard error of the mean.}\tablefoottext{h}{The ratio between the 3~GHz radio size and the average UV size tabulated in column (8).}\tablefootmark{i}{The source was found in the Z{\"u}rich catalogue, but no size was available (value$=-99$).}\tablefoottext{j}{The source was not detected at 3~GHz.}}
\end{minipage} 
}
\end{table*}

\subsection{Size evolution over cosmic time}

\begin{figure}[!htb]
\centering
\resizebox{\hsize}{!}{\includegraphics{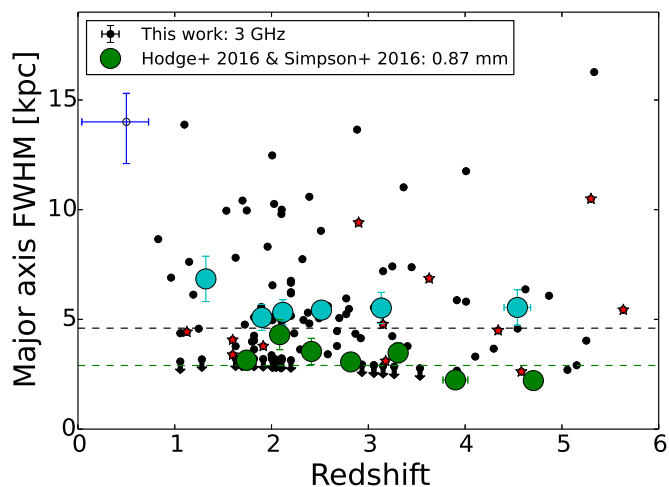}}
\caption{Linear major axis FWHM sizes [kpc] of our SMGs at 3~GHz plotted as a function of redshift (black points). 
The upper size limits are indicated by downwards pointing arrows. Those AzTEC SMGs from M15 that are common with the present sample 
are highlighted by red star symbols. The cyan filled circles represent the mean values of the binned 
version of the data (each bin contains 19 SMGs, except the highest redshift bin that contains 20 sources). 
For the sake of clarity, individual error bars are not shown for our data, but a representative (median) error bar is shown in the top left corner. 
Also shown are the 870~$\mu$m sizes from Hodge et al. (2016) and Simpson et al. (2016) discussed in Sect.~4.2.2; the green filled circles represent the binned averages of their combined data sets. The horizontal black and green dashed lines show the corresponding full sample median major axis FWHM values of 4.6~kpc and 2.9~kpc, respectively.}
\label{figure:corr}
\end{figure}

In Fig.~\ref{figure:corr}, we plot our radio major axis FWHM sizes as a function of redshift. 
For comparison, we also show the ALMA 870~$\mu$m based rest-frame FIR sizes 
from Hodge et al. (2016) and Simpson et al. (2016). 

The derived 3~GHz radio sizes exhibit a large scatter in Fig.~\ref{figure:corr}, but the binned data (the cyan filled circles) show that the average radio size is fairly constant as a function of redshift, with the exception of the lowest redshift bin ($\langle z\rangle = 1.32\pm0.07$), which shows the largest nominal radio size ($6.8\pm1.0$~kpc). However, owing to its large standard error of the mean radio size, the lowest redshift bin is not a statistically significant outlier. Based on a much smaller (by a factor of 7.7) sample of 3~GHz detected COSMOS SMGs with available redshifts, M15 found a hint of larger 3~GHz radio sizes at redshifts of $z\sim2.5-5$ than those outside this redshift interval (Fig.~7 therein). Curiously, a comparable increase in rest-frame FIR sizes was recognised among the Ikarashi et al. (2015) SMGs at $z\sim3.5-5$, but as mentioned in Sect.~4.2.2 most of their sources had only lower redshift limits available. However, our new, large sample of SMGs shows that the more extended radio sizes seen at $z\sim2.5-5$ by M15 was merely a result of a small number statistics. A linear regression through all the binned data points yields $\theta_{\rm 3\, GHz}^{\rm maj}[{\rm kpc}] \propto  -(0.04 \pm 0.24)\times z$ with a Pearson correlation coefficient of $r= -0.35$ ($r$ ranges from $-1$ to 1, and $r=0$ implies no (linear) correlation). Hence, because we could improve upon our earlier work, we can now conclude that the average radio size of an SMG does not appear to evolve as a function of redshift. This is consistent with Murphy et al. (2017), who found that the 10~GHz radio sizes of their sample of star-forming galaxies in GOODS-N do not exhibit any obvious evolution with redshift.

We found that the FIR sizes of the $z=1.51-4.76$ SMGs from Hodge et al. (2016) exhibit a trend of sources being more compact at higher redshifts; a linear fit to this sample yields $\theta_{\rm 870 \mu m}^{\rm maj}[{\rm kpc}] \propto -(0.46 \pm 0.13)\times z$ ($r=-0.76$). We also checked if the FIR sizes from Simpson et al. (2016) show a similar trend, but this was not the case. Instead, a linear fit to their data yields $\theta_{\rm 870 \mu m}^{\rm maj}[{\rm kpc}] \propto (0.10 \pm 0.37)\times z$ ($r=-0.21$), where the one upper size limit was omitted from the fit. Because Hodge et al. (2016) and Simpson et al. (2016) both used observations at 870~$\mu$m, and the redshifts of the Simpson et al. (2016) target sources, $z=1.68-4.91$, are similar to those from Hodge et al. (2016), the rest-frame wavelengths probed in these two studies are comparable. Hence, we augmented the data set by combining their data to see whether the FIR size shows any trend as a function of redshift. A linear fit to the full combined sample yields $\theta_{\rm 870 \mu m}^{\rm maj}[{\rm kpc}] \propto -(0.36 \pm 0.13)\times z$ ($r=-0.49$), while that to the combined, binned data shown in Fig.~\ref{figure:corr} yields $\theta_{\rm 870\, \mu m}^{\rm maj}[{\rm kpc}] \propto -(0.53 \pm 0.16)\times z$ ($r=-0.77$). This suggests that contrary to our radio sizes, the rest-frame FIR dust-emitting sizes of SMGs might be more compact at higher redshifts. Hence, besides the different spatial extents of radio and dust continua, their relative evolution over cosmic time might also be different. However, if the observed-frame 870~$\mu$m emission is dominated by the cold dust component (see Sect.~4.5.2 and Appendix~E), it is possible that the diminishing intensity contrast against the cosmic microwave background (CMB) radiation affects the dust emission size measurements at high redshifts (e.g. \cite{dacunha2013}; \cite{zhang2016}). If this is the case, there is a possible observational bias towards smaller dust-emitting FWHM sizes at higher redshifts, which might explain the anti-correlation seen in  Fig.~\ref{figure:corr}.

We also checked whether the rest-frame UV sizes tabulated in Table~\ref{table:optical} exhibit any trend with redshift. We divided the 39 data points into four redshift bins (three bins of ten sources plus one bin composed of nine sources), and applied a survival analysis to calculate the average half-light radii. Over the redshift range thus obtained, $\langle z \rangle=1.47-3.95$, a linear fit yielded $r_{\rm UV}[{\rm kpc}] \propto -(0.42 \pm 0.03)\times z$ ($r=-0.995$), which implies a size growth towards lower redshifts (from $r_{\rm UV}=1.86\pm0.42$~kpc to $2.86\pm0.38$~kpc for our average redshift interval). Because the rest-frame UV radiation is a tracer of unobscured high-mass star formation, one might expect that also the radio size evolves with redshift in a similar fashion as the galaxy size in the UV. However, similarly to our full sample of 3~GHz detections, the average radio size of the sources listed in Table~\ref{table:optical} shows a flat trend as a function of redshift (the 36 radio detected sources from Table~\ref{table:optical} were split into four bins of nine sources, and a linear fit over the redshift range $\langle z \rangle=1.49-3.75$ yielded $\theta_{\rm 870\, \mu m}^{\rm maj}[{\rm kpc}] \propto (0.04 \pm 0.38)\times z$ with $r=0.05$). The non-evolution of the average radio size of our SMG sample as a function of redshift is discussed further in Sect.~4.5.4.

\subsection{Does the radio-emitting size depend on the galaxy morphology ?}

In this subsection, we explore the 3~GHz radio size dependence on the galaxy morphology as seen in the rest-frame UV. As described in Sect.~4.2.4, the COSMOS field benefits from multiple publicly available morphological catalogues. Here, we make use of three of them, each of which provides a morphological classification into different types of galaxies (rather than just morphology-related parameters). These are the Cassata, Tasca, and Z{\"u}rich catalogues. The morphological classification of the sources found in these catalogues ($1\arcsec$ search radius; Sect.~4.2.4) is summarised in Table~\ref{table:morphology}. 

\begin{figure*}[!htb]
\begin{center}
\includegraphics[width=0.15\textwidth]{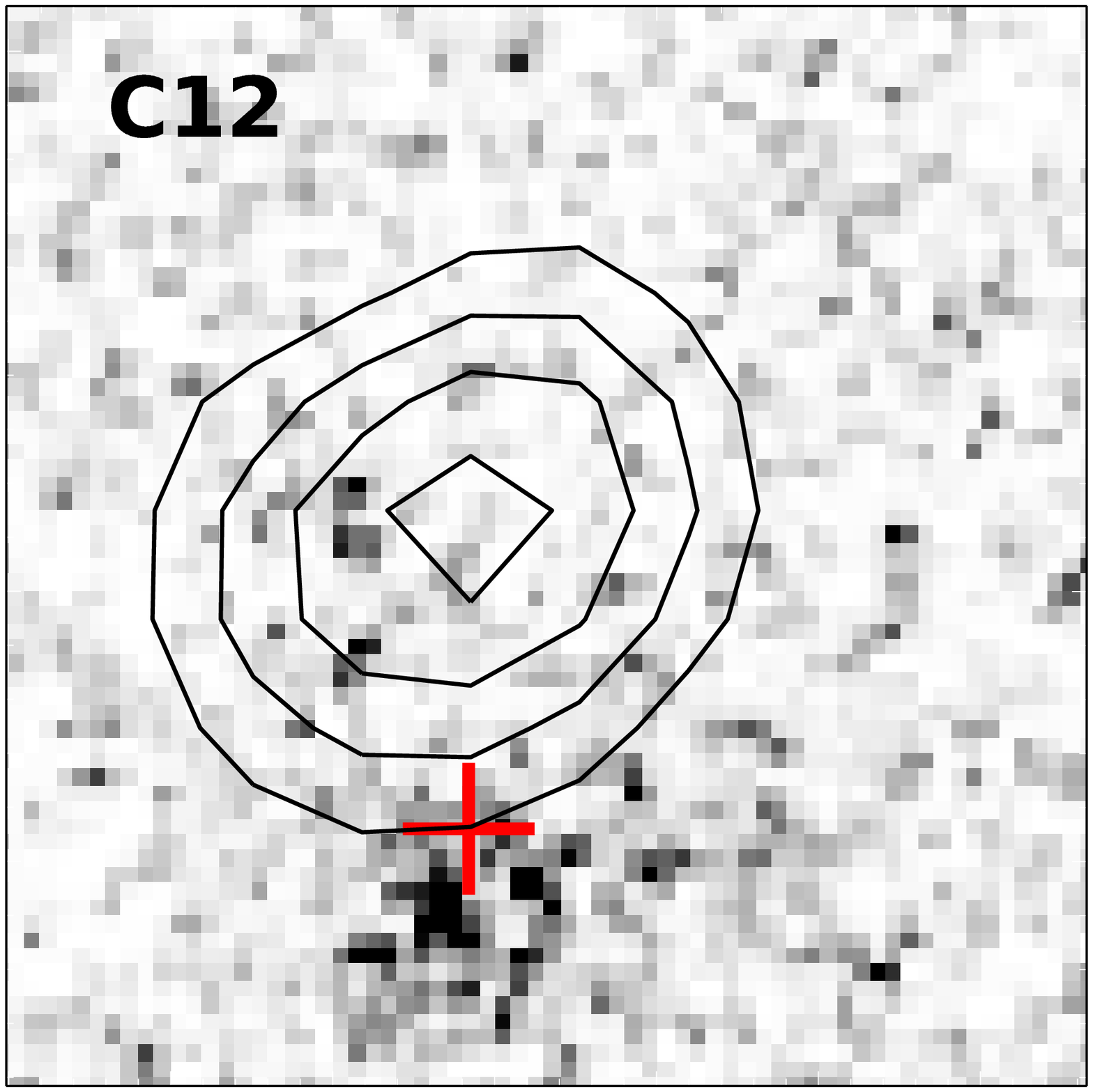}
\includegraphics[width=0.15\textwidth]{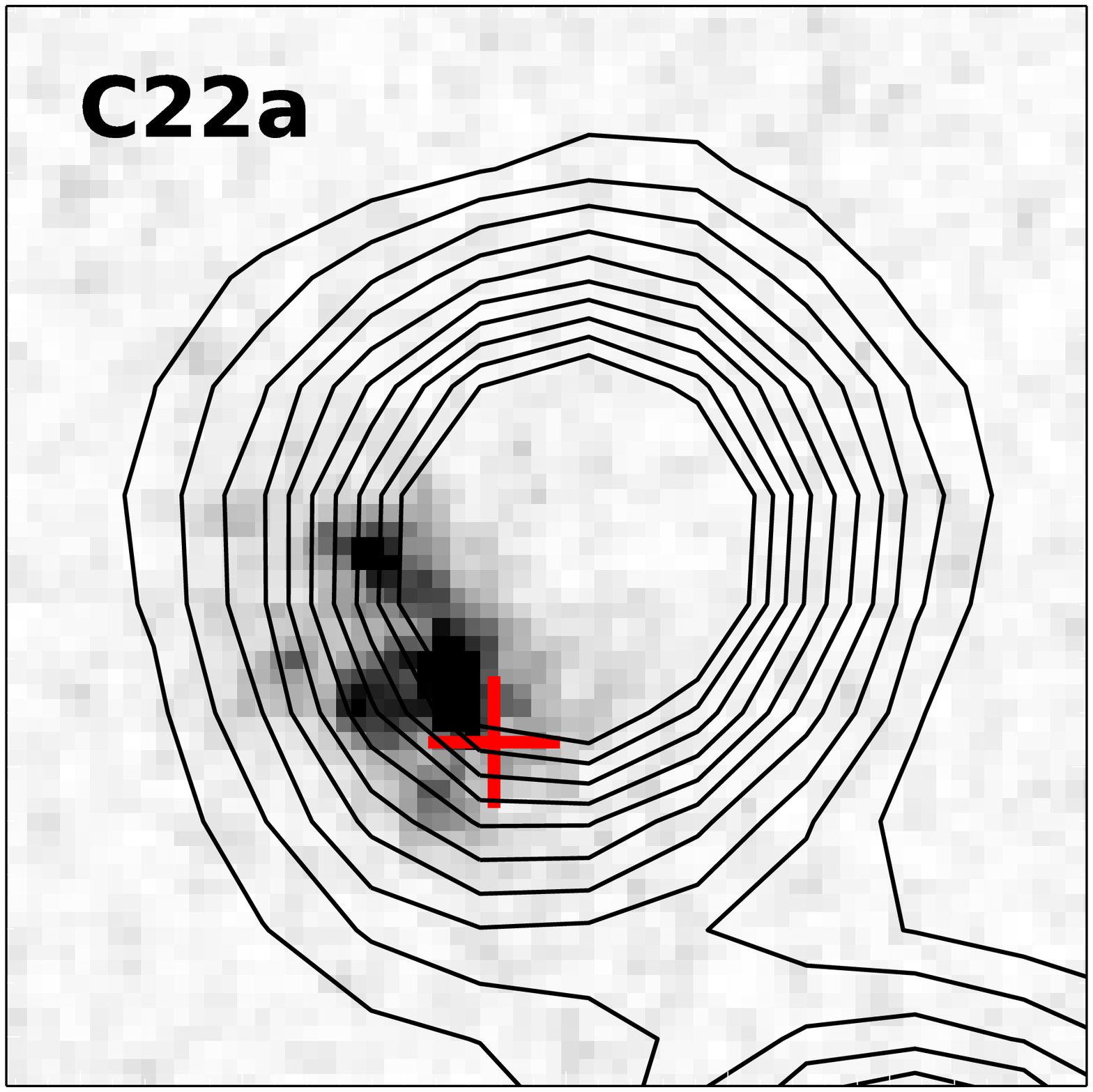}
\includegraphics[width=0.15\textwidth]{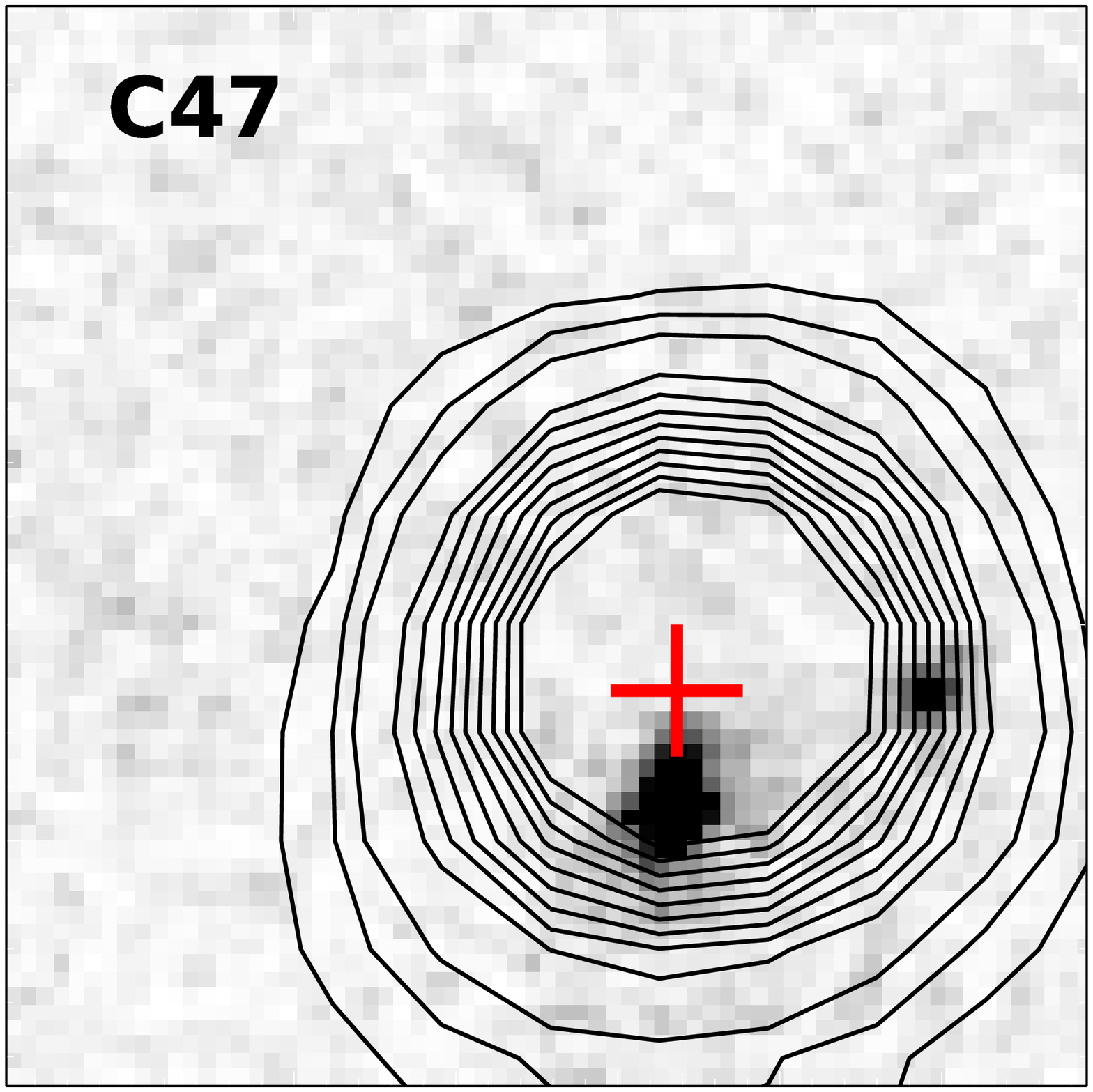}
\includegraphics[width=0.15\textwidth]{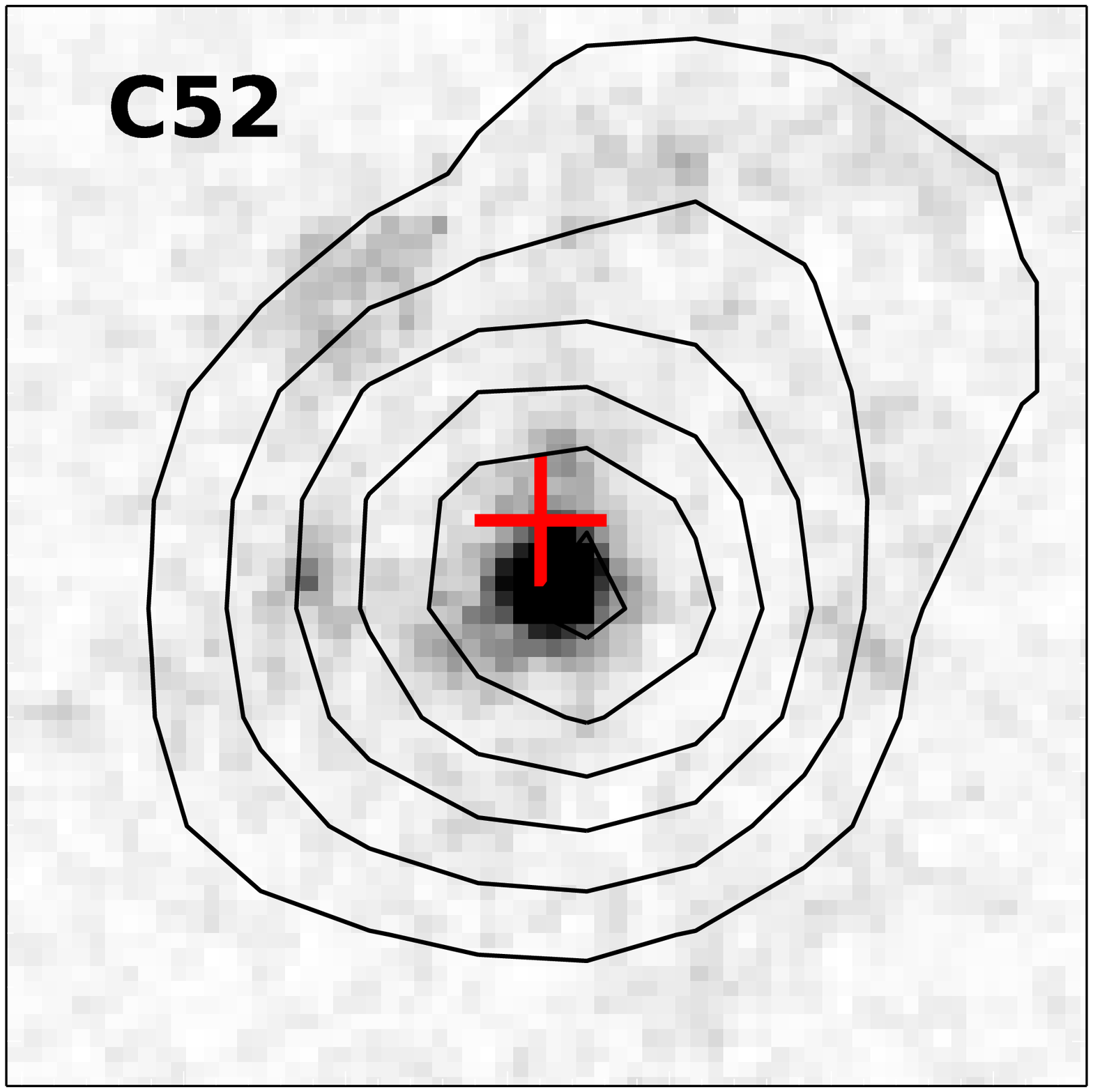}
\includegraphics[width=0.15\textwidth]{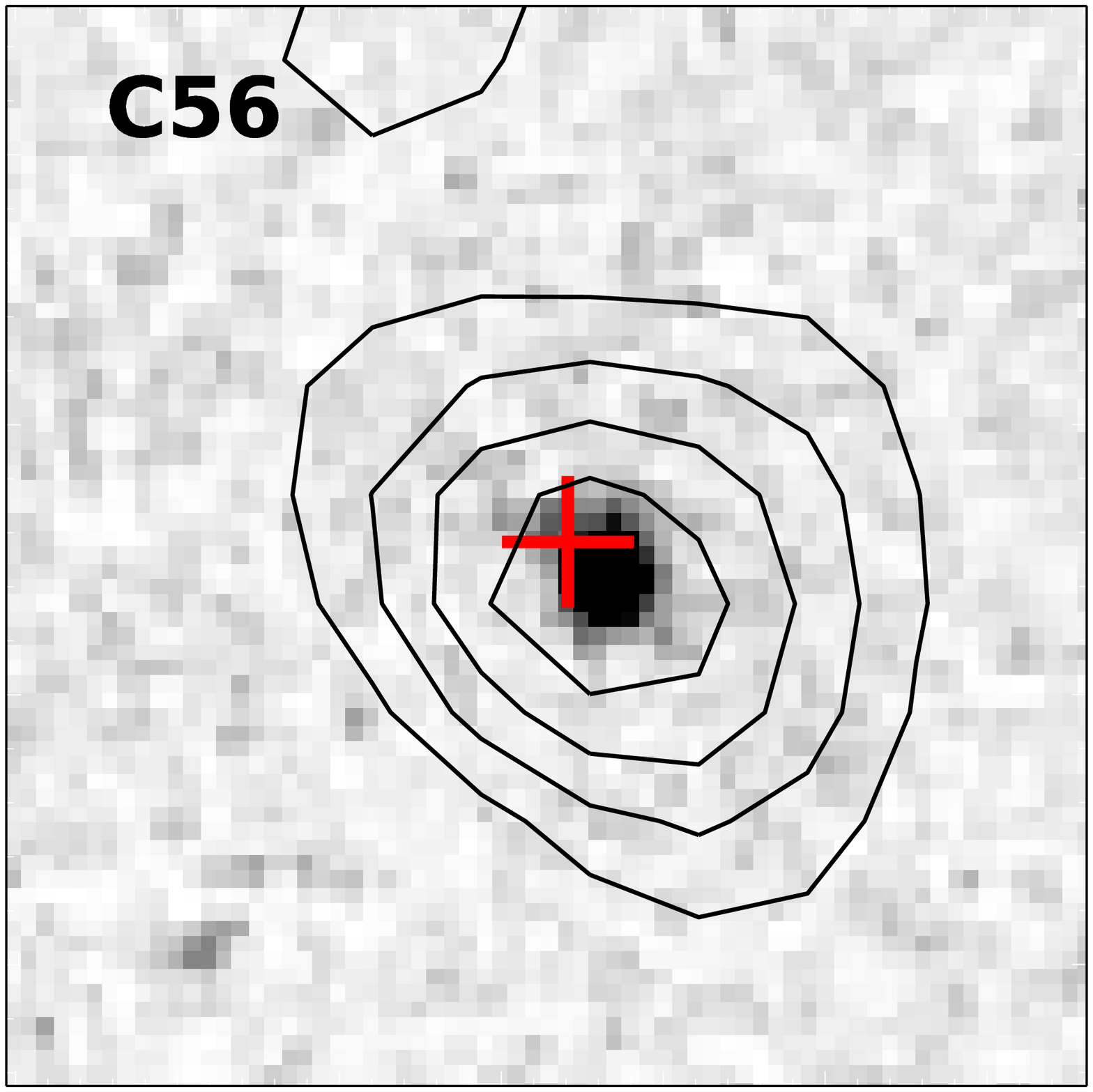}
\includegraphics[width=0.15\textwidth]{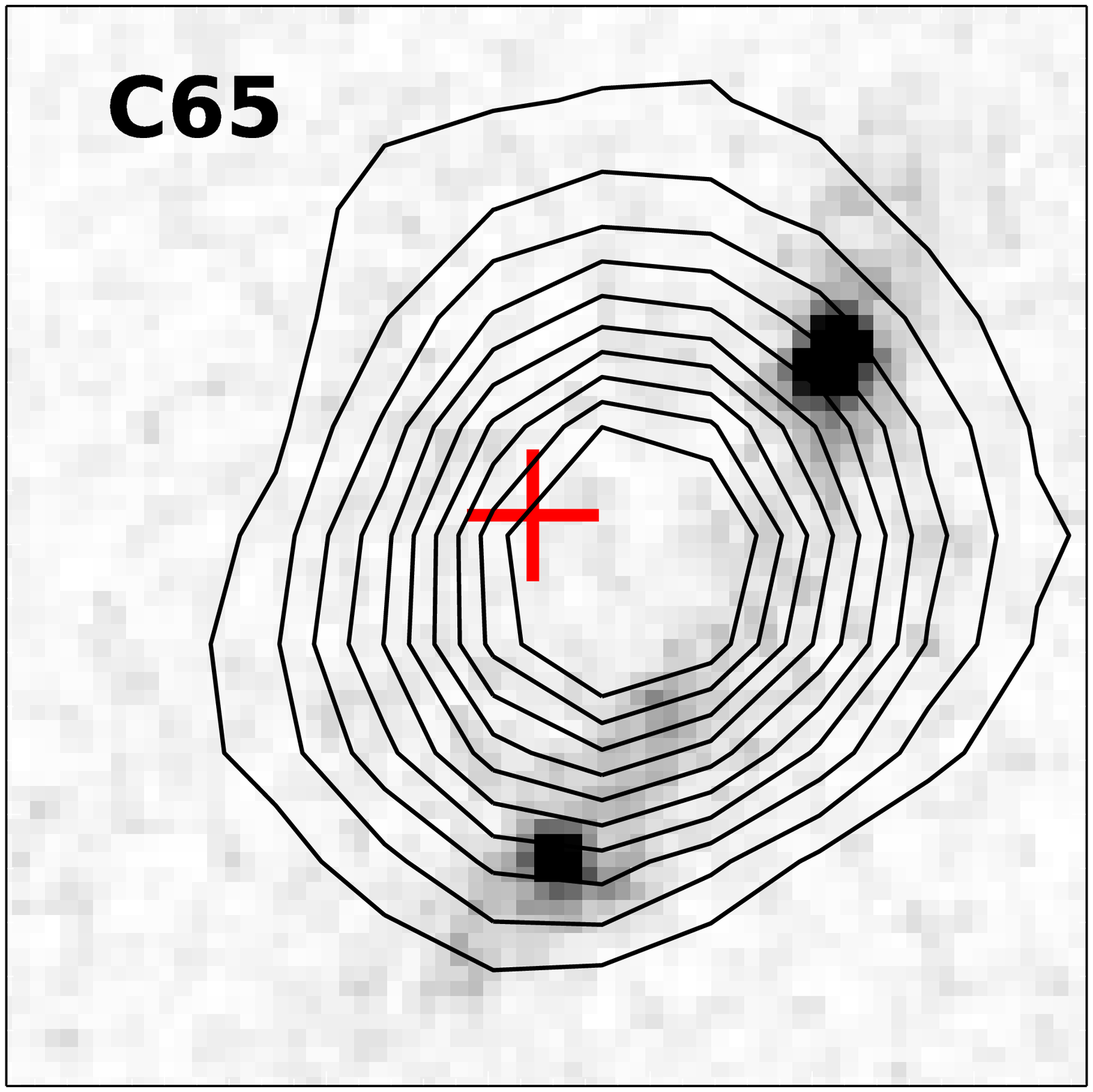}
\includegraphics[width=0.15\textwidth]{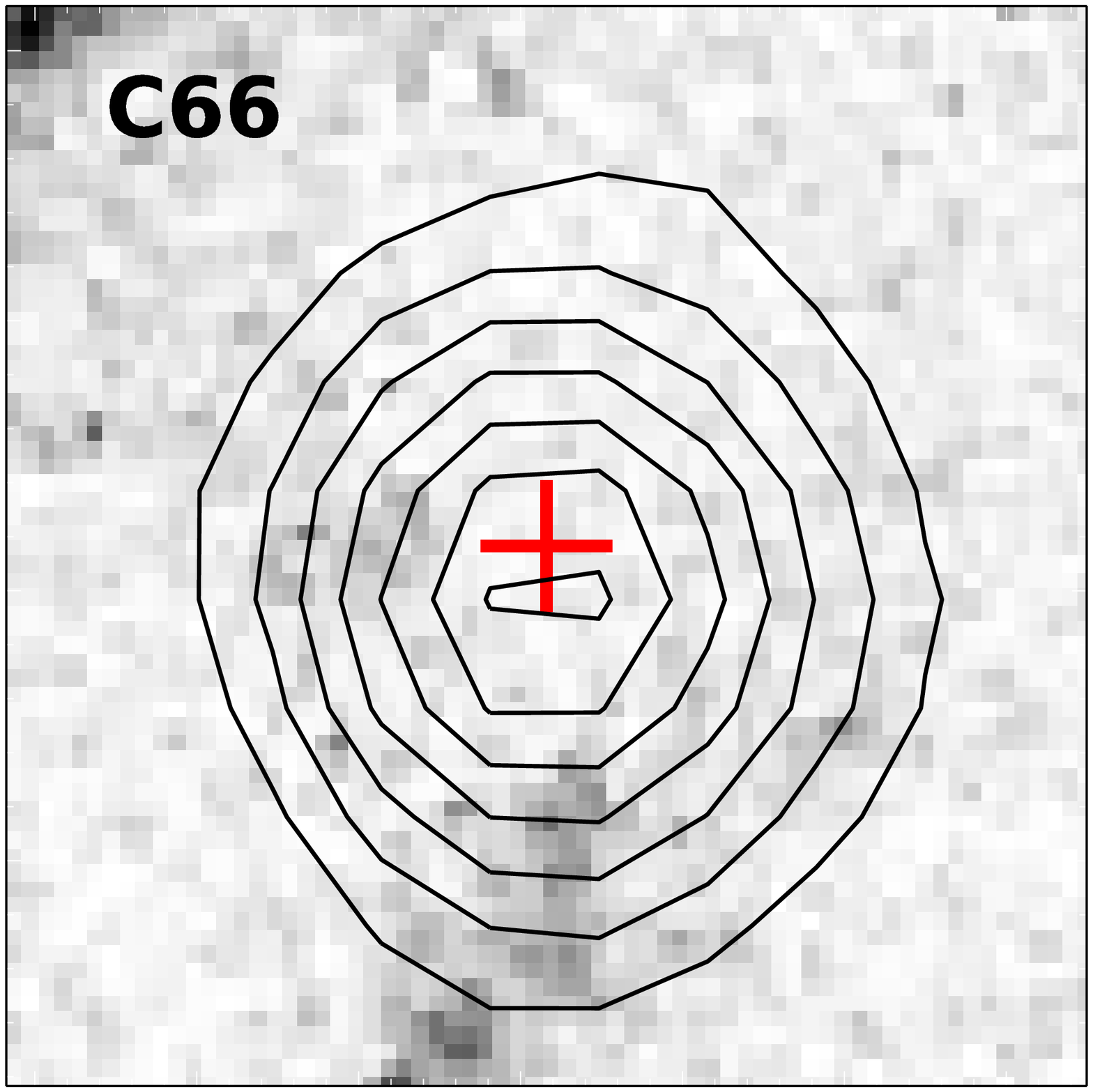}
\includegraphics[width=0.15\textwidth]{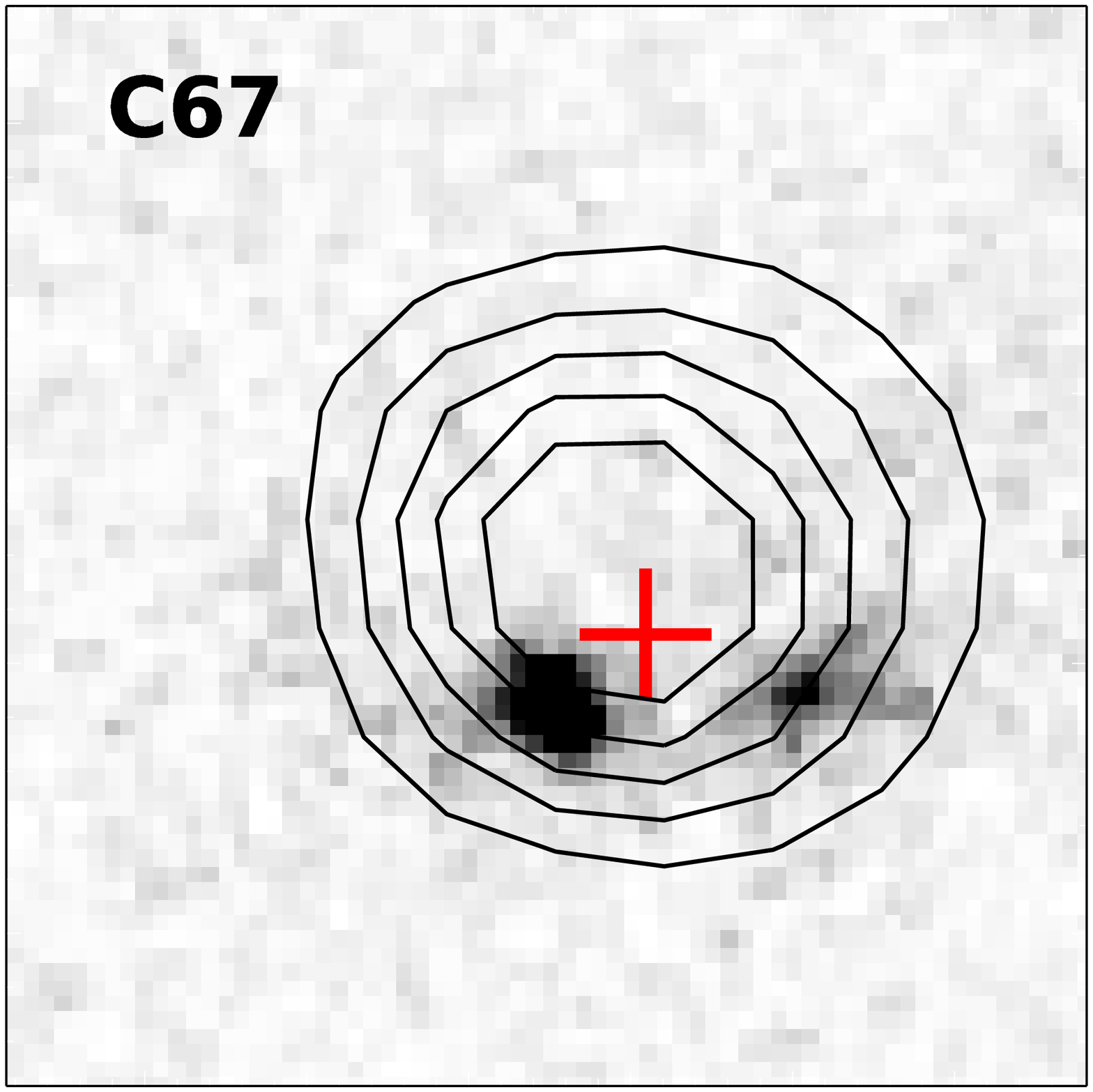}
\includegraphics[width=0.15\textwidth]{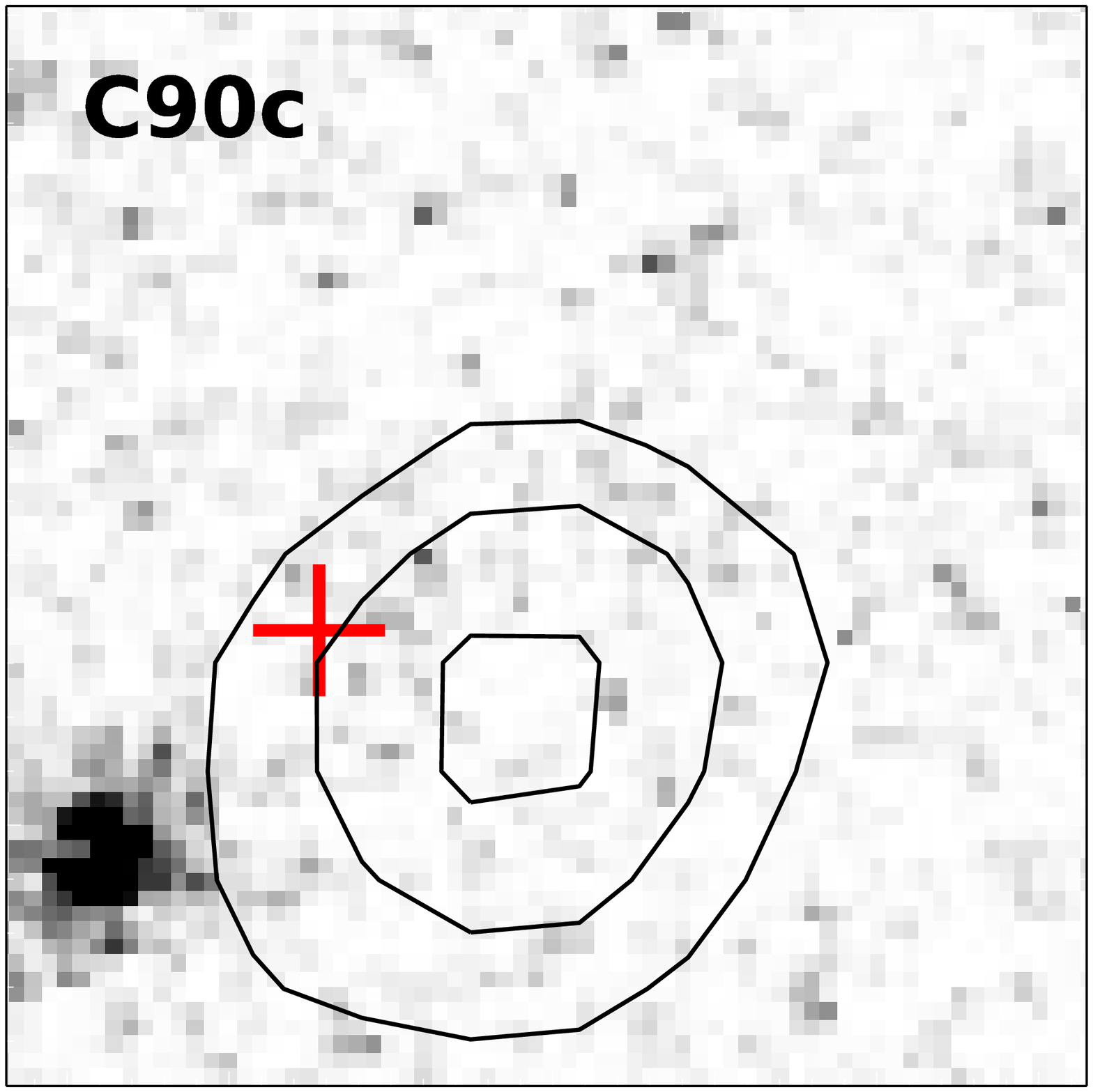}
\includegraphics[width=0.15\textwidth]{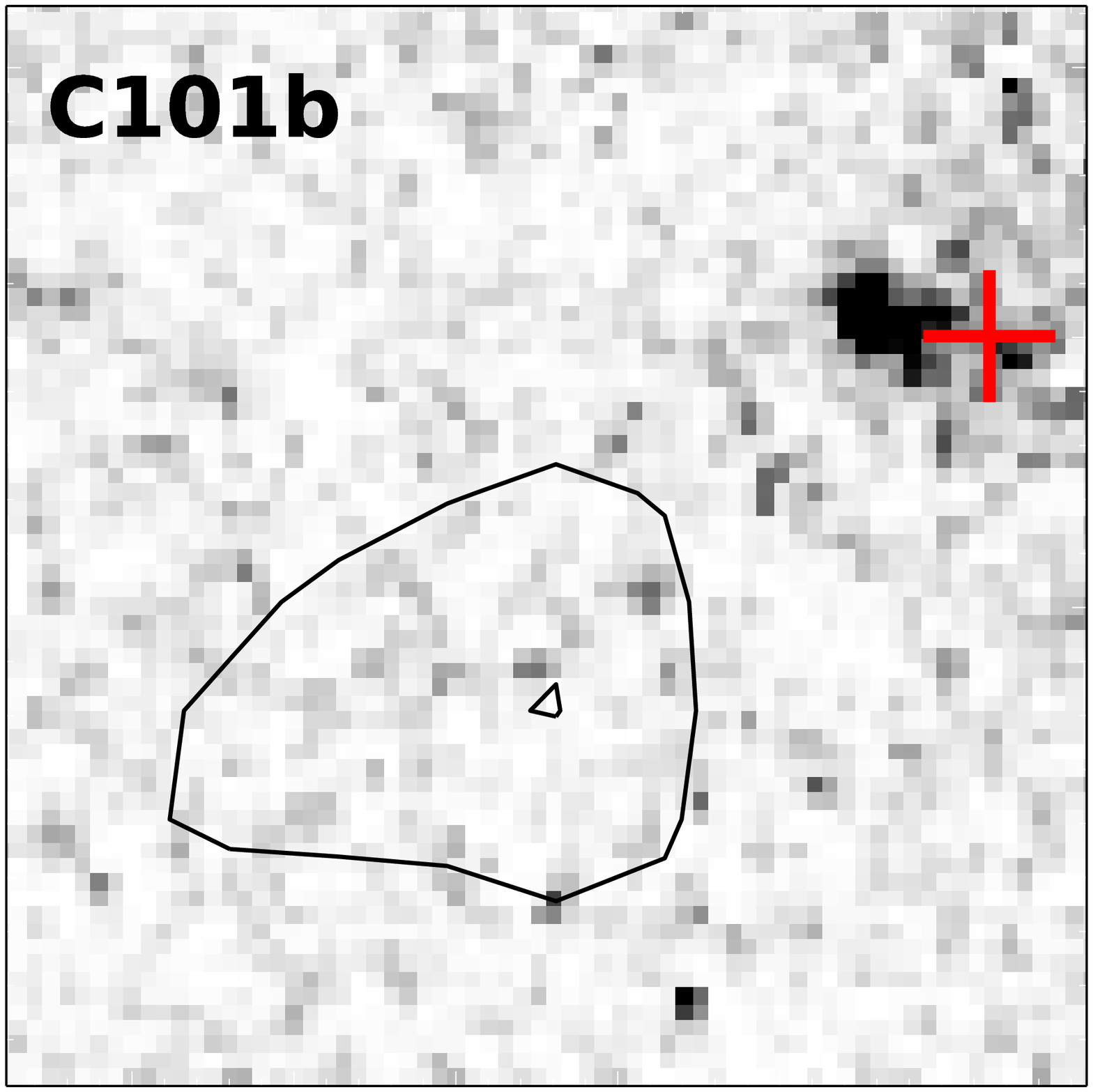}
\includegraphics[width=0.15\textwidth]{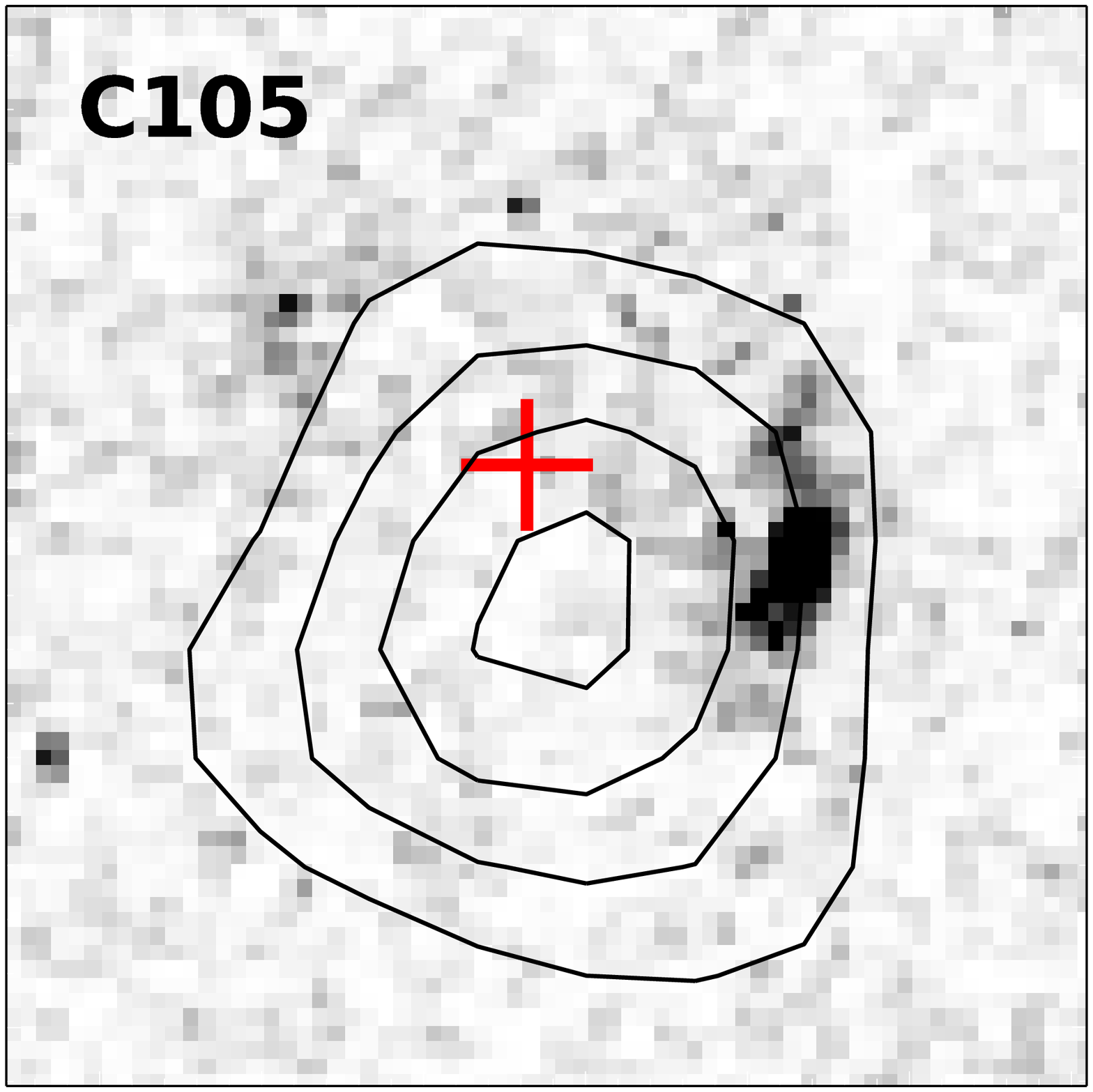}
\includegraphics[width=0.15\textwidth]{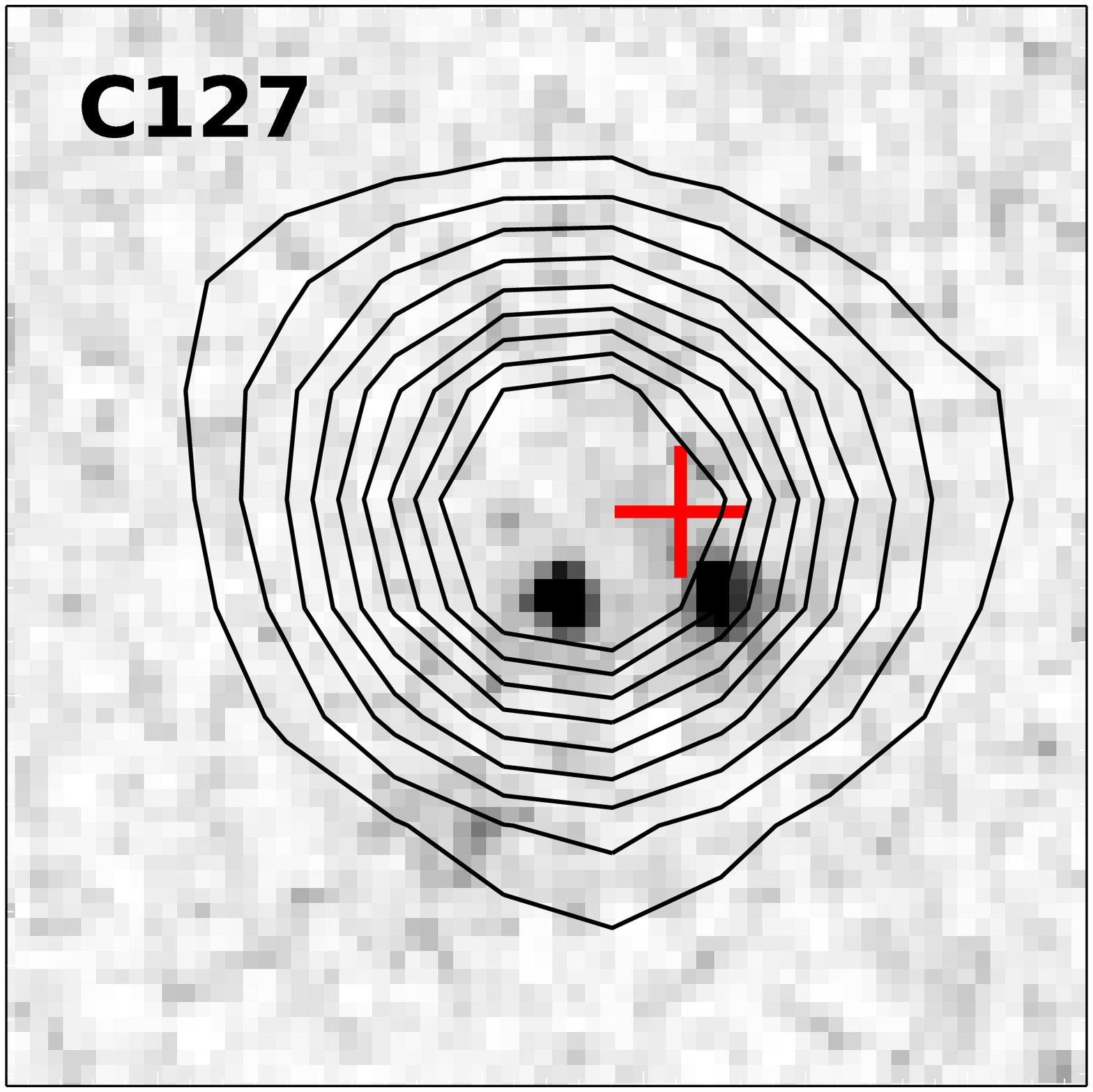}
%\phantom{\includegraphics[width=0.15\textwidth]{ACS_images/C127_disk.eps}}
%\phantom{\includegraphics[width=0.15\textwidth]{ACS_images/C127_disk.eps}}
%\phantom{\includegraphics[width=0.15\textwidth]{ACS_images/C127_disk.eps}}
\caption{\textit{Hubble}/ACS $I$-band images of the 3~GHz detected SMGs whose morphological classification in the rest-frame UV is uncertain as discussed in Sect.~4.4. The greyscale is inverted, and displayed using a power-law stretch to better show the low and high surface brightness features. The black contours show the 3~GHz radio emission as in Fig.~\ref{figure:maps}. Each image is centred on the ALMA 1.3~mm peak position, is $2\arcsec \times 2\arcsec$ in size, and displayed with north up and east left. The red plus signs mark the positions of the optical-NIR counterparts for which the photometric redshifts were derived by Brisbin et al. (2017), except towards C66 where the plus sign marks the ALMA position from which the optical-NIR photometry was manually extracted owing to blending issues.}
\label{figure:acs}
\end{center}
\end{figure*}

To assess a morphological type for each of our sources, we used a ladder approach; for example, if the source is classified as a disk in each of the three catalogues, we classified the source as a disk (e.g. AzTEC/C19), and if two out of the three catalogues gives the same morphology, this most common class is adopted as the source morphology (e.g. AzTEC/C36 is classified as an irregular in both the Cassata and Tasca catalogues but as a disk in the Z{\"u}rich catalogue), and so on. This approach had the advantage of increasing the number of galaxies for which a morphological classifcation is available (30 in total). However, the sources AzTEC/C12, 47, 56, 101b, and 105 could not be classified this way owing to the number of different classifications (e.g. C12 was classified as an irregular galaxy in the Cassata catalogue, and as a disk in the Tasca catalogue, while the source was not found in the Z{\"u}rich catalogue). Hence, we performed visual inspection of these galaxies on the \textit{Hubble}/ACS $I$-band images to try to determine their morphological types (see Fig.~\ref{figure:acs}). In the case of C12, C101b, and C105, the angular offset between the SMG position and the $I$-band source, and the fairly irregular shape of the latter one, led us to classify these sources as irregular galaxies. AzTEC/C47 shows two components in the $I$-band image separated by $0\farcs5$ in projection, which might indicate an interacting galaxy pair, and hence we classified it as an irregular as well. In general, if an SMG is a merger system, its optical appearance can be expected to be asymmetric or clumpy, in which case the source is likely to be classified as an irregular (cf.~\cite{conselice2003}). Finally, a visual inspection of the $I$-band image of C56 suggests that this source could be a compact disk (the Cassata classification), rather than an early-type galaxy (ETG) as classified in the Tasca catalogue. 

\begin{figure}[!htb]
\centering
\resizebox{\hsize}{!}{\includegraphics{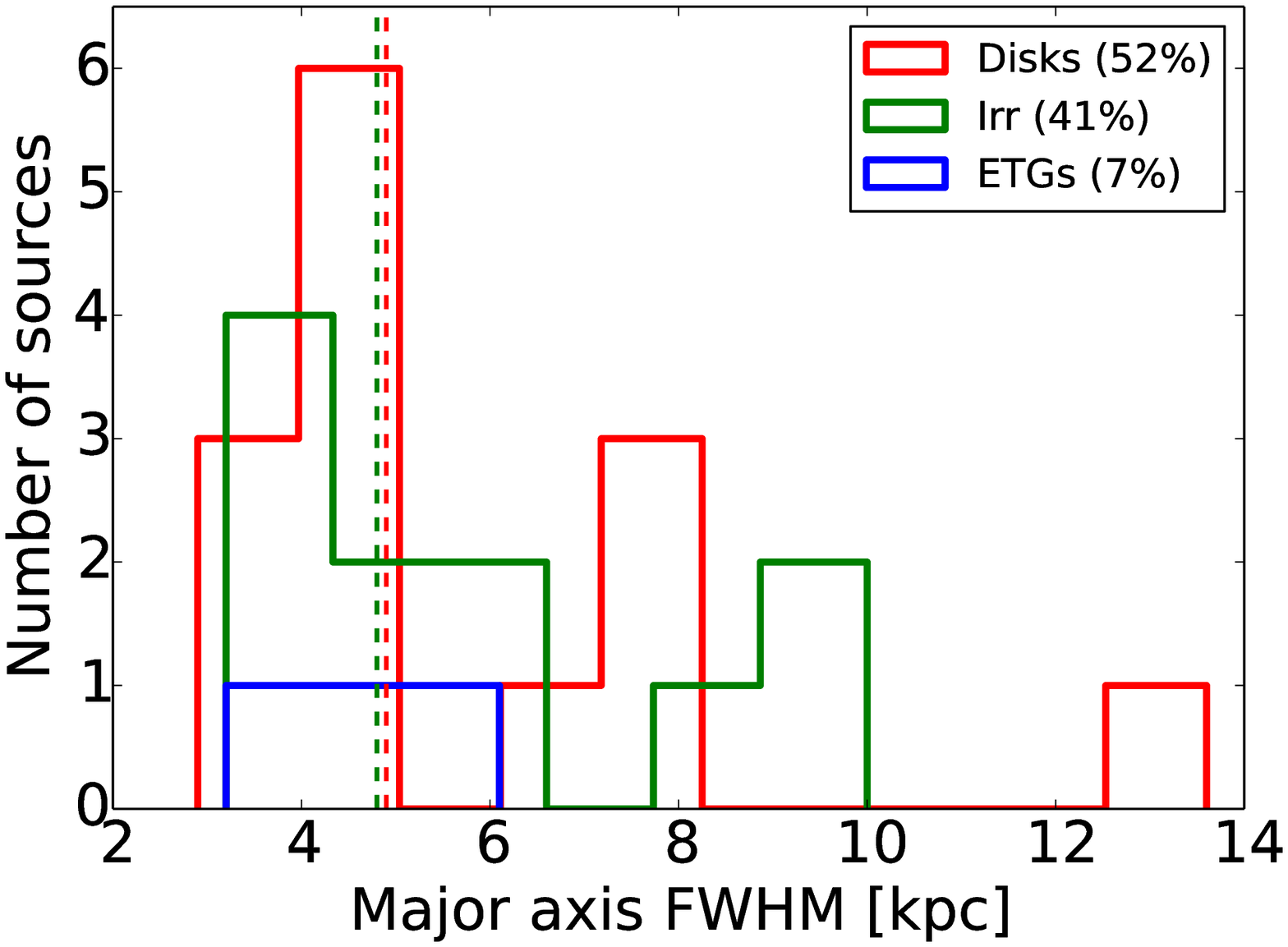}}
\resizebox{\hsize}{!}{\includegraphics{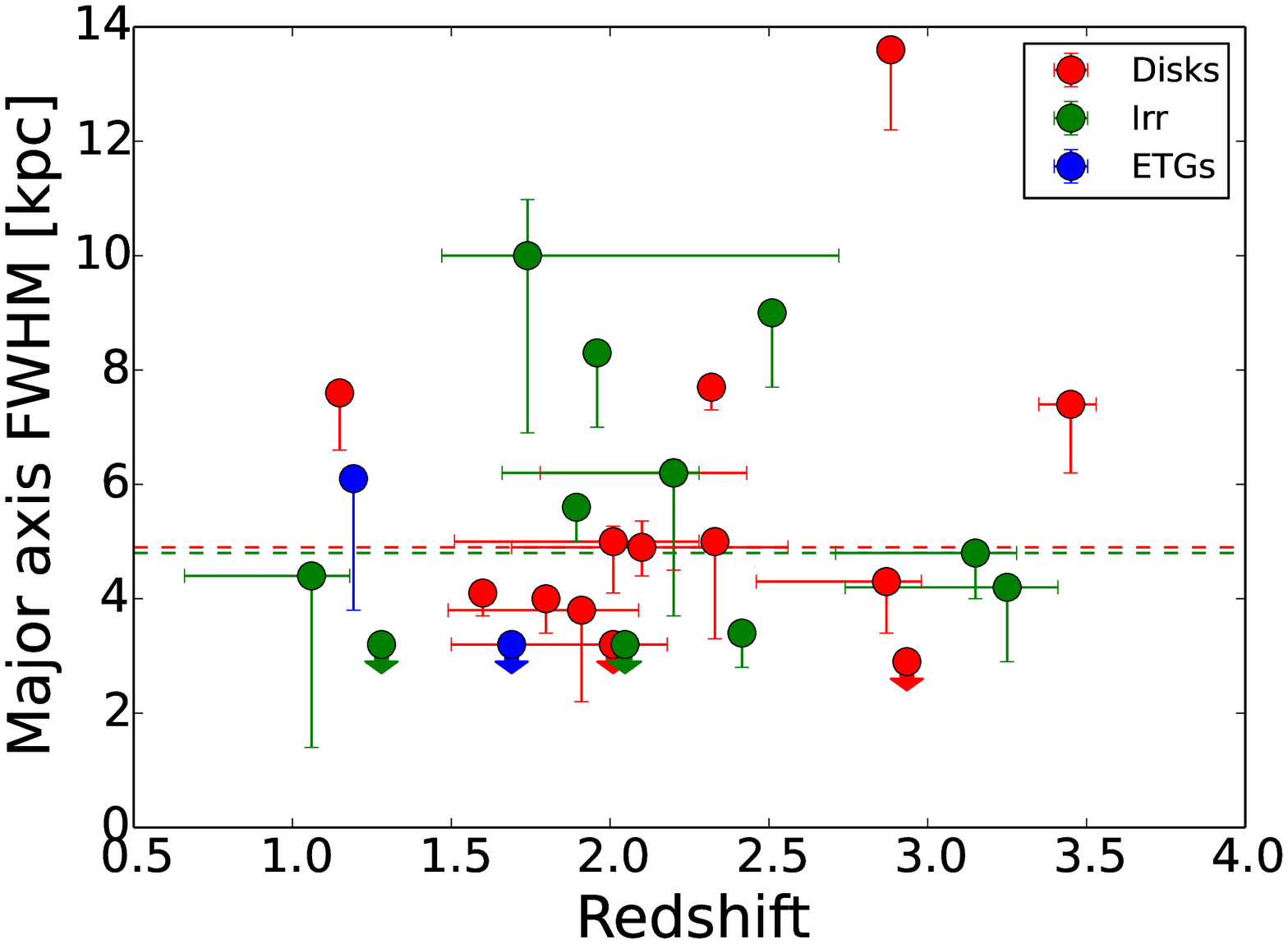}}
\caption{\textbf{Top:} Radio size (physical major axis FWHM at $\nu_{\rm obs}=3$~GHz) distributions of our SMGs that are classified as disks (red histogram), irregular galaxies (green histogram), or early-type galaxies (blue histogram). The bin size is 1.0~kpc, and the upper size limits were placed in the bins corresponding to those values. The vertical dashed lines mark the median values of $4.9\pm0.9$~kpc for disks, and $4.8\pm0.9$~kpc for irregulars; survival analysis was used to take the upper size limits into account when calculating the median sizes. \textbf{Bottom:} Same as above but as a function of redshift. The upper size limits are indicated by downwards pointing arrows, and the horizontal dashed lines show the aforementioned median sizes. }
\label{figure:morpho}
\end{figure}

Out of the aforementioned 30 sources, three (C51b, C97a, and C126) were not detected at 3~GHz, and hence these sources were left out from the physical radio size comparison. Among the remaining 27 sources, we classified 14 sources as disks (52\%), 11 as irregulars (41\%), and two as ETGs (7\%). The 3~GHz radio size distributions of these sources are shown in the top panel in Fig.~\ref{figure:morpho}, and the same information as a function of redshift is depicted in the bottom panel. The first two classes have survival analysis-based median (mean) radio sizes of $4.9\pm0.9$~kpc ($5.8\pm0.7$~kpc) and $4.8\pm0.9$~kpc ($5.7\pm0.7$~kpc), while no median size could be computed for the two ETGs (the other ETG is unresolved). However, the radio sizes of these two ETGs are similar to those of most of the disks and irregulars. Hence, we found no evidence for a statistically significant difference in the radio size between different galaxy morphologies, or any redshift trend among different morphological types, but our analysis can be subject to small number statistics. Another caveat in this analysis is that besides the aforementioned five sources, AzTEC/C12, C47, C56, C101b, and C105, there are seven additional sources whose true morphology is uncertain as described below.

While C22a is classified as a disk in all three catalogues we employed, the ACS $I$-band emission appears clumpy, is offset from the 3~GHz emission maximum by $0\farcs34$ to the south-east (the brightest $I$-band peak), and the SMG system is likely driven by a galaxy merger (Sect.~4.5.4). Hence, the source could be deemed irregular. The source AzTEC/C52 was taken to be a disk following the Tasca and Z{\"u}rich catalogues, but the ACS $I$-band image shown in Fig.~\ref{figure:acs} shows multiple low-surface brightness features, which explains why the source was classified as irregular in the Cassata catalogue. Hence, this SMG could be a clumpy disk or an irregular galaxy, and possibly associated with a merger. The 3~GHz emission extended towards north-west could be an indication of galaxy interaction (Sect.~4.5.4). We also took AzTEC/C65 to be disk-like, but the ACS $I$-band image shows two bright sources with the 3~GHz emission peaking in between them. Hence, this source could also be a merger system, and the irregular morphology from the Cassata catalogue might be more appropriate. The source AzTEC/C66 shows only faint ACS $I$-band emission, and hence it is difficult to say whether the source is irregular (Cassata) or disk-like (Tasca and Z{\"u}rich). The source AzTEC/C67 is also classified as a disk in all three catalogues, but the $I$-band image shows the presence of two sources offset from the radio peak, and hence it could actually be an irregular. The $I$-band source seen towards AzTEC/C90c lies $0\farcs78$ to the south-east of the radio peak, and this large offset makes it unclear whether the observed-frame optical source is physically associated with the SMG. Similarly, the $I$-band source seen towards C101b lies $1\farcs02$ ($0\farcs75$) to the north-west of the 3~GHz (ALMA) peak position, and hence the physical association is uncertain, but this $I$-band source is coincident with the optical-NIR counterpart from Brisbin et al. (2017). Finally, AzTEC/C127 is classified as a disk in both the Cassata and Tasca catalogues (not found in the Z{\"u}rich one), but the $I$-band image reveals the presence of two sources, and the source could actually be an irregular merger system. Hence, besides the small number statistics mentioned above, we conclude that the radio size comparison between different source morphologies presented in this section can also be affected by misclassified morphologies on an individual basis. For instance, a clumpy disk can be difficult to distinguish from an ongoing merger if only continuum imaging observations are available (\cite{immeli2004}), and spectral line observations would be required to probe the gas kinematics and dynamics.

%C12: offset from the radio position, hard to say; may irr bcs of this; cassata assymetry 0.406; tasca asym 0.0126
%47; two components, so maybe classify as Irr; cassata assymetry 0.31; tasca asym 0.12384
%56: maybe a compact disk, so therefore misclassified as ETG by tasca; cassata assymetry 0.091; tasca asym 0.13518
%101b: offset from the SMG and Irr shape; cassata assymetry 0.371; tasca asym 0.07414
%105: offset from the SMG and Irr shape; cassata assymetry 0.26; tasca asym 0.11858

%the asymmetry index $A$ original image-image rotated by $180\degr$/original image

\begin{table}
\renewcommand{\footnoterule}{}
\caption{Morphological classification of a subset of 30 target SMGs.}
{\small
\begin{minipage}{1\columnwidth}
\centering
\label{table:morphology}
\begin{tabular}{ccccc}
\hline\hline 
ID & CC\tablefootmark{a} & TC\tablefootmark{b} & ZSMC\tablefootmark{c} & Morphology\tablefootmark{d} \\
\hline
C9b & disk & disk & \ldots & disk \\[1ex]%& \includegraphics[width=1.5cm, height=1.5cm]{ACS_images/C9b_disk.eps} \\[1ex]
C12 & Irr & disk & \ldots & Irr\tablefootmark{e} \\[1ex]%&  \includegraphics[width=1.5cm, height=1.5cm]{ACS_images/C12_irr.eps} \\[1ex]
C18 & Irr & \ldots & \ldots & Irr \\[1ex]%& \includegraphics[width=1.5cm, height=1.5cm]{ACS_images/C18_irr.eps}\\[1ex]
C19 & disk & disk & disk & disk \\[1ex]%&  \includegraphics[width=1.5cm, height=1.5cm]{ACS_images/C19_disk.eps}\\[1ex]
C22a & disk & disk & disk & disk \\[1ex]%&  \includegraphics[width=1.5cm, height=1.5cm]{ACS_images/C22a_disk.eps}\\[1ex]
C23 & disk & disk & \ldots & disk\\[1ex] %& \includegraphics[width=1.5cm, height=1.5cm]{ACS_images/C23_disk.eps}\\[1ex]
C25 & \ldots & Irr & \ldots  & Irr \\[1ex]%& \includegraphics[width=1.5cm, height=1.5cm]{ACS_images/C25_irr.eps}\\[1ex]
C28a & disk & disk & \ldots & disk \\[1ex]%& \includegraphics[width=1.5cm, height=1.5cm]{ACS_images/C28a_disk.eps}\\[1ex]
C36 & Irr & Irr & disk & Irr \\[1ex]  %\includegraphics[width=1.5cm, height=1.5cm]{ACS_images/C36_irr.eps}\\[1ex]
C44b & ETG & ETG & x\tablefootmark{f} & ETG \\[1ex]%&  \includegraphics[width=1.5cm, height=1.5cm]{ACS_images/C44b_etg.eps}\\[1ex]
C45 & disk & Irr & disk & disk \\[1ex]%&  \includegraphics[width=1.5cm, height=1.5cm]{ACS_images/C45_disk.eps}\\[1ex]
C47 & Irr & disk & \ldots & Irr\tablefootmark{e} \\[1ex]%&  \includegraphics[width=1.5cm, height=1.5cm]{ACS_images/C47_irr.eps}\\[1ex]
C48a & disk & disk & \ldots & disk \\[1ex] %&  \includegraphics[width=1.5cm, height=1.5cm]{ACS_images/C48a_disk.eps}\\[1ex]
C51b\tablefootmark{g} & Irr & disk & disk & disk\\[1ex] %&  \includegraphics[width=1.5cm, height=1.5cm]{ACS_images/C51b_disk.eps}\\[1ex]
C52 & Irr & disk & disk & disk \\[1ex]%&  \includegraphics[width=1.5cm, height=1.5cm]{ACS_images/C52_disk.eps}\\[1ex]
C56 & disk & ETG & \ldots & disk\tablefootmark{e}\\[1ex] %& \includegraphics[width=1.5cm, height=1.5cm]{ACS_images/C56_disk.eps} \\[1ex]
C59 & Irr & Irr & disk & Irr\\[1ex] %&  \includegraphics[width=1.5cm, height=1.5cm]{ACS_images/C59_irr.eps}\\[1ex]
C65 & Irr & disk & disk & disk\\[1ex] %&  \includegraphics[width=1.5cm, height=1.5cm]{ACS_images/C65_disk.eps}\\[1ex]
C66 & Irr & disk & disk & disk \\[1ex]%&  \includegraphics[width=1.5cm, height=1.5cm]{ACS_images/C66_disk.eps}\\[1ex]
C67 & disk & disk & disk & disk\\[1ex] %&  \includegraphics[width=1.5cm, height=1.5cm]{ACS_images/C67_disk.eps}\\[1ex]
C84b & Irr & disk & Irr & Irr\\[1ex]  %&  \includegraphics[width=1.5cm, height=1.5cm]{ACS_images/C84b_irr.eps} \\[1ex]
C86 & ETG & ETG & x\tablefootmark{f} & ETG \\[1ex]%&  \includegraphics[width=1.5cm, height=1.5cm]{ACS_images/C86_etg.eps}\\[1ex]
C90c & \ldots & disk & \ldots & disk \\[1ex]%& \includegraphics[width=1.5cm, height=1.5cm]{ACS_images/C90c_disk.eps} \\[1ex]
C97a\tablefootmark{g} & Irr & Irr & disk & Irr\\[1ex] %&  \includegraphics[width=1.5cm, height=1.5cm]{ACS_images/C97a_irr.eps}\\[1ex]
C101b & Irr & disk & \ldots & Irr\tablefootmark{e}\\[1ex] %&  \includegraphics[width=1.5cm, height=1.5cm]{ACS_images/C101b_irr.eps}\\[1ex]
C105 & Irr & disk & \ldots & Irr\tablefootmark{e}\\[1ex] %&  \includegraphics[width=1.5cm, height=1.5cm]{ACS_images/C105_irr.eps}\\[1ex]
C112 & Irr & Irr & \ldots & Irr\\[1ex] %&  \includegraphics[width=1.5cm, height=1.5cm]{ACS_images/C112_irr.eps}\\[1ex]
C122a & Irr & Irr & disk & Irr \\[1ex] %&  \includegraphics[width=1.5cm, height=1.5cm]{ACS_images/C122a_irr.eps}\\[1ex]
C126\tablefootmark{g} & Irr & Irr & \ldots & Irr \\[1ex]%&  \includegraphics[width=1.5cm, height=1.5cm]{ACS_images/C126_irr.eps}\\[1ex]
C127 & disk & disk & \ldots & disk \\[1ex]%&  \includegraphics[width=1.5cm, height=1.5cm]{ACS_images/C127_disk.eps}\\[1ex]
\hline 
\end{tabular} 
\tablefoot{In the Cassata and Tasca catalogues the term spiral is used instead of disk, the term used in the Z{\"u}rich catalogue; we adopt the term disk. The term Irr refers to an irregular galaxy, while ETG stands for an early-type galaxy. An ellipsis means that the source was not found in the catalogue (within a search radius of $1\arcsec$).\tablefoottext{a}{Cassata's morphological catalogue (\cite{cassata2007}).}\tablefoottext{b}{Tasca's morphological catalogue v1.0 (\cite{tasca2009}). This catalogue provides three different morphological classifications, but following the Tasca et al. preference, we employed the class named class$_{-}$int.}\tablefoottext{c}{Z{\"u}rich structure \& morphology catalogue v1.0 (\cite{scarlata2007}; \cite{sargent2007}).}\tablefoottext{d}{The adopted morphological classification.}\tablefoottext{e}{See text for details on how the morphology was chosen among the different classifications.}\tablefoottext{f}{The source was found in the Z{\"u}rich catalogue, but no morphological classification was given.}\tablefoottext{g}{The source was not detected at 3~GHz.}  }
\end{minipage} 
}
\end{table}

\subsection{Why does the radio emission from submillimetre galaxies appear to be more spatially extended than the dust-emitting region ?}

\subsubsection{The importance of cosmic-ray electron diffusion}

As already discussed in M15, the diffusion of CR electrons in the galactic magnetic field, which was suggested by S15 to lead to a more spatially extended radio emission than the rest-frame FIR emission, appears unlikely to be an important process in SMGs. In Appendix~D, we derive the CR electron cooling time and diffusion length scale in five of our SMGs that benefit from both sub-arcsecond resolution submm interferometric imaging and the physical parameters needed in the analysis. Most notably, a typical diffusion length is found to be only of the order of ten parsecs (mean $\sim 11$~pc and median $\sim 9$~pc), which is not sufficient by orders of magnitude to explain the discrepancy between the radio and dust-emitting sizes.

Considering the nominal values of the median sizes of the dust-emitting regions from Ikarashi et al. (2015), Hodge et al. (2016), and Simpson et al. (2016), that is, $1.4\pm0.3$~kpc, $3.1\pm0.3$~kpc, and $2.7\pm0.4$~kpc, respectively, a (projected) diffusion length of $\sim0.75-1.6$~kpc from the outskirts of the dust-emitting region would be required to explain the median 3~GHz radio size we derived ($4.6\pm0.4$~kpc). A more quantitative comparison could be performed for our sources AzTEC/C4, C5, C17, and C42, where the radio emission appears more extended than the dust emission, and where the required CR electron diffusion lengths are 3.4~kpc, 1.0~kpc, 1.2~kpc, and 2.3~kpc, respectively, but the estimated diffusion length scales are 243, 111, 150, and 329 times shorter (Appendix~D, and Table~\ref{table:cooling} therein). Hence, the extended radio-emitting sizes of SMGs appear to necessitate some other physical mechanism(s) than the diffusion of CR electrons. Yet, the higher the CR electron energy is, the more rapid will be its energy loss due to synchrotron radiation and inverse Compton cooling ($\tau \propto E_{\rm e}^{-1}$ for both processes). Put differently, lower energy electrons can emit non-thermal radio emission for a longer period, and travel further away from their injection sites (e.g. \cite{clemens2010}). Another complicating factor is that, besides diffusion, relativistic electrons in highly star-forming galaxies such as SMGs can be transported out of the galactic disk into the halo by advection in the large-scale galactic winds (e.g. \cite{zirakashvili2006}; \cite{yoasthull2013}). Shocks associated with galactic outflows or superwinds can also (re-)accelerate CRs, in which case radio synchrotron emission can be detected further away from the central parts of the galaxy (e.g. \cite{varenius2016}).

\subsubsection{Bias caused by a radial dust temperature gradient}

As a working hypothesis, an SMG can be thought to be composed of a warm, central starburst region superposed 
on a colder, more extended disk. Such a configuration would exhibit a radial, negative dust temperature gradient 
($T_{\rm dust}$ decreasing inside-out). In principle, a $T_{\rm dust}$ gradient could cause the rest-frame FIR size measurements of SMGs 
to be biased towards the warm central region, and hence yield the compact sizes found 
in the ALMA studies (\cite{ikarashi2015}; S15; \cite{hodge2016}; \cite{simpson2016}).

To test the aforementioned scenario, in Appendix~E we investigate the dust temperature gradient in AzTEC/C5, which is 
the best-studied SMG in our sample. Under the premise that $\beta=1.5$ across the galaxy, we find that 
the angular $T_{\rm dust}$ profile derived using circular annuli can be well fitted by a Plummer-like function (\cite{plummer1911}) of the form 

\begin{equation} 
\label{eq:T}
T_{\rm dust}(r)=\frac{24.3~{\rm K}}{1+\left(\frac{r}{0\farcs436}\right)^{2.378}}\,.
\end{equation} 

A fundamental minimum value to the dust temperature at all redshifts is set by the CMB radiation (e.g. \cite{dacunha2013}). The CMB temperature at the redshift of AzTEC/C5 is $T_{\rm CMB}(z=4.3415)\simeq14.6$~K. Equation~(\ref{eq:T}) suggests that the temperature of the cold dust component drops to the CMB temperature level at $r=0\farcs367$, which corresponds to 2.46~kpc. This is comparable to the radius of the radio-emitting region (half the major axis FWHM), namely $0\farcs335$ or 2.25~kpc. Undoubtedly, the monochromatic ($\lambda_{\rm obs}=870$~$\mu$m) analysis of the intrinsic dust emission and transmitted CMB presented in Appendix~E, combined with the assumptions of uniform dust optical thickness, emissivity, and opacity (i.e. no radial dependence), is too simplistic to yield truly physical $T_{\rm dust}$ values. The formal $T_{\rm dust}$ values derived through panchromatic SED fits are typically dominated by the warm dust component, which is related to the peak of the SED. In contrast, the observed-frame (sub-)mm flux densities are dominated by the colder, less optically thick dust component (out to $z\lesssim4$), which does not contribute much to the peak of the source SED. Nevertheless, if the radial dust temperature profile follows Eq.~(\ref{eq:T}), that is the temperature drops from $\sim24$~K in the nuclear region to a value comparable to the CMB temperature at a radius of about 2.5~kpc, then the effective $T_{\rm dust}(r)$ gradient would be $\sim4$~K~kpc$^{-1}$. Because the CMB temperature appears to be reached within the so-called flat inner region (i.e. $r=0\farcs367 < r_0=0\farcs436$ in Eq.~(\ref{eq:T})), our $T_{\rm dust}(r)$ analysis suggests that the rest-frame FIR-emitting region is likely to be physically compact, rather than an observational bias caused by a temperature gradient. However, high-resolution rest-frame FIR or submm continuum imaging at least at one additional wavelength would be required to construct a spatially resolved $T_{\rm dust}$ map, and explore the temperature gradient in more detail (e.g. \cite{shetty2009}).

There is at least one important supporting argument against $T_{\rm dust}$ gradient being the cause for compact dust-emitting sizes of SMGs: if the outskirts of an SMG, which are emitting in radio but not strongly in the rest-frame FIR, are characterised by very cold dust and gas (as compared to $T_{\rm CMB}(z)$ at the source redshift), one would not expect those parts to exhibit emission from relatively high-excitation CO transitions (e.g. $J=7-6$ with $E_{\rm up}/k_{\rm B}=154.87$~K), unless the rotational levels are significantly sub-thermally populated. This is in contradiction with the observation that the median CO-emitting size from Tacconi et al. (2006) is comparable to our median radio size. Hence, the radio-emitting parts outside the central (starburst) region are expected to have fairly high gas temperatures, and hence also warm dust if the collisional gas-dust coupling is strong enough. On the other hand, the very extended CO$(1-0)$ emission sizes of SMGs (e.g. \cite{riechers2011a},b) reveal the presence of spatially more extended, colder ISM reservoirs in the outermost parts.

Considering the dust temperature, it is noteworthy that the compact FIR-emitting sizes of SMGs found in the ALMA studies discussed in Sect.~4.2.2 are fully consistent with the $L_{\rm IR}-T_{\rm dust}$ relationship seen among dusty starbursts, which from a theoretical point of view suggests a maximum radius of only $\sim2$~kpc for the starburst region (\cite{yan2016}). To give an illustrative example, we can use the Stefan-Boltzmann law to connect the IR luminosities and luminosity-weighted dust temperatures of AzTEC/C5 and C17 ($L_{\rm IR}=(1.4^{+0.1}_{-0.0})\times10^{13}$~L$_{\sun}$ and $T_{\rm dust}=42.8^{+2.0}_{-2.1}$~K, and $L_{\rm IR}=(7.8^{+2.4}_{-2.2})\times10^{12}$~L$_{\sun}$ and $T_{\rm dust}=38.1^{+9.9}_{-3.7}$~K, respectively; O.~Miettinen et al., in prep.; Appendix~D herein) to their size (radius $R$) given by 

\begin{equation} 
R=\sqrt{\frac{L_{\rm IR}}{4\pi \sigma_{\rm SB} T_{\rm dust}^4}}\,,
\end{equation} 
where $\sigma_{\rm SB}$ is the Stefan-Boltzmann constant (very similar to the generalised modified blackbody 
($\beta=1.5$) equivalent; see Eq.~(5) in \cite{yan2016}). For AzTEC/C5, we obtain $R\simeq1.5^{+0.3}_{-0.1}$~kpc, 
and for AzTEC/C17, $R\simeq1.4\pm0.6$~kpc.
The radius of the dust-emitting region based on this simple analysis of a spherical 
blackbody agrees within the uncertainties with the observationally determined rest-frame FIR size ($R_{\rm FIR}=1.3$~kpc for C5 and $R_{\rm FIR}=1.2$~kpc for C17; Appendix~D); see also \cite{hodge2016}. For comparison, the radius (half the major axis FWHM) of the radio-emitting region of AzTEC/C5 is about 2.3~kpc, and 2.4~kpc for C17. The cases of AzTEC/C5 and C17 demonstrate how the $L_{\rm IR}-T_{\rm dust}$ relationship of starbursts is intimately linked to their FIR-emitting sizes.

%(projected angular distance about $0\farcs26$) 

%For comparison, Miettinen et al. (2016) derived a luminosity-weighted 
%dust temperature of $T_{\rm dust}=41.4^{+0.9}_{-1.2}$~K.

%the typically negative outward dust temperature gradient in galaxies (\cite{hunt2015} )

%T gradient 1994MNRAS.270..641H; Scoville 2012 is better here

\subsubsection{Metallicity gradient bias}

In principle, another factor behind the dust emission gradient (and dust-radio size discrepancy) could be a large radial metallicity gradient in the galactic disk. This is based on the positive correlation between the dust-to-gas mass ratio and the gas-phase metallicity (e.g. \cite{franco1986}; \cite{draine2007}). A metallicity gradient can arise from a differential chemical enrichment of the ISM through SNe, stellar winds, and planetary nebulae, and hence is related to the star formation history of the galaxy. On top of this, the dynamical processes, such as gas infall or accretion from the (pristine) circumgalactic and intergalactic medium, outflow activity (e.g. metal-enriched superwind), galaxy interactions, and gas stripping can contribute to the metallicity gradient of a galaxy. The other way round, if the dust-to-gas ratio is constant across the galaxy, then the dust distribution would be expected to follow that of the gas, which is in tension with observations.

A strong radial metallicity gradient in SMGs might be disfavoured by their large CO$(1-0)$-emitting sizes (see Sect.~1, and references therein). This is because in the case of a strong metallicity gradient, one would expect a steep radial gradient in the CO-to-H$_2$ conversion factor as well (the conversion factor increases with decreasing metallicity; e.g. \cite{narayanan2012}; see \cite{bolatto2013} for a review). However, one potential observational support for SMGs having a radial metallicity gradient is that local and low to intermediate-redshift ($z\lesssim 1$) elliptical galaxies are found to exhibit colour gradients, which can be attributed to stellar metallicity gradients (the metallicity being higher in the central parts; e.g. \cite{franx1990}; \cite{peletier1990}; \cite{tamura2000}). If SMGs are indeed the early precursors of the present-day gas-poor ellipticals (e.g. \cite{swinbank2006}; \cite{fu2013}; \cite{toft2014}; \cite{simpson2014}), then the observed colour gradients in ellipticals could be (in part) an imprint of a metallicity gradient in the early SMG stage of this evolutionary pathway. A potential caveat is that if SMGs tend to show gravitational disk instabilities, which could also trigger their high observed SFRs (e.g. \cite{dekel2009}), the associated turbulent radial metal mixing can flatten or smooth out the metallicity gradient in the disk (e.g. \cite{ceverino2016}). Galaxy interactions can also redistribute the gas component in such a way that the metallicity gradient becomes flattened. Moreover, if the present-day ellipticals were formed through minor mergers from $z\sim2$ compact, quiescent galaxies, which themselves might be the descendants of $z\gtrsim3$ SMGs (\cite{toft2014}), then the difference in stellar metallicities in nearby ellipticals could just be the result of an inside-out size growth through accretion of different stellar populations via minor mergers (\cite{bezanson2009}; \cite{forbes2011}).

In the case of a negative metallicity gradient, the opacity of the dust grains is expected to be lower in the outer parts of the galactic disk (e.g. \cite{bate2014}, and references therein). Hence, the reprocessing of the UV-optical radiation from high-mass stars into the IR regime would be less efficient than in the inner parts which, in concordance with observations, which would cause the FIR-emitting region to be confined in the central area. 

High-resolution imaging of forbidden nebular line emission (e.g. [\ion{O}{ii}], [\ion{O}{iii}], and [\ion{N}{ii}]) from SMGs would be helpful to quantitatively investigate the spatial distribution of the gas-phase metallicity, and to compare it with the dust emission size scale. Although very difficult in practice (e.g. \cite{troncoso2014}), the future metallicity measurements with the \textit{James Webb Space Telescope} (\textit{JWST}; see \cite{gardner2006} for a review)\footnote{{\tt http://www.jwst.nasa.gov}.} will be very useful for these types of studies.

%pitfalls: the gas metallicity gradient can be stronger than the stellar metallicity gradient (2004, 606, 32) 

\subsubsection{The role of interacting galaxy pairs: Is a typical submillimetre galaxy a Taffy-like system ?}

In M15, we discussed a scenario where the spatially extended radio emission from SMGs might be the result of a dynamical interaction between two disk galaxies. Such a process has the potential to drag out the magnetic fields and CRs from the interacting, and ultimately interpenetrating disks (e.g. \cite{condon1993}; \cite{murphy2013}). As a result, and as has been observed in the Taffy systems UGC~12914/5 and UGC~813/6 (\cite{condon1993}, 2002), a synchrotron-emitting bridge can be formed between the two galaxies. Because this non-thermal radio emission is not linked to SNe (and hence to high-mass star formation), but rather is produced via the shock acceleration of CR electrons in the supersonic galaxy collision, its spatial scale is naturally different (larger) than that of active star formation which is bright in the rest-frame FIR continuum. 

Interestingly, Braine et al. (2003) and Zhu et al. (2007) found that the bridge region in the after-head-on-collision Taffy system 
UGC~12914/5 is very rich in molecular gas through their CO line observations (though the bridge gas is predominantly atomic). The origin of this dense gas might be in the disks of the colliding galaxies, from which the giant molecular clouds were being pulled out during the interaction (\cite{zhu2007}). At least qualitatively, this is consistent with our median radio size to be comparable to the median mid-$J$ CO-emitting size from Tacconi et al. (2006). 

\begin{figure}[!htb]
\centering
\resizebox{\hsize}{!}{\includegraphics{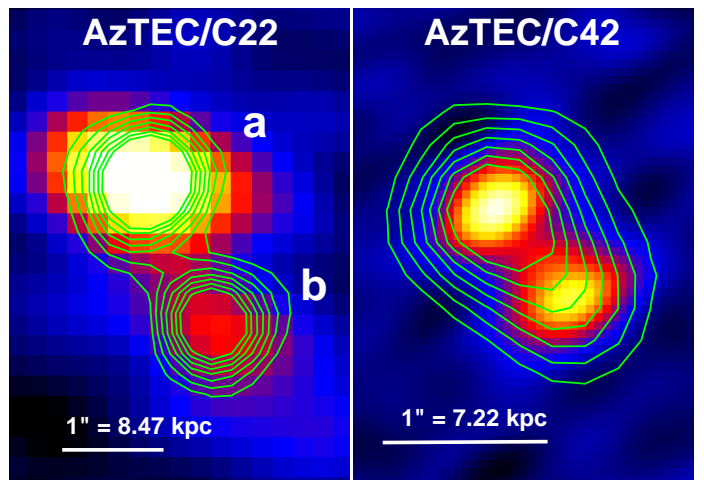}}
\caption{\textbf{Left:} ALMA 1.3~mm image of AzTEC/C22, overlaid with contours of 3~GHz radio emission. The ALMA image has an angular resolution of $1\farcs58 \times 0\farcs93$ (FWHM). \textbf{Right:} ALMA 994~$\mu$m image of AzTEC/C42, overlaid with contours of 3~GHz radio emission. The ALMA image has an angular resolution of $0\farcs52 \times 0\farcs30$ (FWHM). In both panels, the contours start at $3\sigma$, and are incremented in $2\sigma$ steps. A scale bar of $1\arcsec$ projected length is shown in both panels. }
\label{figure:taffy}
\end{figure}

Among our target SMGs, AzTEC/C22 and AzTEC/C42 are good candidates for Taffy-like systems, as illustrated in Fig.~\ref{figure:taffy}. The ALMA 1.3~mm image of AzTEC/C22 shows the presence of two dust-emitting bodies separated by 13.8~kpc in projection, both associated with 3~GHz radio emission with a radio-emitting bridge connecting them. The SMG AzTEC/C42 is unresolved in our ALMA 1.3~mm image, but was resolved into two components with ALMA at about three times higher resolution at $\lambda_{\rm obs}=994$~$\mu$m (Cycle~1 ALMA project 2012.1.00978.S; PI: A.~Karim). The detected 3~GHz radio emission encompasses both the dust-emitting components, which are separated by 5.3~kpc. A well-separated binary nature of AzTEC/C22 and C42 suggests an incomplete, or an early-stage SMG-SMG merger, and part of the radio emission from these systems is likely to arise from a magnetised medium between them. 

Iono et al. (2016) found that AzTEC/C4 is resolved into two components at $0\farcs064 \times 0\farcs057$ resolution with ALMA at $\lambda_{\rm obs}=860$~$\mu$m, which indicates a mid-stage major merger with a projected separation of $\sim1.5$~kpc. At still higher resolutions of $0\farcs017 \times 0\farcs014$ and $0\farcs026 \times 0\farcs018$, the authors found that AzTEC/C2a and C5 both exhibit a double nucleus structure with a separation of $\sim200$~pc and $\sim150$~pc between the nuclei, respectively. This suggests that the latter two SMGs are observed near the final stages of merging. For AzTEC/C2a, C4, and C5, we found that the radio-emitting size is $0.7^{+0.6}_{-0.4}$, $2.9^{+2.3}_{-1.2}$, and $1.7^{+0.5}_{-0.5}$ times the rest-frame FIR-emitting size (Appendix~D), which might reflect the different merger stages. 

For his sample of 31 local starbursts with $L_{\rm IR}\sim1.6\times10^{11}-3.7\times10^{12}$~L$_{\sun}$, Murphy (2013) found 
that the sources classified as ongoing mergers or post-merger systems exhibit spectral index flattening at $\nu<10$~GHz as 
a result of high free-free optical thickness. Among the aforementioned systems, we derived a flat rest-frame spectral index of 
$\alpha_{\rm 3.6\, GHz}^{\rm 7.8\, GHz}=-0.37\pm0.27$ for AzTEC/C22a (the northern component), and a fairly flat value of 
$\alpha_{\rm 6.5\, GHz}^{\rm 13.9\, GHz}=-0.50\pm0.24$ for AzTEC/C42. For AzTEC/C4, we derived a lower limit of $\alpha_{\rm 8.8\, GHz}^{\rm 18.9\, GHz}>-0.91$. For AzTEC/C2a and C5, where Iono et al. (2016) found two closely separated nuclei, we found no evidence of spectral index flattening ($\alpha_{\rm 5.9\, GHz}^{\rm 12.5\, GHz}=-0.95\pm0.32$ and $\alpha_{\rm 7.5\, GHz}^{\rm 16.0\, GHz}=-0.69\pm0.61$, respectively). Hence, our candidate Taffy systems, AzTEC/C22 and C42, could potentially be high-redshift manifestations of the trend found by Murphy (2013). However, Murphy (2013) also found that those starbursts that exhibit the steepest spectral indices at high frequencies (defined as $\nu>4$~GHz; $\sim12$~GHz on average) are found among the ongoing-merger systems in which the two components are separable and they either share a common radio-continuum envelope or display strong stellar tidal features (or both), that is in Taffy-like systems.

Owing to the excess radio emission from a synchrotron bridge, a dynamically interacting galaxy pair is expected to show a lowered IR-to-radio luminosity ratio with respect to the IR-radio correlation of star-forming galaxies (e.g. \cite{murphy2013}; \cite{donevski2015}). This is possibly exemplified by AzTEC/C22a and C42 for which the total-IR-radio correlation parameters are $q_{\rm TIR}=2.19\pm0.19$ and $q_{\rm TIR}=2.49\pm0.09$, respectively (revised from \cite{miettinen2017} (see Eqs.~(1) and (2) therein) owing to the updated redshift for C42 (\cite{brisbin2017}) and improved $L_{\rm IR}$ and $S_{\rm 325\, MHz}$ estimates for both sources from O.~Miettinen et al., in prep.). Although these values are only $1.21\pm0.10$ and $1.06\pm0.04$ times lower than the local universe median value of $q_{\rm TIR}=2.64$ (\cite{bell2003}; \cite{sargent2010}), this could be an indication of a weak excess radio emission not related to star formation and SN activity. For AzTEC/C2a, C4, and C5, which show a mismatch between the radio and rest-frame FIR-emitting sizes (see above), the $q_{\rm TIR}$ values revised from Miettinen et al. (2017) are $2.37\pm0.07$, $<2.32$, and $2.24\pm0.13$, respectively. These lower than local $q_{\rm TIR}$ parameters could also be potential indications of radio excess emission caused by process(es) not linked to the evolution of high-mass stars (there is also no evidence for a buried AGN in these SMGs). Curiously, the highest $q_{\rm TIR}$ source among these three sources is AzTEC/C2a, which appears to have a smaller radio size compared to its dust-emitting region, but this is not a significant outlier owing to the measurement uncertainties. 

Because the aforementioned low-$q_{\rm TIR}$ SMGs lie at high redshifts (C2a at $z_{\rm spec}=3.179$, C4 at $z=5.30^{+0.70}_{-1.10}$, C5 at $z_{\rm spec}=4.3415$, C22a at $z_{\rm spec}=1.599$, and C42 at $z_{\rm phot}=3.63^{+0.37}_{-0.56}$), their low $q_{\rm TIR}$ values might reflect a possible decreasing evolution of $q_{\rm TIR}$ as a function of redshift (e.g. \cite{ivison2010}; \cite{magnelli2015}; \cite{delhaize2017}). If the evolution has a functional form of $q_{\rm TIR}(z)=(2.88\pm0.03)\times (1+z)^{-0.19\pm0.01}$ derived for star-forming galaxies in COSMOS (\cite{delhaize2017}), the expected values at the nominal redshifts of AzTEC/C2a, C4, C5, C22a, and C42 would be $q_{\rm TIR}=2.19^{+0.06}_{-0.05}$, $q_{\rm TIR}= 2.03^{+0.06}_{-0.06}$, $q_{\rm TIR}= 2.09^{+0.06}_{-0.05}$, $q_{\rm TIR}= 2.40^{+0.05}_{-0.05}$, and $q_{\rm TIR}= 2.15^{+0.06}_{-0.05}$. The observed $q_{\rm TIR}$ values for these SMGs are mostly in fairly good agreement with the aforementioned redshift evolution ($q_{\rm TIR}^{\rm obs}/q_{\rm TIR}^{\rm predicted}= 1.08^{+0.06}_{-0.06}$ for C2a, $<1.18$ for C4, $1.07^{+0.09}_{-0.09}$ for C5, $0.91^{+0.10}_{-0.09}$ for C22a, and $1.16^{+0.07}_{-0.07}$ for C42). If the observed redshift dependence of $q_{\rm TIR}$ is physical, one possible physical explanation for this is an excess radio emission arising from a process(es) not linked to star formation; it is tempting to associate a decreasing $q_{\rm TIR}(z)$ trend with an enhanced galaxy interaction rate, and hence a higher synchrotron radio bridge occurrence at higher redshifts. The finding that most of our $q_{\rm TIR}$ values appear to be slightly higher than predicted by the Delhaize et al. (2017) relationship could simply reflect the fact that SMGs are very bright IR emitters (i.e. they exhibit an IR excess emission), but also a relative radio deficit is possible if the magnetic field strength is enhanced as a result of galaxy interaction (as opposed to a radio excess emission from the bridge in Taffy-like systems), which would cause the CR electrons to cool faster via synchrotron emission (see Appendix~D.2).

Similarly, one might think of a scenario where galaxy interactions lead to expanded radio-emitting regions of SMGs, and hence erase the potential radio size evolution over cosmic time, yielding an apparent flat trend between the average radio size and redshift we see in Fig.~\ref{figure:corr}. However, it is unclear whether the merger rate among our SMGs is high enough to account for the required radio size growth of the high redshift sources. Considering the merger systems discussed in this section, the number of irregular sources tabulated in Table~\ref{table:morphology}, and the occurrence of disturbed or clumpy morphologies based on visual inspection of multiwavelength images of our target SMGs (see \cite{brisbin2017}), we obtain a rough estimate of the merger percentage among our 152 target SMGs of $\sim25\%\pm4\%$. The latter percentage is likely a lower limit because it is based on morphological information only, while kinematic analysis could reveal a larger number of potential mergers (e.g. \cite{engel2010}). We note that the percentage of starbursts among our target SMGs is $\sim40\%$ (defined as sources, which lie above the main sequence by a factor of $>3$; O.~Miettinen et al., in prep.), and if starbursts tend to be driven by mergers, we might expect the merger percentage to be closer to 40\%. Nevertheless, even if the merger occurrence among our SMGs were relatively high, it is unlikely that every merger event or stage would be associated with an increased radio-emitting size. In fact, the 3~GHz radio size of our starburst SMGs (be they mergers or not) appears to be somewhat smaller on average than the radio-emitting region of the main-sequence SMGs (O.~Miettinen et al., in prep.). Hence, it remains to be established what is the possible physical reason(s) behind the non-evolution of the average radio size of our SMGs shown in Fig.~\ref{figure:corr}.

To conclude, our present results point towards the possibility that some of the extended radio-emitting sizes of SMGs are the result of galaxy interactions that can also eventually lead to a merger. This conforms to the observed high rates of dust-enshrouded star formation in our SMGs (\cite{miettinen2017}; O.~Miettinen et al., in prep.), which can be powered by such gravitational interactions. Although Taffy-type systems are rare in the local universe, collisions between gas-rich galaxies are believed to be much more frequent at high redshifts (where the gas fraction is higher and intergalaxy distances are shorter), and hence the appearance of Taffies is expected to be more common among high-redshift star-forming galaxies (\cite{gao2003}). That we found no dependence of the radio-emitting size on the source flux density (Fig.~\ref{figure:fluxsize}) might, however, be at odds with the aforementioned galaxy interaction scenario if brighter SMGs are more strongly clustered than their fainter counterparts (e.g. \cite{williams2011}; \cite{shimizu2012}). Hence, sources like AzTEC/C22 and C42 could also be unique SMG systems. The galaxy environments of our ALMA SMGs, including the possible trends with source brightness, will be investigated in a dedicated forthcoming paper.

%but also a radio deficit is possible in the case of a very recent ($\lesssim10$~Myr) burst of high-mass star formation when the SNe have not yet led to strong radio synchrotron emission relative to the thermal IR.

%sterile gas component in the outer parts of the SMGs

%the B field can be stronger due to a galaxy interaction, which would cause the CRS to lose thier energies on a shorter timescale (Mulcahy et al. 2016)

%factor-of-two variation

%on the journey

%larger distnaces thanasociated dust-heating photons

\section{Summary and conclusions}

We explored the $\nu_{\rm obs}=3$~GHz radio-emitting sizes of a large, flux density-limited sample of SMGs in the COSMOS field. The target SMGs were originally uncovered in a 1.1~mm continuum survey carried out with the AzTEC bolometer, and followed up with ALMA continuum imaging at $\lambda_{\rm obs}=1.3$~mm. Our main results are summarised as follows:

\begin{enumerate}
\item The radio detection rate is high: out of the 152 SMGs detected with ALMA at a ${\rm S/N}_{\rm 1.3\, mm}$ ratio of $\geq5$, 115 ($\sim76\%$) were found to have a $\geq 4.2\sigma$ radio counterpart at 3~GHz. The redshift distribution of the 3~GHz detected SMGs appears to be fairly similar to that of 3~GHz non-detections. 
\item In angular units, the median major axis (deconvolved FWHM) of the radio-emitting size was found to be $0\farcs59\pm0\farcs05$, while that of the linear major axis FWHM was derived to be $4.6\pm0.4$~kpc. The linear radio sizes appear to roughly follow a log-normal distribution. On average, the radio size shows no statistically significant evolutionary trend with redshift or galaxy morphology.
\item The median radio spectral index between the observed-frame 1.4~GHz and 3~GHz was found to be $\alpha_{\rm 1.4\, GHz}^{\rm 3\, GHz}=-0.67$, which is consistent with optically thin non-thermal synchrotron emission. The full sample median 3~GHz brightness temperature was found to be only $12.6\pm2$~K, which shows that the observed radio emission is predominantly powered by star formation and supernova activity. However, three of our SMGs  that have been detected with the VLBA at 1.4~GHz (AzTEC/C24b, 61, and 77a) exhibit elevated brightness temperatures. Most notably, AzTEC/C61 has $T_{\rm B}(3\,{\rm GHz})>10^{4.03}$~K, which is an indication of the presence of a compact AGN core in this X-ray detected SMG.
\item The median radio size we have derived is $\sim1.5-3$ times larger than the typical rest-frame FIR dust-emitting regions of SMGs found in high-resolution ALMA studies. On the other hand, a typical radio-emitting region of an SMG appears to extend over similar physical scales as the molecular gas component giving rise to mid-$J$ ($J=3-2$, $7-6$) CO rotational line emission. The observed spatial scale of stellar emission from SMGs shows a broad range of values, either larger or smaller than the radio-emitting region, which can be attributed to 
pre-existing stellar populations and differential dust obscuration.
\item The diffusion of cosmic-ray electrons in an SMG-type, strongly star-forming galaxy appears to be unable to explain the observed extended radio sizes as compared with the compact, dusty star-forming nuclear portions. Our case study of AzTEC/C5 suggests that the mean radial dust temperature profile can be fitted by a Plummer-like function with $T_{\rm dust}(r)\propto r^{-2.38}$, but the importance of $T_{\rm dust}(r)$ gradient in AzTEC/C5 as being the root physical cause of the compact FIR-emitting size seems unlikely on the basis of its SED properties ($L_{\rm IR}$ and $T_{\rm dust}$), although this
remains inconclusive owing to the simplistic monochromatic analysis allowed by the currently available data. Instead, our results bolster a scenario where SMG starbursts are triggered by galaxy interactions and mergers, because a supersonic galaxy-galaxy collision can create an extended, synchrotron-emitting bridge between the interacting pair. As the radio emission from such a dynamically interacting system is not solely arising from processes related to star formation, a deviation (radio excess) from the well-established IR-radio correlation can be understood, while simultaneously providing an explanation for the spatially extended radio emission with respect to that of the active star formation. 

Our ALMA Cycle~4 observations of 870~$\mu$m emission towards a significant subsample of the present target SMGs will allow us to make the dust and radio size comparison in a source-by-source fashion, and to reach better statistics and understanding of the occurrence of extended radio emission associated with compact dust-emitting SMGs. 
\end{enumerate}

\begin{acknowledgements}

We appreciate the exceptionally constructive and positive feedback from the referee. 
This research was funded by the European Union's Seventh Framework programme 
under grant agreement 337595 (ERC Starting Grant, 'CoSMass'). M.~A. acknowledges 
partial support from FONDE-CYT through grant 1140099. A.~K. acknowledges support by the Collaborative Research Council 956,
sub-project A1, funded by the Deutsche Forschungsgemeinschaft (DFG). E.~S. acknowledges funding from the European Research Council 
(ERC) under the European Union's Horizon 2020 research and innovation programme (grant agreement No. 694343). 
D.~R. acknowledges support from the National Science Foundation under grant number AST-1614213 to Cornell University. 

This work was performed in part at the Aspen Center for Physics, which is supported by National Science Foundation grant PHY-1066293. 
This work was partially supported by a grant from the Simons Foundation. The Flatiron Institute is supported by the Simons Foundation. 
This paper makes use of the following ALMA data: ADS/JAO.ALMA\#2012.1.00978.S and 
ADS/JAO.ALMA\#2013.1.00118.S. ALMA is a partnership of ESO (representing its member states), NSF (USA) and NINS (Japan), 
together with NRC (Canada), NSC and ASIAA (Taiwan), and KASI (Republic of Korea), in cooperation with the Republic of Chile. 
The Joint ALMA Observatory is operated by ESO, AUI/NRAO and NAOJ. 
This research has made use of NASA's Astrophysics Data System, and the NASA/IPAC Infrared Science Archive, 
which is operated by the JPL, California Institute of Technology, under contract with the NASA. This research has made use of 
the SIMBAD database, operated at CDS, Strasbourg, France. This research has made use of the NASA/IPAC Extragalactic Database (NED) 
which is operated by the Jet Propulsion Laboratory, California Institute of Technology, under contract with the National Aeronautics 
and Space Administration. This research made use of Astropy, a community-developed core Python package for Astronomy 
(\cite{astropy2013}). We gratefully acknowledge the contributions of the entire COSMOS collaboration consisting of more 
than 100 scientists. More information on the COSMOS survey is available at {\tt http://cosmos.astro.caltech.edu}. 

\end{acknowledgements}

\appendix

\section{3~GHz maps} 

The VLA 3~GHz contour maps of our 3~GHz detected SMGs are shown in Fig.~\ref{figure:maps}.

\begin{figure*}[!htb]
\begin{center}
\includegraphics[width=0.95\textwidth]{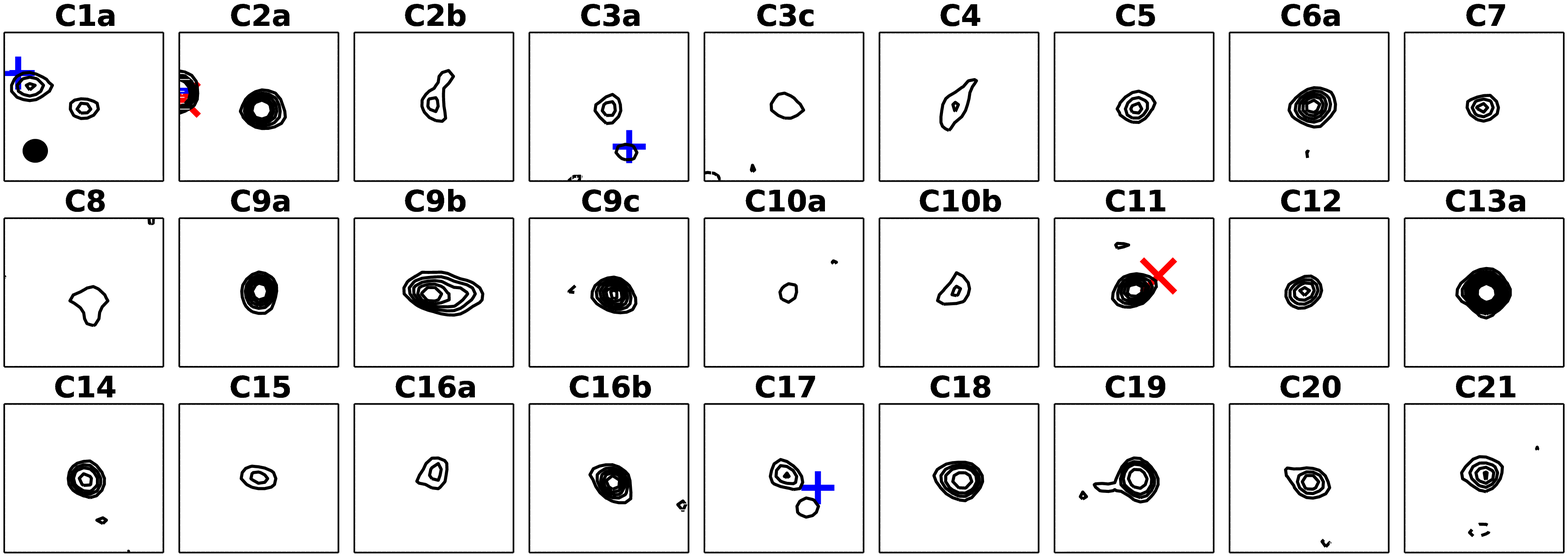}\\[-45pt]
\includegraphics[width=0.95\textwidth]{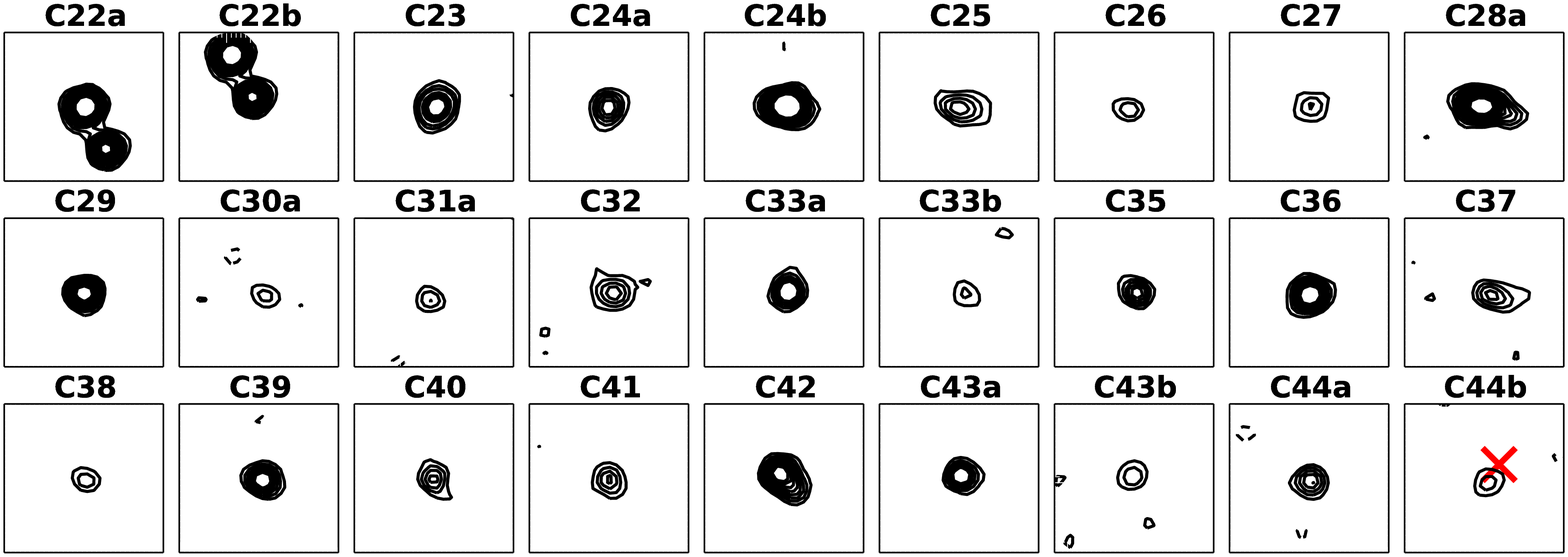}\\[-45pt]
\includegraphics[width=0.95\textwidth]{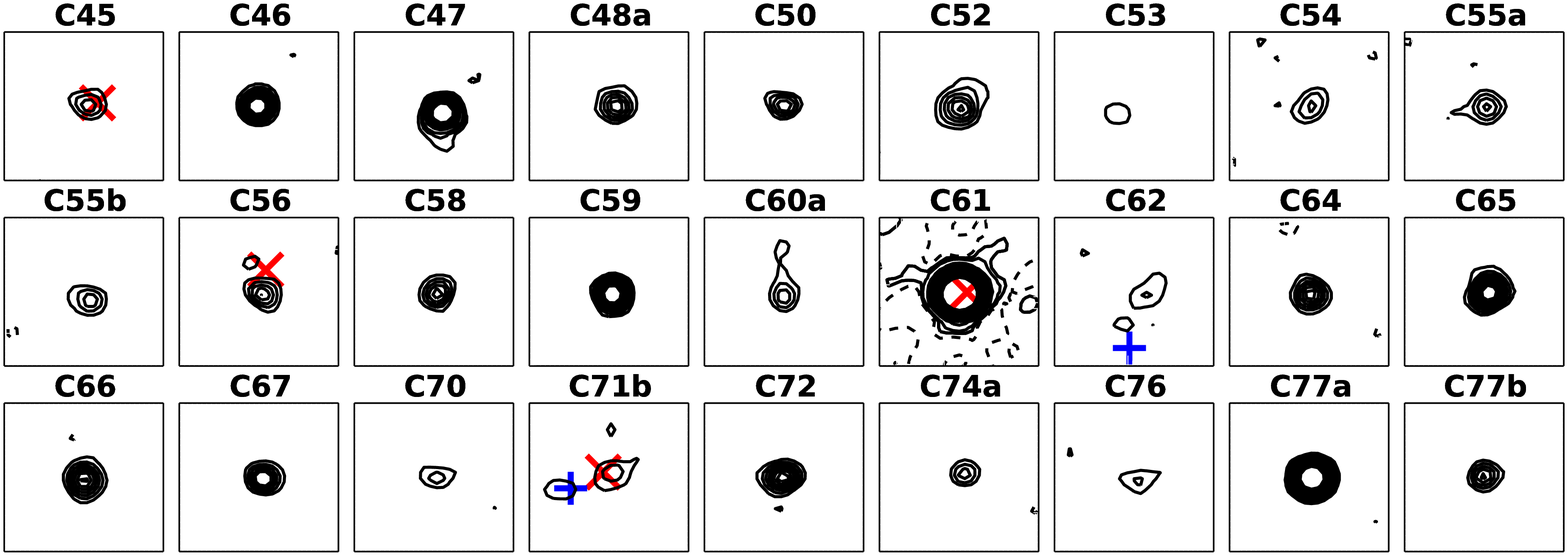}\\[-45pt]
\caption{VLA 3~GHz contour maps of our 3~GHz detected SMGs. Each image is centred on the ALMA 1.3~mm peak position, is $5\arcsec \times 5\arcsec$ in size, and displayed with north up and east left. The contour levels start from $3\sigma$, and progress in steps of $2\sigma$ except for AzTEC/C61 where the level step is $15\sigma$. The dashed contours show the negative features at the $-3\sigma$ level. The blue plus signs and red crosses mark, respectively, the positions of the optical-NIR and X-ray sources seen towards those 3~GHz sources that are discussed in Appendix~B and Sect.~2.1. The black filled circle in the first panel (AzTEC/C1a) shows the synthesised beam size of $0\farcs75$ (FWHM). The projected angular offset between the ALMA source position (i.e. the image centre position) and the 3~GHz peak position is given in column~(11) in Table~\ref{table:results}.}
\label{figure:maps}
\end{center}
\end{figure*}

\addtocounter{figure}{-1}
\begin{figure*}[!htb]
\begin{center}
\includegraphics[width=0.95\textwidth]{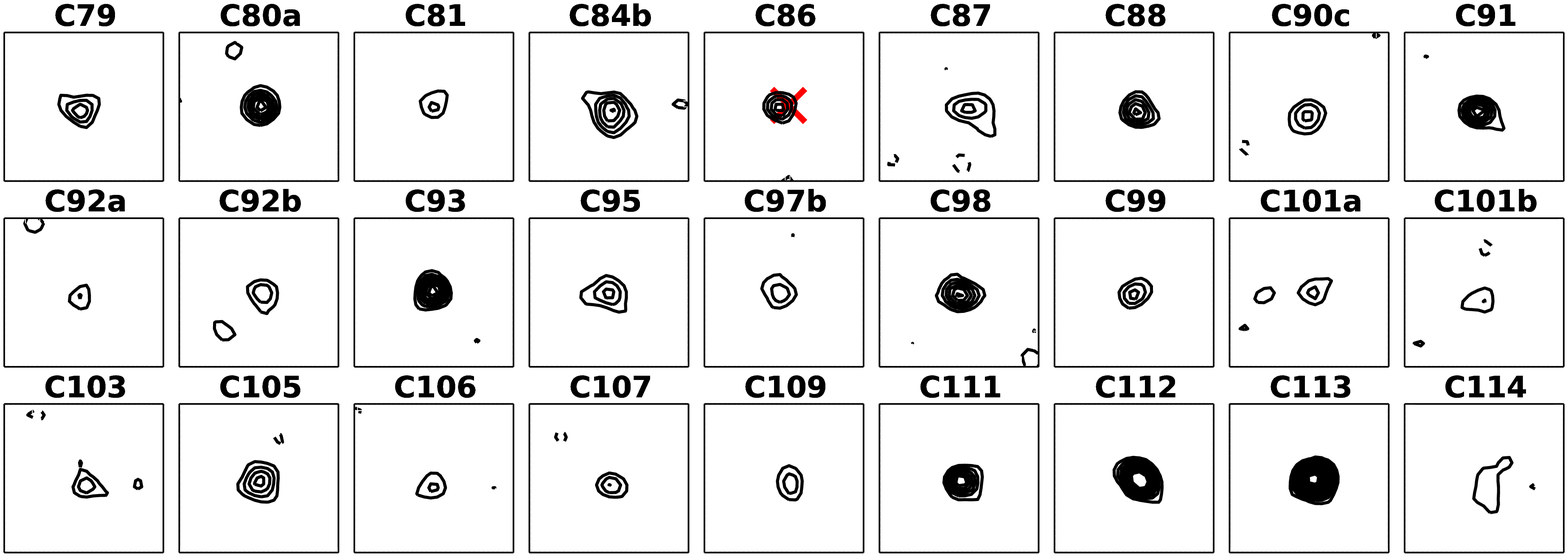}\\[-45pt]
\includegraphics[width=0.95\textwidth]{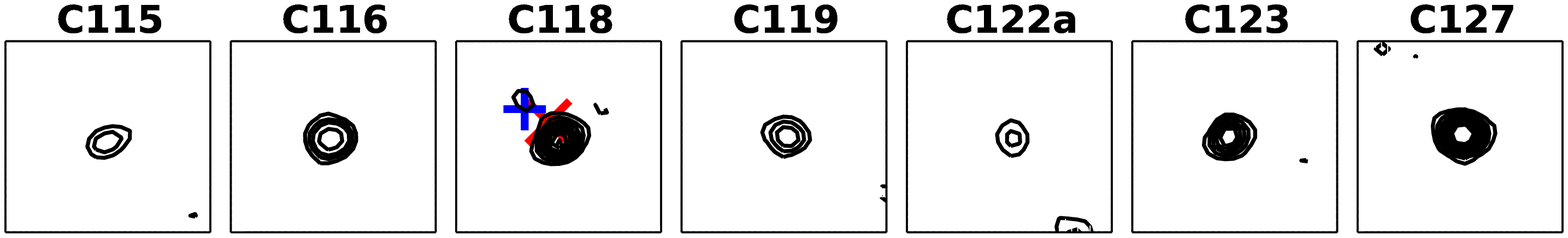}\\[-150pt]
\caption{continued.}
\label{figure:maps}
\end{center}
\end{figure*}

\section{Notes on additional 3~GHz sources seen near the ALMA detected submillimetre galaxies}

The following 12 3~GHz detected SMGs have an additional 3~GHz source ($\geq 3\sigma$) within the area shown in Fig.~\ref{figure:maps}: AzTEC/C1a, C2a, C3a, C17, C56, C62, C71b, C80a, C92a, C92b, C101a, and C118. We found a potential or possible counterpart for eight of these sources, and briefly describe them below. 

\textit{AzTEC/C1a}: A counterpart was found from the COSMOS2015 catalogue (ID 761831; \cite{laigle2016}), with an angular offset of $0\farcs58$ to the north-east from the 3~GHz position. It has a stellar mass of $M_{\star}=1.2\times10^8$~M$_{\sun}$ as reported in the COSMOS2015 catalogue, and a very secure spectroscopic redshift of $z_{\rm spec}=0.17236$ derived from a high quality spectrum (quality flag 4; M.~Salvato et al., in prep.). Hence, it is likely to be a foreground galaxy with respect to the $z_{\rm spec}=4.7$ SMG AzTEC/C1a. The very low stellar mass of the foreground galaxy suggests only a minimal lensing effect on our target SMG.
% UV coord: 150.4245537626, 2.4539209338 

\textit{AzTEC/C2a}: As we already discussed in M15, the additional radio source seen $2\farcs67$ to the north-east of AzTEC/C2a ($=$AzTEC~8) is associated with the X-ray detected galaxy CXOC J095959.5+023441 (\cite{salvato2011}). This source has an ID of CID-1787 in the \textit{Chandra} COSMOS Legacy Survey (\cite{civano2016}), and its position is indicated by a red cross in Fig.~\ref{figure:maps} ($0\farcs26$ offset from the 3~GHz position). The source was also detected with the VLBA at 1.4~GHz ($S_{\rm 1.4\, GHz}=83.8$~$\mu$Jy; N.~Herrera Ruiz et al., in prep.). The COSMOS2015 catalogue (\cite{laigle2016}) gives a redshift of $z_{\rm phot}=2.40^{+0.16}_{-0.10}$ for this source (ID 842595; $0\farcs14$ offset from the 3~GHz position), and a CO-based spectroscopic redshift of $z_{\rm spec}=1.950$ was derived by M.~S.~Yun et al. (in prep.; see \cite{iono2016}). Hence, the source in question appears to be a foreground galaxy with respect to the $z_{\rm spec}=3.179$ SMG AzTEC/C2a.
% Chandra Legacy: CID-1917 at 149.997910235, 2.57814865447 (0.26'' from the 3 GHz pos.); UV coord: 149.9979681785, 2.5782276345

\textit{AzTEC/C3a}: The additional 3~GHz source lies $0\farcs24$ away from the $z_{\rm phot}=0.45$ galaxy COSMOS 2006577 (\cite{capak2007}), and could therefore be associated with a foreground galaxy if the SMG AzTEC/C3a ($=$AzTEC~2) lies at $z_{\rm spec}=1.125$; the redshift of AzTEC/C3a is uncertain, and it could lie at a considerably higher redshift (see \cite{miettinen2017}; E.~F.~Jim{\'e}nez Andrade, in prep.). In M15, we also mentioned the possibility that the radio feature in question is related to a jet emanating from the SMG (i.e. a radio lobe). 
%from NED, COSMOS 2006577 lies at 150.033333, 2.436389

\textit{AzTEC/C17}: The COSMOS2015 catalogue contains a source (ID 841161; \cite{laigle2016}) that lies $0\farcs77$ north-west from the additional 3~GHz source, and $1\farcs10$ south-west from the SMG's 3~GHz position. The redshift of this source is $z_{\rm phot}=0.60^{+0.03}_{-0.07}$, and if related to the additional 3~GHz source, it is likely to be a foreground galaxy with respect to the $z_{\rm spec}=4.542$ SMG AzTEC/C17. The 3~GHz source configuration appears similar to that seen towards AzTEC/C3a, and a radio jetted SMG could be another explanation. 
%UV coord: 150.2267638236, 2.5766208663

\textit{AzTEC/C56}: The additional 3~GHz feature lies $0\farcs47$ to the north-east of the \textit{Chandra} COSMOS Legacy Survey X-ray source CID-2518, which has a reported redshift of $z_{\rm phot}=3.45^{+0.08}_{-0.10}$ (\cite{civano2016}; see also \cite{salvato2011} for the source CXOC J095905.0+022157), and which we have assumed to be associated with our SMG (the X-ray source lies $0\farcs86$ north of the SMG's 3~GHz position). Hence, the nature of the additional 3~GHz feature remains uncertain. 
%Chandra Legacy: CID-2518 at 149.770999429, 2.36597414625; 0.47'' from the 3 GHz pos.; photo-z is said to be 3.447

\textit{AzTEC/C62}: The additional 3~GHz feature lies $0\farcs72$ north from the COSMOS2015 catalogue source ID 879292, 
which has a redshift of $z_{\rm phot}=4.42^{+1.07}_{-1.34}$ (\cite{laigle2016}). If these two sources are physically 
related, it could be a background galaxy seen towards the $ z_{\rm FIR}=3.36^{+0.97}_{-0.97}$ SMG AzTEC/C62, but the large redshift 
uncertainties do not allow any firm conclusions to be drawn.  
%UV coord: 150.2520707184, 2.6344654836

\textit{AzTEC/C71b}: The additional 3~GHz feature lies $1\farcs73$ ($1\farcs61$) to the south-east of the 3~GHz (1.3~mm) peak position of AzTEC/C71b. Although the additional 3~GHz feature has an angular offset of $0\farcs32$ from the COSMOS2015 catalogue source 669348 ($z_{\rm phot}=0.85\pm0.01$; \cite{laigle2016}), the latter source was associated with our SMG that lies $1\farcs26$ to the north-west of this optical-NIR source (\cite{brisbin2017}). A spectroscopic redshift of $z_{\rm spec}=0.8521$ has been measured for this source, which is in excellent agreement with the COSMOS2015 photometric value (M.~Salvato et al., in prep.). The \textit{Chandra} COSMOS Legacy Survey X-ray source CID-1522 ($z_{\rm spec}=0.829$; \cite{civano2016}; see also \cite{salvato2011} for the source CXOC J095953.8+021854), which can be associated with our SMG ($0\farcs25$ from the ALMA position), lies $1\farcs51$ to the north-west of the additional 3~GHz feature. The nature of the latter source remains unclear, but it is possible that it is associated with the aforementioned COSMOS2015 source, and physically linked to the SMG.
%UV coord: 149.9747345203, 2.3149032265; X-ray at 149.974167, 2.315000; C71b has an X-ray counterpart CID-1522 at 149.974449978, 2.3150549065 (0.25'' from the ALMA position); 

\textit{AzTEC/C118}: The additional 3~GHz feature has a nearby ($0\farcs33$ offset) counterpart in the COSMOS2015 catalogue, with a reported photo-$z$ of $z_{\rm phot}=0.94\pm0.02$ (ID 531461; \cite{laigle2016}). The projected offset of the additional 3~GHz feature from the \textit{Chandra} COSMOS Legacy Survey X-ray source CID-838 (\cite{civano2016}; see also \cite{salvato2011} for the source CXOC J095959.9+020633, and \cite{civano2012}; \cite{johnson2013}) we have associated with the SMG is $0\farcs83$. Because the redshift of the SMG is $z_{\rm spec}=2.2341$, the additional radio feature could belong to a foreground galaxy. 
%Chandra Legacy: CID-838 at 149.9998400115, 2.1093708927, 0.83'' from the 3 GHz pos.; UV coord: 150.0000004331, 2.1094658625

\section{Radio properties of the target submillimetre galaxies at the observed frequency of 3~GHz}

The radio emission parameters derived in Sect.~3 are listed in Table~\ref{table:results}.

\begin{landscape}
\begin{table}
\caption{Radio continuum characteristics of the 3~GHz sources.}
{\tiny
\centering
\renewcommand{\footnoterule}{}
\label{table:results}
\begin{tabular}{c c c c c c c c c c c c c c}
\hline\hline 
Source ID & $z$\tablefootmark{a} & $\alpha_{2000.0}$ & $\delta_{2000.0}$ & $I_{\rm 3\, GHz}$\tablefootmark{b} & $S_{\rm 3\, GHz}$\tablefootmark{b} & S/N & \multicolumn{2}{c}{\underline{FWHM size}\tablefootmark{c}} & P.A.\tablefootmark{c} & Offset & $\alpha_{\rm 1.4\, GHz}^{\rm 3\, GHz}$\tablefootmark{d} & $T_{\rm B}$\tablefootmark{d} & Comments\tablefootmark{e} \\
       & & [h:m:s] & [$\degr$:$\arcmin$:$\arcsec$] & [$\mu$Jy~beam$^{-1}$] & [$\mu$Jy] & & [$\arcsec$] & [kpc] & [$\degr$] & [$\arcsec$] & & [K] &\\ 
\hline
AzTEC/C1a\tablefootmark{e} & $z_{\rm spec}=4.7$ & 10 01 41.75 & +02 27 12.92 & $13.7\pm2.2$ & $13.7\pm2.2$ & 6.2 & $<0.38$ & $<2.5$ & \ldots & 0.06 & $>-1.76$ & $>10.8$ & \\ [1ex]
%         & $z_{\rm spec}=4.7$ & 10 01 41.87 & +02 27 13.65 & $16.5\pm2.2$ & $25.7\pm5.1$ & 7.5 & $0.80^{+0.18}_{-0.22} \times < 0.38$ & $5.2^{+1.1}_{-1.4} \times <2.5$ & $90.7^{+16.8}_{-16.8}$ & 1.89 & & & \\ [1ex]
AzTEC/C2a & $z_{\rm spec}=3.179$ & 09 59 59.33 & +02 34 41.05 & $38.8\pm2.4$ & $49.4\pm4.8$ & 16.2 & $0.41^{+0.10}_{-0.17} \times < 0.38$ & $3.1^{+0.8}_{-1.3} \times < 2.9$ & $46.4^{+0}_{-0}$ & 0.12 & $-0.95\pm0.32$ & $40.0\pm26.6$ & AzTEC8\\ [1ex]
AzTEC/C2b & $z_{\rm synth}=1.10^{+2.60}_{-1.10}$ & 09 59 58.80 & +02 34 57.90 & $10.0\pm2.1$ & $51.1\pm12.6$ & 4.8 & $2.51^{+0.57}_{-0.58} \times 0.81^{+0.29}_{-0.37}$ & $20.5^{+4.6}_{-4.7}\times 6.6^{+2.4}_{-4.7}$ & $164.3^{+8.6}_{-8.6}$ & 0.08 & $>-0.28$ & $1.1\pm0.6$ & \\ [1ex]
AzTEC/C3a & $z_{\rm spec}=1.125$ & 10 00 08.04 & +02 26 12.26 & $15.0\pm2.4$ & $18.9\pm2.4$ & 6.3 & $0.54^{+0.22}_{-0.21} \times < 0.38$ & $4.4^{+1.8}_{-1.7} \times <3.1$ & $170.9^{+32.6}_{-32.7}$ & 0.09 & $-1.83\pm0.77$ & $8.8\pm7.4$ & AzTEC2\\ [1ex]
AzTEC/C3c & $z_{\rm 3\, GHz}=2.03^{+1.19}_{-0.31}$ & 10 00 08.01 & +02 26 10.81 & $10.0\pm2.4$ & $14.0\pm5.0$ & 4.2 & $0.72^{+0.31}_{-0.44} \times < 0.38$ & $10.3^{+3.1}_{-3.4} \times <6.8$ & $72.6^{+28.8}_{-28.8}$ & 0.63 & $>-1.28$ & $2.4\pm1.7$ & \\ [1ex]
AzTEC/C4 & $z_{\rm synth}=5.30^{+0.70}_{-1.10}$ & 09 59 31.70 & +02 30 43.96 & $11.4\pm2.3$ & $31.5\pm7.7$ & 5.0 & $1.72^{+0.37}_{-0.39} \times < 0.38$ & $10.5^{+2.3}_{-2.4} \times < 2.3$ & $150.6^{+8.3}_{-8.3}$ & 0.10 & $>-0.91$ & $1.4\pm0.7$ & AzTEC4 \\ [1ex]
AzTEC/C5 & $z_{\rm spec}=4.3415$ & 09 59 42.86 & +02 29 38.20 & $18.3\pm2.4$ & $28.3\pm5.2$ & 7.6 & $0.67^{+0.17}_{-0.20} \times 0.43^{+0.19}_{-0.30}$ & $4.5^{+1.1}_{-1.3} \times 2.9^{+1.3}_{-1.3}$ & $118.1^{+31.8}_{-31.8}$ & 0.03 & $-0.69\pm0.61$ & $8.9\pm5.0$ & AzTEC1\\ [1ex]
AzTEC/C6a & $z_{\rm spec}=2.494$ & 10 00 56.95 & +02 20 17.31 & $29.5\pm2.3$ & $43.6\pm5.2$ & 12.8 & $0.63^{+0.11}_{-0.13} \times 0.41^{+0.12}_{-0.17}$ & $5.1^{+0.9}_{-1.0} \times 3.3^{+1.0}_{-1.0}$ & $125.2^{+22.9}_{-23.0}$ & 0.11 & $-0.76\pm0.39$ & $15.2\pm6.0$ & Cosbo-3\\ [1ex]
%         & $1.90^{+0.22}_{-0.91}$ & 10 00 57.28 & +02 20 12.60 & $23.9\pm2.3$ & $28.8\pm4.6$ & 10.4 & $0.51^{+0.15}_{-0.18} \times <0.38$ & $4.3^{+1.2}_{-1.5} \times < 3.2$ & $53.8^{+20.4}_{-20.5}$ & 0.32 & $-0.66\pm0.32$ & $20.2\pm7.0$ &   \\ [1ex]
AzTEC/C7 & $3.06^{+1.88}_{-1.76}$ & 10 00 15.62 & +02 15 48.99 & $19.4\pm2.4$ & $19.4\pm2.4$ & 8.1 & $<0.38$ & $<2.9$ & $91.6^{+41.3}_{-41.2}$ & 0.08 & $>-1.18$ & $>16.0$ & Cosbo-1\\ [1ex]
AzTEC/C8b & $z_{\rm synth}=1.10^{+0.30}_{-0.20}$ & 10 00 13.85 & +01 56 39.10 & $9.7\pm2.1$ & $42.8\pm11.3$ & 4.6 & $1.70^{+0.43}_{-0.46} \times 1.10^{+0.34}_{-0.38}$ & $13.9^{+3.6}_{-3.7} \times 9.0^{+2.8}_{-3.7}$ & $178.6^{+26.4}_{-26.5}$ & 0.41 & $>-0.57$ & $2.0\pm1.2$ & COSLA9-S     \\ [1ex]
AzTEC/C9a & $2.68^{+0.24}_{-0.51}$ & 10 01 22.96 & +02 20 05.92 & $34.3\pm2.3$ & $39.1\pm4.4$ & $14.9$ & $0.48^{+0.10}_{-0.13} \times <0.38$ & $3.8^{+0.8}_{-1.0} \times < 3.0$ & $178.4^{+13.0}_{-13.1}$ & 0.004 & $-0.89\pm0.39$ & $23.4\pm11.4$ & \\ [1ex]
%          & $$ & $-0.89\pm0.39$ & $23.4\pm11.4$ & \\ [1ex]
AzTEC/C9b & $z_{\rm spec}=2.8837$ & 10 01 22.36 & +02 20 02.73 & $25.4\pm2.1$ & $83.8\pm8.9$ & $12.0$ & $1.75^{+0.17}_{-0.17} \times 0.62^{+0.12}_{-0.14}$ & $13.6^{+1.3}_{-1.4} \times 4.8^{+0.9}_{-1.4}$ & $85.2^{+4.7}_{-4.7}$ & 0.06 & $-0.70\pm0.33$ & $3.7\pm0.8$ & \\ [1ex]
AzTEC/C9c & $z_{\rm spec}=2.9219$ & 10 01 23.18 & +02 20 05.64 & $33.1\pm2.3$ & $40.3\pm4.5$ & $14.5$ & $0.53^{+0.11}_{-0.12} \times <0.38$ & $4.1^{+0.8}_{-0.9} \times <2.9$ & $55.0^{+13.1}_{-13.1}$ & 0.18 & $>-0.19$ & $19.3\pm8.3$ & \\ [1ex]
AzTEC/C10a & $z_{\rm 3\, GHz}=3.40^{+3.60}_{-0.59}$ & 10 00 13.52 & +02 34 23.78 & $10.2\pm2.3$ & $11.4\pm4.4$ & $4.4$ & $0.51^{+0.32}_{-0.51} \times <0.38$ & $3.8^{+2.3}_{-3.8} \times <2.8$ & $159.7^{+34.4}_{-34.4}$ & 0.16 & $>-2.66$ & $6.0\pm9.9$ & \\ [1ex]
AzTEC/C10b & $z_{\rm synth}=2.90^{+0.30}_{-0.90}$ & 10 00 12.95 & +02 34 34.92 & $12.2\pm2.4$ & $27.9\pm6.9$ & 5.1 & $1.21^{+0.29}_{-0.32} \times 0.50^{+0.26}_{-0.50}$ & $9.4^{+2.3}_{-2.5}\times 3.9^{+2.0}_{-2.5}$ & $146.3^{+31.0}_{-15.5}$ & 0.21 & $>-0.93$ & $2.6\pm1.5$ & AzTEC15\\ [1ex]
AzTEC/C11 & $4.30^{+0.07}_{-3.33}$ & 10 01 41.04 & +02 04 04.97 & $32.3\pm2.3$ & $37.6\pm4.4$ & 14.0 & $0.54^{+0.11}_{-0.12} \times <0.38$ & $3.7^{+0.7}_{-0.8}\times <2.6$ & $116.2^{+10.5}_{-10.5}$ & 0.02 & $>-0.30$ & $17.3\pm7.5$ & X-ray   \\ [1ex] 
AzTEC/C12 & $3.25^{+0.16}_{-0.51}$ & 10 01 36.82 & +02 11 10.06 & $23.4\pm2.3$ & $28.7\pm4.5$ & 10.2 & $0.56^{+0.14}_{-0.16} \times <0.38$ & $4.2^{+1.1}_{-1.3}\times <2.9$ & $125.8^{+16.0}_{-16.0}$ & 0.16 & $>-0.77$ & $12.3\pm7.0$ & COSLA17-N\\ [1ex] 
AzTEC/C13a & $2.01^{+0.15}_{-0.49}$ & 09 58 37.97 & +02 14 08.43 & $82.9\pm2.4$ & $82.9\pm2.4$ & 34.5 & $<0.38$ & $<3.2$ & \ldots & 0.07 & $-0.73\pm0.13$ & $>75.8$ &      \\ [1ex] 
AzTEC/C14 & $4.58^{+0.25}_{-0.68}$ & 09 59 57.29 & +02 27 30.54 & $29.4\pm2.2$ & $33.3\pm4.3$ & 13.4 & $0.40^{+0.13}_{-0.17} \times <0.38$ & $2.6^{+0.9}_{-1.1}\times <2.5$ & $33.2^{+24.1}_{-24.1}$ & 0.04 & $-0.94\pm 0.44$ & $28.3\pm21.5$ & AzTEC9\\ [1ex] 
AzTEC/C15 & $3.91^{+0.28}_{-2.35}$ & 10 01 31.55 & +02 25 16.15 & $13.9\pm2.3$ & $20.6\pm5.2$ & 6.1 & $0.84^{+0.23}_{-0.28} \times <0.38$ & $5.9^{+1.7}_{-1.9}\times <2.7$ & $75.3^{+15.7}_{-15.7}$ & 0.03 & $>-1.50$ & $4.0\pm2.6$ & \\ [1ex] 
AzTEC/C16a & $3.15^{+0.62}_{-1.54}$ & 09 58 53.69 & +02 16 52.88 & $12.6\pm2.2$ & $23.6\pm5.9$ & 5.8 & $0.95^{+0.25}_{-0.28} \times 0.44^{+0.25}_{-0.44}$ & $7.2^{+1.9}_{-2.2} \times 3.4^{+1.9}_{-2.2}$ & $150.8^{+21.0}_{-21.0}$ & 0.16 & $>-1.31$ & $3.6\pm2.2$ &  \\ [1ex] 
AzTEC/C16b & $2.39^{+0.27}_{-0.56}$ & 09 58 54.19 & +02 16 45.95 & $30.7\pm2.3$ & $38.5\pm4.6$ & 13.6 & $0.59^{+0.11}_{-0.12} \times <0.38$ & $4.8^{+0.9}_{-1.0} \times <3.1$ & $40.3^{+11.4}_{-11.4}$ & 0.18 & $-0.99\pm0.40$ & $15.0\pm6.1$ & \\ [1ex] 
AzTEC/C17 & $z_{\rm spec}=4.542$ & 10 00 54.49 & +02 34 36.24 & $16.0\pm2.2$ & $21.9\pm4.8$ & 7.2 & $0.70^{+0.19}_{-0.24} \times <0.38$ & $4.6^{+1.3}_{-1.5} \times <2.5$ & $54.0^{+17.7}_{-17.7}$ & 0.08 & $>-1.31$ & $6.1\pm4.0$ & \\ [1ex] 
AzTEC/C18 & $3.15^{+0.13}_{-0.44}$ & 10 00 35.30 & +02 43 53.27 & $36.5\pm2.5$ & $52.5\pm5.2$ & 14.6 & $0.63^{+0.10}_{-0.10} \times <0.38$ & $4.8^{+0.8}_{-0.8}\times <2.9$ & $78.1^{+14.8}_{-14.8}$ & 0.05 & $-0.82\pm0.34$ & $18.0\pm6.0$ & AzTEC12\\ [1ex] 
AzTEC/C19 & $2.87^{+0.11}_{-0.41}$ & 09 59 50.28 & +01 53 36.35 & $34.0\pm2.2$ & $47.0\pm4.8$ & 15.3 & $0.56^{+0.09}_{-0.11} \times <0.38$ & $4.3^{+0.8}_{-0.9} \times <3.0$ & $22.9^{+22.6}_{-22.6}$ & 0.18 & $>0.04$ & $20.6\pm7.9$ & \\ [1ex] 
AzTEC/C20 & $3.06^{+0.13}_{-0.54}$ & 10 01 14.54 & +02 27 05.34 & $19.1\pm2.2$ & $27.6\pm4.9$ & 8.7 & $0.72^{+0.17}_{-0.18} \times <0.38$ & $5.5^{+1.3}_{-1.4} \times <2.9$ & $64.2^{+16.0}_{-15.9}$ & 0.16 & $>-0.88$ & $7.2\pm3.7$ &  \\ [1ex] 
AzTEC/C21 & $z_{\rm synth}=2.70^{+1.30}_{-0.40}$ & 09 59 21.43 & +02 22 40.05 & $20.9\pm2.3$ & $31.7\pm5.2$ & 9.3 & $0.64^{+0.15}_{-0.18} \times 0.44^{+0.16}_{-0.25}$ & $5.1^{+1.2}_{-1.4} \times 3.5^{+1.3}_{-1.4}$ & $82.0^{+34.2}_{-34.3}$ & 0.08 & $>-0.66$ & $10.6\pm5.8$ &   \\ [1ex] 
AzTEC/C22a & $z_{\rm spec}=1.599$ & 10 00 08.94 & +02 40 10.90 & $77.5\pm2.3$ & $99.6\pm4.8$ & 33.7 & $0.48^{+0.04}_{-0.05} \times <0.38$ & $4.1^{+0.3}_{-0.4} \times <3.2$ & $163.2^{+14.1}_{-14.1}$ & 0.04 & $-0.37\pm0.27$ & $58.8\pm11.4$ & AzTEC11-S\\ [1ex] 
AzTEC/C22b & $z_{\rm spec}=1.599$ & 10 00 08.90 & +02 40 09.52 & $57.0\pm2.3$ & $67.5\pm4.6$ & 24.8 & $0.40^{+0.07}_{-0.08} \times <0.38$ & $3.4^{+0.6}_{-0.7} \times <3.2$ & $31.3^{+20.5}_{-20.5}$ & 0.44 & $-0.94\pm0.30$ & $57.4\pm21.9$ & AzTEC11-N\\ [1ex] 
AzTEC/C23 & $2.10^{+0.46}_{-0.41}$ & 10 01 42.36 & +02 18 35.88 & $53.0\pm2.2$ & $71.5\pm4.6$ & 24.4 & $0.59^{+0.06}_{-0.07} \times <0.38$ & $4.9^{+0.5}_{-0.5} \times <3.2$ & $155.4^{+9.1}_{-9.2}$ & 0.06 & $-0.73\pm0.22$ & $28.0\pm6.3$ \\ [1ex] 
AzTEC/C24a & $2.01^{+0.19}_{-0.46}$ & 10 00 10.36 & +02 22 24.42 & $32.7\pm2.4$ & $43.9\pm5.1$ & 13.7 & $0.61^{+0.11}_{-0.12} \times <0.38$ & $5.1^{+0.9}_{-1.0} \times <3.2$ & $164.8^{+13.8}_{-13.8}$ & 0.06 & $-1.07\pm0.37$ & $16.0\pm6.2$ &\\ [1ex] 
AzTEC/C24b & $2.10^{+0.08}_{-0.63}$ & 10 00 09.49 & +02 22 19.49 & $114.3\pm2.4$ & $142.9\pm4.9$ & 47.4 & $0.51^{+0.03}_{-0.04} \times <0.38$ & $4.2^{+0.3}_{-0.3} \times <3.2$ & $80.2^{+5.5}_{-5.4}$ & 0.03 & $-1.15\pm0.14$ & $75.2\pm10.4$ & VLBA-det.\\ [1ex] 
\hline
\end{tabular} }
\end{table}
\end{landscape}

\addtocounter{table}{-1}

\begin{landscape}
\begin{table}
\caption{continued.}
{\tiny
\centering
\renewcommand{\footnoterule}{}
\label{table:results}
\begin{tabular}{c c c c c c c c c c c c c c}
\hline\hline 
Source ID & $z$\tablefootmark{a} & $\alpha_{2000.0}$ & $\delta_{2000.0}$ & $I_{\rm 3\, GHz}$\tablefootmark{b} & $S_{\rm 3\, GHz}$\tablefootmark{b} & S/N & \multicolumn{2}{c}{\underline{FWHM size}\tablefootmark{c}} & P.A.\tablefootmark{c} & Offset & $\alpha_{\rm 1.4\, GHz}^{\rm 3\, GHz}$\tablefootmark{d} & $T_{\rm B}$\tablefootmark{d} & Comments\tablefootmark{e} \\
       & & [h:m:s] & [$\degr$:$\arcmin$:$\arcsec$] & [$\mu$Jy~beam$^{-1}$] & [$\mu$Jy] & & [$\arcsec$] & [kpc] & [$\degr$] & [$\arcsec$] & & [K] &\\ 
\hline
AzTEC/C25 & $z_{\rm spec}=2.51$ & 10 01 21.95 & +01 56 43.97 & $22.0\pm2.1$ & $46.7\pm6.3$ & 10.4 & $1.12^{+0.15}_{-0.16} \times 0.48^{+0.14}_{-0.19}$ & $9.0^{+1.2}_{-1.3} \times 3.8^{+1.2}_{-1.3}$ & $80.7^{+9.0}_{-9.1}$ & 0.06 & $-0.53\pm0.44$ & $5.1\pm1.6$ \\ [1ex] 
AzTEC/C26 & $5.06^{+0.08}_{-0.90}$ & 10 01 32.30 & +02 32 09.14 & $16.4\pm2.3$ & $16.9\pm4.12$ & 7.1 & $0.43^{+0.21}_{-0.43} \times <0.38$ & $2.7^{+1.3}_{-2.7} \times <2.4$ & $78.8^{+22.5}_{-22.7}$ & 0.17& $>-1.73$ & $12.3\pm18.6$ & \\ [1ex] 
AzTEC/C27 & $2.77^{+0.88}_{-0.47}$ & 09 59 38.07 & +02 06 56.28 & $15.1\pm2.2$ & $27.1\pm5.7$ & 6.8 & $0.76^{+0.20}_{-0.46} \times 0.57^{+0.39}_{-0.27}$ & $6.0^{+1.6}_{-3.6} \times 4.5^{+3.1}_{-3.6}$ & $122.7^{+41.1}_{-41.0}$ & 0.18 & $>-1.01$ & $6.4\pm5.8$ \\ [1ex] 
AzTEC/C28a & $z_{\rm spec}=2.319$ & 09 58 49.28 & +02 13 01.64 & $76.4\pm2.3$ & $128.3\pm5.8$ & 33.2 & $0.95^{+0.04}_{-0.05} \times <0.38$ & $7.7^{+0.4}_{-0.4} \times <3.1$ & $79.3^{+1.8}_{-2.8}$ & 0.08 & $-1.43\pm0.19$ & $19.5\pm2.1$ & \\ [1ex] 
AzTEC/C29 & $1.82^{+0.35}_{-0.54}$ & 09 59 18.38 & +02 01 07.13 & $60.5\pm2.3$ & $60.5\pm2.3$ & 26.3 & $<0.38$ & $<3.2$ & \ldots & 0.05 & $-1.03\pm0.14$ & $>54.8$ &\\ [1ex] 
AzTEC/C30a & $z_{\rm 3\, GHz}=2.01^{+1.16}_{-0.31}$ & 10 00 26.67 & +02 31 26.22 & $11.7\pm2.1$ & $25.1\pm6.3$ & 5.6 & $1.49^{+0.33}_{-0.35} \times <0.38$ & $12.5^{+2.8}_{-2.9} \times <3.2 $ & $75.5^{+7.9}_{-7.9}$ & 0.24 & $>-1.24$ & $1.5\pm0.8$ & \\ [1ex] 
AzTEC/C31a & $z_{\rm synth}=2.10^{+3.20}_{-0.10}$ & 10 01 47.30 & +02 24 49.32 & $16.2\pm2.3$ & $16.2\pm2.3$ & 7.0 & $<0.38$ & $<3.2$ & \ldots & 0.24 & $>-1.47$ & $>13.1$ & \\ [1ex] 
AzTEC/C32 & $1.63^{+0.20}_{-0.47}$ & 10 00 12.53 & +02 01 24.22 & $21.9\pm2.2$ & $145.3\pm6.5$ & 10.0 & $0.92^{+0.15}_{-0.16}\times 0.63^{+0.14}_{-0.17}$ & $7.8^{+1.4}_{-1.3} \times 5.3^{+1.2}_{-1.4}$ & $91.4^{+21.0}_{-21.0}$ & 0.10 & $-0.28\pm0.47$ & $7.2\pm2.7$ & Cosbo-33\\ [1ex] 
AzTEC/C33a & $2.30^{+0.16}_{-0.46}$ & 10 00 27.14 & +02 31 40.77 & $41.1\pm2.4$ & $46.4\pm4.4$ & 17.1 & $0.44^{+0.09}_{-0.11} \times <0.38$ & $3.6^{+0.8}_{-0.9}\times <3.1$ & $166.5^{+13.0}_{-13.1}$ & 0.17 & $-0.49\pm0.34$ & $32.3\pm15.0$ & \\ [1ex] 
AzTEC/C33b & $z_{\rm 3\, GHz}=2.38^{+1.68}_{-0.35}$ & 10 00 26.67 & +02 31 26.22 & $12.9\pm2.3$ & $17.6\pm5.0$ & 5.6 & $0.65^{+0.25}_{-0.33} \times <0.38$ & $5.3^{+2.0}_{-2.7} \times <3.1$ & $42.3^{+28.3}_{-28.3}$ & 0.20 & $>-1.79$ & $5.7\pm5.3$ & \\ [1ex] 
AzTEC/C35 & $3.91^{+0.18}_{-0.50}$ & 10 00 08.97 & +02 20 26.69 & $30.4\pm2.3$ & $30.4\pm2.3$ & 13.2 & $<0.38$ & $<2.7$ & \ldots & 0.14 & $-0.93\pm0.30$ & $>26.5$ & \\ [1ex] 
AzTEC/C36 & $z_{\rm spec}=2.415$ & 09 58 40.29 & +02 05 14.58 & $66.3\pm2.5$ & $179.1\pm5.0$ & 26.5 & $0.41^{+0.07}_{-0.07} \times <0.38$ & $3.4^{+0.5}_{-0.6}\times <3.1$ & $126.3^{+17.4}_{-17.4}$ & 0.09 & $-0.99\pm0.21$ & $63.0\pm21.3$ & \\ [1ex] 
AzTEC/C37 & $z_{\rm synth}=1.70^{+0.70}_{-0.30}$ & 10 01 21.82 & +02 31 29.44 & $21.0\pm2.3$ & $42.6\pm6.5$ & 9.1 & $1.23^{+0.18}_{-0.19} \times <0.38$ & $10.4^{+1.5}_{-1.6}\times <3.2$ & $79.0^{+6.9}_{-6.9}$ & 0.35 & $-0.26\pm0.49$ & $3.8\pm1.3$\\ [1ex] 
AzTEC/C38 & $1.91^{+0.53}_{-0.46}$ & 10 00 23.65 & +02 21 55.34 & $15.8\pm2.3$ & $15.8\pm2.3$ & 6.9 & $<0.38$ & $<3.2$ & \ldots & 0.13 & $-1.32\pm0.52$ & $>12.7$ & COSLA35\\ [1ex] 
AzTEC/C39 & $z_{\rm synth}=2.00^{+0.20}_{-0.40}$ & 10 01 26.53 & +02 00 05.97 & $38.8\pm2.2$ & $45.5\pm4.3$ & 17.6 & $0.40^{+0.10}_{-0.11}\times <0.38$ & $3.4^{+0.8}_{-1.0}\times < 3.2$ & $54.6^{+24.0}_{-24.0}$ & 0.13 & $>0.02$ & $38.2\pm20.3$ \\ [1ex] 
AzTEC/C40 & $z_{\rm FIR}=5.25^{+1.11}_{-1.11}$ & 09 59 35.36 & +02 19 20.14 & $21.3\pm2.2$ & $28.5\pm4.7$ & 9.7 & $0.66^{+0.15}_{-0.17}\times <0.38$ & $4.0^{+0.9}_{-1.1}\times <2.3$ & $17.7^{+15.1}_{-15.0}$ & 0.07 & $-1.05\pm0.53$ & $9.0\pm4.6$\\ [1ex] 
AzTEC/C41 & $1.25^{+0.18}_{-0.34}$ & 10 01 48.20 & +02 21 32.34 & $21.5\pm2.3$ & $27.8\pm4.7$ & 9.3 & $0.55^{+0.15}_{-0.19}\times <0.38$ & $4.6^{+1.3}_{-1.6}\times < 3.2$ & $21.1^{+24.9}_{-24.9}$ & 0.09 & $>-0.85$ & $12.6\pm8.2$ & \\ [1ex] 
AzTEC/C42 & $3.63^{+0.37}_{-0.56}$ & 10 00 19.75 & +02 32 04.29 & $49.2\pm2.4$ & $85.8\pm5.8$ & 20.5 & $0.95^{+0.07}_{-0.07}\times <0.38$ & $6.9^{+0.5}_{-0.5}\times < 2.7$ & $41.3^{+4.5}_{-4.5}$ & 0.06 & $-0.50\pm0.24$ & $12.9\pm2.1$ & AzTEC5 \\ [1ex] 
AzTEC/C43a & $2.01^{+0.23}_{-0.47}$ & 10 00 03.12 & +02 02 01.53 & $37.0\pm2.4$ & $37.0\pm2.4$ & 15.4 & $<0.38$ & $<3.2$ & \ldots & 0.12 & $-2.15\pm0.32$ & $>32.7$ & Cosbo-4\\ [1ex] 
AzTEC/C43b & $1.82^{+0.29}_{-0.36}$ & 10 00 03.41 & +02 02 04.24 & $16.4\pm2.3$ & $20.7\pm4.7$ & 7.1 & $0.48^{+0.04}_{-0.48}\times <0.38$ & $4.0^{+1.7}_{-4.0}\times < 3.2$ & $149.4^{+36.8}_{-36.8}$ & 0.11 & $>-1.42$ & $12.3\pm17.8$ & \\ [1ex] 
AzTEC/C44a & $2.01^{+0.29}_{-0.44}$ & 10 00 34.25 & +01 48 57.64 & $26.8\pm2.3$ & $33.1\pm4.7$ & 11.7 & $0.38^{+0.14}_{-0.32}\times <0.38$ & $3.2^{+1.2}_{-2.7}\times <3.2$ & $78.2^{+0}_{-0}$ & 0.13 & $>-0.54$ & $31.0\pm38.2$ & \\ [1ex] 
AzTEC/C44b & $z_{\rm spec}=1.192$ & 10 00 33.89 & +01 49 10.89 & $14.0\pm2.3$ & $22.0\pm5.4$ & 6.1 & $0.74^{+0.23}_{-0.28}\times <0.38$ & $6.1^{+1.9}_{-2.3}\times < 3.1$ & $127.5^{+29.0}_{-29.0}$ & 0.26 & $>-1.39$ & $5.5\pm4.0$ & X-ray \\ [1ex] 
AzTEC/C45 & $z_{\rm spec}=2.330$ & 10 00 06.57 & +02 32 59.79 & $19.2\pm2.3$ & $27.2\pm5.1$ & 8.3 & $0.61^{+0.17}_{-0.21}\times <0.38$ & $5.0^{+1.4}_{-1.7}\times <3.1$ & $78.2^{+31.2}_{-31.2}$ & 0.16 & $>-0.93$ & $10.0\pm6.5$ & X-ray\\ [1ex] 
AzTEC/C46 & $1.06^{+1.07}_{-0.41}$ & 10 01 14.71 & +02 35 18.26 & $56.5\pm2.3$ & $56.5\pm2.3$ & 24.6 & $<0.38$ & $<3.1$ & \ldots & 0.04 & $-1.01\pm0.16$ & $>51.1$\\ [1ex] 
AzTEC/C47 & $z_{\rm spec}=2.0468$ & 09 59 40.87 & +02 01 13.25 & $105.7\pm2.3$ & $105.7\pm2.3$ & 46.0 & $<0.38$ & $<3.2$ & \ldots & 0.27 & $-1.49\pm0.12$ & $>97.4$\\ [1ex] 
AzTEC/C48a & $1.91^{+0.18}_{-0.42}$ & 10 00 39.28 & +02 38 45.14 & $28.0\pm2.2$ & $37.6\pm4.7$ & 12.7 & $0.45^{+0.13}_{-0.19}\times 0.43^{+0.15}_{-0.17}$ & $3.8^{+1.1}_{-1.6}\times 3.6^{+1.3}_{-1.6}$ & $61.3^{+0}_{-0}$ & 0.27 & $-0.67\pm0.43$ & $25.3\pm18.2$ & AzTEC24b\\ [1ex] 
AzTEC/C50 & $3.15^{+0.78}_{-1.32}$ & 09 59 33.29 & +02 08 32.71 & $25.5\pm2.3$ & $25.5\pm2.3$ & 11.1 & $<0.38$ & $<2.9$ & \ldots & 0.04 & $>-0.72$ & $>21.8$ &\\ [1ex] 
AzTEC/C52 & $z_{\rm spec}=1.1484$ & 10 01 56.57 & +02 21 00.93 & $27.7\pm2.2$ & $58.9\pm6.5$ & 12.6 & $0.93^{+0.11}_{-0.13}\times 0.67^{+0.11}_{-0.13}$ & $7.6^{+1.0}_{-1.0}\times 5.5^{+0.9}_{-1.0}$ & $142.6^{+18.8}_{-18.8}$ & 0.07 & $-0.92\pm0.36$ & $9.4\pm2.7$\\ [1ex] 
AzTEC/C53 & $z_{\rm synth}=2.20^{+0.60}_{-0.70}$ & 10 01 22.63 & +02 12 14.79 & $11.0\pm2.3$ & $15.8\pm5.0$ & 4.8 & $0.62^{+0.29}_{-0.62}\times <0.38$ & $5.2^{+2.3}_{-5.2}\times <3.1$ & $69.0^{+40.3}_{-40.2}$ & 0.60 & $>-2.04$ & $5.5\pm8.2$ &\\ [1ex] 
AzTEC/C54 & $3.25^{+0.04}_{-0.52}$ & 10 01 26.02 & +01 57 51.32 & $15.8\pm2.2$ & $30.0\pm5.8$ & 7.2 & $0.99^{+0.20}_{-0.22}\times 0.42^{+0.20}_{-0.42}$ & $7.4^{+1.5}_{-1.7}\times 3.2^{+1.5}_{-1.7}$ & $135.8^{+14.6}_{-14.7}$ & 0.09 & $>-0.83$ & $4.2\pm2.0$ \\ [1ex] 
AzTEC/C55a & $2.49^{+0.33}_{-0.45}$ & 10 00 05.09 & +01 55 18.17 & $20.5\pm2.2$ & $31.2\pm5.1$ & 9.3 & $0.64^{+0.15}_{-0.18}\times 0.44^{+0.16}_{-0.24}$ & $5.2^{+1.2}_{-1.4}\times 3.6^{+1.3}_{-1.4}$ & $99.4^{+34.7}_{-34.8}$ & 0.06 & $>-0.68$ & $10.4\pm5.7$ & \\ [1ex] 
AzTEC/C55b & $2.77^{+0.32}_{-0.41}$ & 10 00 04.40 & +01 55 15.75 & $18.1\pm2.2$ & $27.1\pm5.0$ & 8.2 & $0.67^{+0.17}_{-0.21}\times <0.38$ & $5.2^{+1.4}_{-1.6} \times < 3.0$ & $69.7^{+27.5}_{-27.4}$ & 0.27 & $-1.15\pm0.58$ & $8.3\pm4.9$\\ [1ex] 
AzTEC/C56 & $z_{\rm AGN}=3.45^{+0.08}_{-0.10}$ & 09 59 05.05 & +02 21 56.74 & $21.1\pm2.2$ & $39.6\pm6.0$ & 9.6 & $1.00^{+0.16}_{-0.17}\times <0.38$ & $7.4^{+1.2}_{-1.2}\times <2.9$ & $32.0^{+10.4}_{-10.5}$ & 0.13 & $>-0.33$ & $5.4\pm1.9$ & X-ray\\ [1ex] 
AzTEC/C58 & $4.10^{+0.32}_{-0.79}$ & 10 00 20.05 & +01 45 02.64 & $28.7\pm2.3$ & $34.1\pm4.5$ & 12.5 & $0.48^{+0.12}_{-0.15}\times <0.38$ & $3.3^{+0.8}_{-1.0}\times <2.6$ & $151.4^{+18.9}_{-18.9}$ & 0.09 & $-0.72\pm0.45$ & $20.1\pm11.7$ &\\ [1ex] 
\hline
\end{tabular} }
\end{table}
\end{landscape}

\addtocounter{table}{-1}

\begin{landscape}
\begin{table}
\caption{continued.}
{\tiny
\centering
\renewcommand{\footnoterule}{}
\label{table:results}
\begin{tabular}{c c c c c c c c c c c c c c}
\hline\hline 
Source ID & $z$\tablefootmark{a} & $\alpha_{2000.0}$ & $\delta_{2000.0}$ & $I_{\rm 3\, GHz}$\tablefootmark{b} & $S_{\rm 3\, GHz}$\tablefootmark{b} & S/N & \multicolumn{2}{c}{\underline{FWHM size}\tablefootmark{c}} & P.A.\tablefootmark{c} & Offset & $\alpha_{\rm 1.4\, GHz}^{\rm 3\, GHz}$\tablefootmark{d} & $T_{\rm B}$\tablefootmark{d} & Comments\tablefootmark{e} \\
       & & [h:m:s] & [$\degr$:$\arcmin$:$\arcsec$] & [$\mu$Jy~beam$^{-1}$] & [$\mu$Jy] & & [$\arcsec$] & [kpc] & [$\degr$] & [$\arcsec$] & & [K] &\\ 
\hline
AzTEC/C59 & $z_{\rm spec}=1.2802$ & 10 00 30.14 & +02 37 16.76 & $65.2\pm2.4$ & $65.2\pm2.4$ & 27.2 & $<0.38$ & $<3.2$ & \ldots & 0.15 & $-1.19\pm0.14$ & $>59.2$\\ [1ex] 
AzTEC/C60a & $0.96^{+0.14}_{-0.40}$ & 10 01 28.39 & +02 21 27.49 & $18.2\pm2.2$ & $27.6\pm5.2$ & 8.3 & $0.87^{+0.18}_{-0.20} \times <0.38$ & $6.9^{+1.4}_{-1.6}\times <3.0$ & $2.1^{+11.1}_{-9.0}$ & 0.09 & $>-0.91$ & $5.0\pm2.3$ &\\ [1ex] 
AzTEC/C61 & $z_{\rm spec}=3.2671$ & 10 01 20.06 & +02 34 43.68 & $11\,502.0\pm4.2$ & $11\,502.0\pm4.2$ & 2\,738.6 & $<0.38$ & $<2.8$ & \ldots & 0.11 & $0.11\pm0.01$ & $>10\,832$ & VLBA-det.,\\ [1ex] 
           & &  & & & &  &  &  & & & & & X-ray\\ [1ex]
AzTEC/C62 & $z_{\rm FIR}=3.36^{+0.97}_{-0.97}$ & 10 01 00.46 & +02 38 05.88 & $10.9\pm2.1$ & $30.4\pm7.8$ & 5.2 & $1.48^{+0.36}_{-0.37} \times 0.58^{+0.27}_{-0.45}$ & $11.0^{+2.7}_{-2.8}\times 4.3^{+2.0}_{-2.8}$ & $135.0^{+13.4}_{-13.3}$ & 0.46 & $>-1.00$ & $1.9\pm1.1$ & \\ [1ex] 
AzTEC/C64 & $2.58^{+0.79}_{-0.63}$ & 10 01 39.35 & +02 23 41.43 & $31.6\pm2.2$ & $39.7\pm4.4$ & 14.4 & $0.43^{+0.11}_{-0.29} \times <0.38$ & $3.4^{+0.9}_{-2.3}\times <3.0$ & $10.1^{+36.7}_{-26.6}$ & 0.07 & $>-0.21$ & $29.8\pm28.1$ & \\ [1ex] 
AzTEC/C65 & $z_{\rm spec}=1.798$ & 09 59 42.94 & +02 21 44.91 & $61.5\pm2.3$ & $76.9\pm4.6$ & 26.7 & $0.47^{+0.06}_{-0.06} \times <0.38$ & $4.0^{+0.5}_{-0.6}\times < 3.2$ & $152.2^{+13.8}_{-13.7}$ & 0.14 & $-0.91\pm0.20$ & $47.0\pm12.8$ & \\ [1ex] 
AzTEC/C66 & $2.01^{+0.27}_{-0.50}$ & 10 01 04.64 & +02 26 33.98 & $35.6\pm2.3$ & $52.2\pm5.1$ & 15.5 & $0.59^{+0.10}_{-0.10}\times 0.43^{+0.10}_{-0.13}$ & $5.0^{+0.8}_{-0.9}\times 3.6^{+0.9}_{-0.9}$ & $3.0^{+25.0}_{-22.1}$ & 0.10 & $-0.66\pm0.32$ & $20.2\pm7.0$\\ [1ex] 
AzTEC/C67 & $z_{\rm spec}=2.9342$ & 10 01 19.53 & +02 09 44.67 & $34.7\pm2.3$ & $34.7\pm2.3$ & 15.1 & $<0.38$ & $<2.9$ & \ldots & 0.25 & $-0.92\pm0.26$ & $>30.5$\\ [1ex] 
AzTEC/C70 & $4.01^{+0.09}_{-0.66}$ & 10 00 25.50 & +02 03 12.66 & $12.8\pm2.2$ & $21.9\pm5.6$ & 5.8 & $0.84^{+0.24}_{-0.29} \times 0.41^{+0.25}_{-0.41}$ & $5.8^{+1.7}_{-2.0}\times 2.8^{+1.8}_{-2.0}$ & $103.1^{+25.2}_{-25.3}$ & 0.07 & $-1.06\pm0.81$ & $4.3\pm2.9$ \\ [1ex] 
AzTEC/C71b & $z_{\rm spec}=0.829$ & 09 59 53.85 & +02 18 54.14 & $13.6\pm2.2$ & $32.3\pm7.1$ & 6.2 & $1.14^{+0.26}_{-0.28}\times 0.63^{+0.22}_{-0.30}$ & $8.7^{+2.0}_{-2.1}\times 4.8^{+1.7}_{-2.1}$ & $126.3^{+19.4}_{-19.4}$ & 0.18 & $-1.16\pm0.68$ & $3.4\pm1.8$ & X-ray,\\ [1ex] 
           & &  & & & &  &  &  & & & & & Cosbo-36\\ [1ex] 
AzTEC/C72 & $1.72^{+0.38}_{-0.45}$ & 10 01 58.99 & +02 04 57.67 & $39.6\pm2.3$ & $50.2\pm4.8$ & 17.2 & $0.56^{+0.09}_{-0.09}\times <0.38$ & $4.8^{+0.7}_{-0.8}\times <3.2$ & $102.1^{+11.4}_{-11.3}$ & 0.14 & $-0.84\pm0.32$ & $21.6\pm7.4$\\ [1ex] 
AzTEC/C74a & $2.10^{+0.20}_{-0.67}$ & 10 01 05.08 & +02 21 52.44 & $19.2\pm2.3$ & $19.2\pm2.3$ & 8.3 & $<0.38$ & $<3.2$ & \ldots & 0.20 & $>-1.19$ & $>15.9$ & \\ 
AzTEC/C76 & $4.01^{+0.07}_{-0.57}$ & 10 00 12.93 & +02 12 11.51 & $10.9\pm2.3$ & $32.6\pm8.7$ & 4.7 & $1.69^{+0.41}_{-0.43}\times 0.51^{+0.29}_{-0.51}$ & $11.8^{+2.9}_{-3.0}\times 3.6^{+2.0}_{-3.0}$ & $109.7^{+10.9}_{-10.8}$ & 0.17 & $>-0.94$ & $1.6\pm0.9$ & \\ [1ex] 
AzTEC/C77a & $3.53^{+0.58}_{-1.29}$ & 09 59 35.73 & +01 58 05.41 & $261.1\pm2.3$ & $261.1\pm2.3$ & 113.5 & $<0.38$ & $<2.8$ & \ldots & 0.18 & $-0.99\pm0.03$ & $>243.9$ & VLBA-det.\\ [1ex] 
AzTEC/C77b & $3.06^{+0.59}_{-1.19}$ & 09 59 35.30 & +01 57 59.20 & $26.8\pm2.3$ & $26.8\pm2.3$ & 11.7 & $<0.38$ & $<2.9$ & \ldots & 0.05 & $-1.24\pm0.30$ & $>23.1$ &\\ [1ex] 
AzTEC/C79 & $2.20^{+0.33}_{-0.96}$ & 09 59 43.66 & +02 13 40.18 & $16.8\pm2.1$ & $30.1\pm5.6$ & 8.0 & $0.81^{+0.18}_{-0.21}\times 0.52^{+0.18}_{-0.25}$ & $6.7^{+1.5}_{-1.7}\times 4.3^{+1.7}_{-1.5}$ & $72.5^{+27.2}_{-27.2}$ & 0.13 & $>-0.80$ & $6.3\pm3.2$ & \\ [1ex] 
AzTEC/C80a & $2.10^{+0.66}_{-0.43}$ & 10 00 33.35 & +02 26 01.64 & $33.7\pm2.3$ & $39.8\pm4.5$ & 14.7 & $0.39^{+0.11}_{-0.15}\times <0.38$ & $3.2^{+0.9}_{-1.2}\times <3.2$ & $9.1^{+38.3}_{-29.3}$ & 0.08 & $>-0.21$ & $36.0\pm24.7$ & COSLA47\\ [1ex] 
AzTEC/C81 & $z_{\rm FIR}=4.62^{+1.48}_{-1.48}$ & 10 00 06.28 & +01 52 48.03 & $11.6\pm2.2$ & $23.2\pm6.3$ & 5.3 & $0.98^{+0.28}_{-0.33}\times 0.52^{+0.27}_{-0.52}$ & $6.4^{+2.1}_{-1.9}\times 3.4^{+1.8}_{-2.1}$ & $152.9^{+25.8}_{-25.7}$ & 0.03 & $>-1.40$ & $3.3\pm2.3$ & \\ [1ex] 
AzTEC/C84b & $z_{\rm spec}=1.959$ & 09 59 42.58 & +01 55 01.49 & $22.7\pm2.1$ & $50.3\pm6.4$ & 10.8 & $0.99^{+0.14}_{-0.15}\times 0.66^{+0.14}_{-0.15}$ & $8.3^{+1.2}_{-1.3}\times 5.6^{+1.1}_{-1.3}$ & $21.9^{+17.3}_{-17.2}$ & 0.13 & $-0.65\pm0.41$ & $7.0\pm2.2$\\ [1ex] 
AzTEC/C86 & $z_{\rm AGN}=1.69^{+0.02}_{-0.02}$ & 10 01 09.91 & +02 17 27.63 & $25.7\pm2.5$ & $25.7\pm2.5$ & 10.3 & $<0.38$ & $<3.2$ & \ldots & 0.05 & $>-0.74$ & $>21.8$ & X-ray\\ [1ex] 
AzTEC/C87 & $2.39^{+0.20}_{-0.45}$ & 10 02 04.93 & +02 17 01.26 & $15.5\pm2.2$ & $41.4\pm7.8$ & 7.0 & $1.30^{+0.24}_{-0.25}\times 0.66^{+0.20}_{-0.24}$ & $10.6^{+2.0}_{-2.1}\times 5.4^{+1.6}_{-2.1}$ & $73.6^{+13.7}_{-13.7}$ & 0.33 & $>-0.38$ & $3.3\pm1.4$ & \\ [1ex] 
AzTEC/C88 & $1.82^{+0.38}_{-0.47}$ & 09 59 37.48 & +02 04 24.19 & $27.5\pm2.3$ & $34.6\pm4.7$ & 12.0 & $0.50^{+0.13}_{-0.15}\times <0.38$ & $4.3^{+1.1}_{-1.3}\times < 3.2$ & $34.7^{+24.2}_{-24.2}$ & 0.21 & $>-0.47$ & $18.5\pm10.7$ & \\ [1ex] 
AzTEC/C90c & $2.20^{+0.23}_{-0.42}$ & 10 01 35.21 & +02 16 49.20 & $16.6\pm2.2$ & $28.6\pm5.4$ & 7.5 & $0.75^{+0.18}_{-0.21}\times 0.52^{+0.19}_{-0.26}$ & $6.2^{+1.5}_{-1.7}\times 4.3^{+1.5}_{-1.7}$ & $145.9^{+33.2}_{-35.0}$ & 0.33 & $>-0.88$ & $6.9\pm3.9$ & \\ [1ex] 
AzTEC/C91 & $1.63^{+0.29}_{-0.41}$ & 10 01 28.49 & +02 23 44.88 & $33.2\pm2.3$ & $37.1\pm4.2$ & 14.4 & $0.45^{+0.10}_{-0.14}\times <0.38$ & $3.8^{+0.9}_{-1.1}\times < 3.2$ & $73.4^{+14.3}_{-14.2}$ & 0.24 & $>-0.30$ & $25.4\pm13.9$ & \\ [1ex] 
AzTEC/C92a & $2.58^{+2.67}_{-0.46}$ & 10 01 40.44 & +02 30 10.43 & $11.2\pm2.2$ & $16.1\pm5.0$ & 5.1 & $0.70^{+0.28}_{-0.34}\times <0.38$ & $5.6^{+2.2}_{-2.9}\times <3.0$ & $162.1^{+5.3}_{-30.6}$ & 0.14 & $>-1.98$ & $4.5\pm4.3$ & \\ [1ex] 
AzTEC/C92b & $4.87^{+0.22}_{-0.98}$ & 10 01 39.17 & +02 30 19.24 & $13.9\pm2.2$ & $25.9\pm5.9$ & 6.3 & $0.95^{+0.26}_{-0.25}\times 0.43^{+0.23}_{-0.43}$ & $6.1^{+1.5}_{-1.6}\times 2.7^{+1.5}_{-2.7}$ & $14.1^{+18.3}_{-9.8}$ & 0.11 & $>-1.12$ & $3.9\pm2.2$ & \\ [1ex] 
AzTEC/C93 & $1.63^{+1.10}_{-0.53}$ & 10 01 31.88 & +02 11 38.77 & $38.8\pm2.3$ & $38.8\pm2.3$ & 16.9 & $<0.38$ & $<3.2$ & \ldots & 0.06 & $-0.57\pm0.26$ & $>34.4$\\ [1ex] 
AzTEC/C95 & $z_{\rm spec}=2.1021$ & 10 00 18.23 & +02 12 42.52 & $14.5\pm2.1$ & $39.9\pm7.7$ & 6.9 & $1.18^{+0.24}_{-0.26}\times 0.81^{+0.21}_{-0.24}$ & $9.8^{+2.0}_{-2.1}\times 6.8^{+1.8}_{-2.1}$ & $70.4^{+25.4}_{-25.5}$ & 0.07 & $>-0.44$ & $3.9\pm1.8$ & \\ [1ex] 
AzTEC/C97b & $2.01^{+0.08}_{-0.48}$ & 10 02 14.50 & +02 19 42.84 & $14.0\pm2.2$ & $24.5\pm5.6$ & 6.4 & $0.78^{+0.22}_{-0.26}\times 0.51^{+0.23}_{-0.35}$ & $6.6^{+1.9}_{-2.2}\times 4.3^{+1.9}_{-2.2}$ & $33.6^{+35.7}_{-31.5}$ & 0.18 & $-1.06\pm0.75$ & $5.4\pm3.6$\\ [1ex] 
AzTEC/C98 & $1.82^{+0.60}_{-0.46}$ & 10 00 43.18 & +02 05 19.03 & $30.8\pm2.3$ & $41.3\pm4.9$ & 13.4 & $0.60^{+0.11}_{-0.12}\times <0.38$ & $5.1^{+0.9}_{-1.0}\times <3.2$ & $74.1^{+14.7}_{-14.6}$ & 0.08 & $-0.83\pm0.40$ & $15.4\pm6.2$ & COSLA18\\ [1ex] 
AzTEC/C99 & $2.68^{+1.37}_{-0.92}$ & 10 00 06.99 & +01 59 57.55 & $18.4\pm2.3$ & $22.5\pm4.6$ & 8.0 & $0.56^{+0.18}_{-0.23}\times <0.38$ & $4.5^{+1.4}_{-1.9}\times <3.0$ & $133.6^{+20.4}_{-20.5}$ & 0.03 & $>-1.24$ & $9.7\pm7.4$ & \\ [1ex] 
AzTEC/C101a & $1.53^{+0.31}_{-0.51}$ & 09 59 45.33 & +02 30 16.81 & $11.1\pm2.2$ & $25.4\pm6.8$ & 5.0 & $1.18^{+0.31}_{-0.34}\times 0.54^{+0.27}_{-0.54}$ & $10.0^{+2.6}_{-2.9}\times 4.5^{+2.3}_{-2.9}$ & $132.9^{+18.5}_{-18.5}$ & 0.16 & $>-1.26$ & $2.5\pm1.5$ & \\ [1ex] 
AzTEC/C101b & $z_{\rm 3\, GHz}=1.74^{+0.98}_{-0.27}$ & 09 59 45.87 & +02 30 24.80 & $10.5\pm2.2$ & $23.4\pm6.7$ & 4.8 & $1.18^{+0.33}_{-0.36}\times 0.50^{+0.29}_{-0.50}$ & $10.0^{+2.8}_{-3.1}\times 4.2^{+2.4}_{-3.1}$ & $106.7^{+18.5}_{-18.5}$ & 0.33 & $>-1.42$ & $2.3\pm1.5$ & \\ [1ex]
 
\hline
\end{tabular} }
\end{table}
\end{landscape}

\addtocounter{table}{-1}

\begin{landscape}
\begin{table}
\caption{continued.}
{\tiny
\centering
\renewcommand{\footnoterule}{}
\label{table:results}
\begin{tabular}{c c c c c c c c c c c c c c}
\hline\hline 
Source ID & $z$\tablefootmark{a} & $\alpha_{2000.0}$ & $\delta_{2000.0}$ & $I_{\rm 3\, GHz}$\tablefootmark{b} & $S_{\rm 3\, GHz}$\tablefootmark{b} & S/N & \multicolumn{2}{c}{\underline{FWHM size}\tablefootmark{c}} & P.A.\tablefootmark{c} & Offset & $\alpha_{\rm 1.4\, GHz}^{\rm 3\, GHz}$\tablefootmark{d} & $T_{\rm B}$\tablefootmark{d} & Comments\tablefootmark{e} \\
       & & [h:m:s] & [$\degr$:$\arcmin$:$\arcsec$] & [$\mu$Jy~beam$^{-1}$] & [$\mu$Jy] & & [$\arcsec$] & [kpc] & [$\degr$] & [$\arcsec$] & & [K] &\\ 
\hline
AzTEC/C103 & $2.10^{+0.33}_{-0.57}$ & 10 01 24.48 & +01 56 15.74 & $12.7\pm2.2$ & $22.4\pm5.8$ & 5.8 & $1.20^{+0.29}_{-0.30}\times 0.62^{+0.24}_{-0.33}$ & $10.0^{+2.4}_{-2.5}\times 5.2^{+2.0}_{-2.5}$ & $28.4^{+18.4}_{-18.3}$ & 0.31 & $>-0.90$ & $2.9\pm1.6$ \\ [1ex]
AzTEC/C105 & $2.20^{+0.08}_{-0.54}$ & 09 58 45.12 & +02 14 30.84 & $21.8\pm2.3$ & $40.3\pm6.0$ & 9.5 & $0.75^{+0.15}_{-0.30}\times 0.63^{+0.27}_{-0.18}$ & $6.2^{+1.2}_{-2.5}\times 5.2^{+2.3}_{-2.5}$ & $167.5^{+15.4}_{-40.4}$ & 0.07 & $>-0.30$ & $9.6\pm5.9$ \\ [1ex] 
AzTEC/C106 & $z_{\rm FIR}=5.63^{+1.37}_{-2.77}$ & 10 00 06.49 & +02 38 37.40 & $12.3\pm2.3$ & $22.4\pm5.8$ & 5.3 & $0.92^{+0.26}_{-0.30}\times 0.43^{+0.26}_{-0.00}$ & $5.4^{+1.5}_{-1.8} \times 2.5^{+1.5}_{-0} $ & $154.3^{+22.5}_{-22.6}$ & 0.32 & $>-1.48$ & $3.6\pm2.4$ & AzTEC6\\ [1ex] 
AzTEC/C107 & $5.15^{+0.93}_{-1.40}$ & 09 59 39.87 & +02 22 33.54 & $16.9\pm2.4$ & $18.9\pm4.5$ & 7.0 & $0.47^{+0.21}_{-0.34}\times <0.38$ & $2.9^{+1.3}_{-2.1}\times <2.4$ & $68.7^{+26.4}_{-26.4}$ & 0.24 & $>-1.55$ & $11.7\pm13.9$ &\\ [1ex] 
AzTEC/C109 & $2.20^{+0.28}_{-0.41}$ & 10 01 11.56 & +02 28 40.89 & $14.0\pm2.2$ & $19.9\pm4.8$ & 6.4 & $0.82^{+0.22}_{-0.26}\times <0.38$ & $6.8^{+1.8}_{-2.1}\times <3.1$ & $5.2^{+14.1}_{-14.1}$ & 0.27 & $-1.43\pm0.75$ & $4.1\pm2.6$ &\\ [1ex] 
AzTEC/C111 & $2.10^{+0.54}_{-0.59}$ & 09 59 29.23 & +02 12 43.97 & $36.1\pm2.4$ & $36.1\pm2.4$ & 15.0 & $<0.38$ & $<3.2$ & \ldots & 0.14 & $-0.81\pm0.27$ & $>31.8$\\ [1ex] 
AzTEC/C112 & $z_{\rm spec}=1.894$ & 10 00 11.03 & +01 53 14.06 & $49.7\pm2.2$ & $65.4\pm4.6$ & 22.6 & $0.66^{+0.07}_{-0.07}\times <0.38$ & $5.6^{+0.5}_{-0.6} \times <3.2$ & $41.8^{+5.9}_{-5.9}$ & 0.19 & $-0.82\pm0.23$ & $20.3\pm4.4$\\ [1ex] 
AzTEC/C113 & $z_{\rm spec}=2.0899$ & 09 59 14.40 & +02 29 60.00 & $107.3\pm2.3$ & $107.3\pm2.3$ & 46.7 & $<0.38$ & $<3.2$ & \ldots & 0.11 & $-0.63\pm0.11$ & $>98.9$\\ [1ex] 
AzTEC/C114 & $z_{\rm FIR}=5.33^{+1.67}_{-3.22}$ & 10 00 24.15 & +02 20 05.22 & $9.3\pm2.1$ & $54.6\pm14.0$ & 4.4 & $2.68^{+0.63}_{-0.65}\times 0.93^{+0.32}_{-0.38}$ & $16.3^{+3.9}_{-3.9} \times 5.7^{+2.0}_{-3.9}$ & $168.5^{+9.5}_{-9.4}$ & 0.33 & $>-0.23$ & $1.0\pm0.6$ & \\ [1ex] 
AzTEC/C115 & $z_{\rm synth}=2.80^{+1.30}_{-0.60}$ & 10 00 15.36 & +02 05 31.72 & $16.6\pm2.3$ & $20.1\pm4.6$ & 7.2 & $0.70^{+0.19}_{-0.24}\times <0.38$ & $5.5^{+1.5}_{-1.8} \times <3.0$ & $119.9^{+12.6}_{-12.7}$ & 0.13 & $>-1.45$ & $5.6\pm3.7$ & \\ [1ex] 
AzTEC/C116 & $2.20^{+1.75}_{-0.43}$ & 10 01 09.85 & +02 03 46.42 & $35.9\pm2.3$ & $35.9\pm2.3$ & 15.6 & $<0.38$ & $<3.1$ & \ldots & 0.06 & $-0.64\pm0.27$ & $>31.7$ & Cosbo-27\\ [1ex] 
AzTEC/C118 & $z_{\rm spec}=2.2341$ & 09 59 59.94 & +02 06 33.26 & $38.0\pm2.2$ & $49.3\pm4.6$ & 17.3 & $0.53^{+0.09}_{-0.10}\times <0.38$ & $4.4^{+0.7}_{-0.8}\times <3.1$ & $140.2^{+16.4}_{-16.4}$ & 0.12 & $-0.98\pm0.30$ & $23.9\pm8.7$ & X-ray,\\ [1ex] 
           & &  & & & &  &  &  & & & & & Cosbo-8\\ [1ex] 
AzTEC/C119 & $3.25^{+0.82}_{-0.62}$ & 09 59 15.34 & +02 07 47.38 & $20.7\pm2.2$ & $25.0\pm4.3$ & 9.4 & $0.45^{+0.16}_{-0.23}\times <0.38$ & $3.4^{+1.2}_{-1.7} \times <2.9$ & $61.7^{+35.4}_{-35.5}$ & 0.08 & $>-1.00$ & $16.8\pm14.8$ &\\ [1ex] 
AzTEC/C122a & $1.06^{+0.12}_{-0.40}$ & 10 02 00.74 & +02 16 38.12 & $13.5\pm2.3$ & $16.1\pm4.4$ & 5.9 & $0.54^{+0.24}_{-0.38}\times <0.38$ & $4.4^{+1.9}_{-3.0} \times 3.1^{+0.2}_{-3.0}$ & $6.9^{+28.5}_{-28.5}$ & 0.08 & $>-1.89$ & $7.5\pm8.8$ & \\ [1ex] 
AzTEC/C123 & $1.82^{+0.20}_{-0.61}$ & 10 00 22.82 & +01 51 40.40 & $32.5\pm2.3$ & $36.3\pm4.4$ & 14.1 & $0.43^{+0.11}_{-0.14}\times <0.38$ & $3.6^{+1.0}_{-1.2} \times <3.2$ & $143.3^{+16.7}_{-16.7}$ & 0.12 & $>-0.35$ & $26.7\pm16.4$ & \\ [1ex] 
AzTEC/C127 & $2.01^{+0.17}_{-0.51}$ & 10 01 25.33 & +02 35 27.32 & $62.1\pm2.4$ & $62.1\pm2.4$ & 25.9 & $<0.38$ & $<3.2$ & \ldots & 0.06 & $-0.98\pm0.15$ & $>56.3$\\ [1ex] 
\hline
\end{tabular} }
\tablefoot{The meaning of columns is as follows: (1): SMG name; (2): redshift; (3) and (4): celestial peak position (equinox J2000.0) of the fitted Gaussian 3~GHz intensity distribution; 
(5): peak surface brightness; (6): total flux density provided by the Gaussian fit; (7): S/N ratio as determined from the peak intensity and local rms noise given in Col.~(5); 
(8) and (9): deconvolved FWHM size ($\theta_{\rm maj}\times \theta_{\rm min}$) in arcsec and physical kpc; (10): major axis position angle of the fitted Gaussian measured from north through east; (11): projected angular offset from the ALMA 1.3~mm peak position; (12): spectral index between the observed-frame frequencies of 1.4~GHz and 3~GHz; (13): brightness temperature at the observed frequency of 3~GHz; and (14): comments on the source.\tablefoottext{a}{Unless otherwise stated, the quoted redshift was derived from optical-MIR photometry (i.e. $z \equiv z_{\rm phot}$; \cite{brisbin2017}). For AzTEC/C1a, C2a, C3a, C5, C6a, and C17 the quoted spectroscopic redshift is taken from M.~S.~Yun et al. (in prep.), D.~A.~Riechers et al. (in prep.), E.~F. Jim{\'e}nez Andrade et al. (in prep.), Yun et al. (2015), Wang et al. (2016), and Schinnerer et al. (2008; their source J1000+0234), respectively. The remaining \textbf{21} spectroscopic redshifts were taken from the COSMOS spectroscopic redshift catalogue (M.~Salvato et al., in prep.). The redshifts of AzTEC/C2b, C4, C8b, C10b, C21, C31a, C37, C39, C53, and C115 are the synthetic values derived by Brisbin et al. (2017), while those of AzTEC/C40, C62, C81, C106, and C144 are based on the peak position of the FIR SED, and those of AzTEC/C3c, C10a, C30a, C33b, and C101b are based on the 3~GHz and submm flux density comparison. The photo-$z$'s of AzTEC/C56 and C86 are based on AGN template fitting (see \cite{brisbin2017} for details).}\tablefoottext{b}{The quoted error in $I_{\rm 3\, GHz}$ and $S_{\rm 3\, GHz}$ represents the formal error returned by {\tt JMFIT}. The uncertainties do not include the absolute calibration uncertainty.}\tablefoottext{c}{The size and P.A. uncertainties represent the minimum and maximum values as returned by {\tt JMFIT}. The P.A. is formally defined to range from $0\degr$ to $180\degr$, but for example for AzTEC/C3a the maximum P.A. value is $203\fdg5$, which is symmetric with respect to an angle of $203\fdg5-180\degr=23\fdg5$. The minimum and maximum P.A. values for AzTEC/C2a, C44a, and C48a are equal to the nominal value, and hence the quoted uncertainties are equal to zero.}\tablefoottext{d}{The spectral index is a lower limit if the source was not detected at 1.4~GHz, and the brightness temperature is a lower limit for the unresolved sources.}\tablefoottext{e}{A note is given if the source has been detected with the VLBA or in the X-rays. The other AzTEC ID refers to the JCMT/AzTEC SMG ID (see \cite{scott2008}), and is quoted for the sources whose 3~GHz radio sizes we studied in M15. The COSLA IDs refer to the LABOCA 870~$\mu$m selected sources (F.~Navarrete et al., in prep.) that Smol{\v c}i{\'c} et al. (2012) detected with the PdBI at 1.3~mm, while the Cosbo IDs refer to the MAMBO-2 1.2~mm sources (\cite{bertoldi2007}). See Smol{\v c}i{\'c} et al. (2012) for more details about the COSLA and Cosbo IDs.}}
\end{table}
\end{landscape}

\section{Cosmic-ray electron cooling time and diffusion length in AzTEC/C2a, C4, C5, C17, and C42}

Five of our target SMGs benefit from high, subarcsecond resolution submm interferometric imaging, namely AzTEC/C2a ($=$AzTEC~8), C4 ($=$AzTEC~4), C5 ($=$AzTEC~1), C17 ($=$J1000+0234), and C42 ($=$AzTEC~5). This provides us with information about their dust-emitting sizes that can be quantitatively compared on a galaxy-by-galaxy basis with the present radio sizes. This, in turn, allows us to examine the length scale that CR electrons would have to travel in case they originate in the dust-emitting region.

\subsection{Rest-frame far-infrared-emitting sizes}

The SMGs AzTEC/C2a and C4 were imaged with the Submillimetre Array (SMA) at 870~$\mu$m by Younger et al. (2010), where the angular resolutions of the data were $0\farcs86 \times 0\farcs55$ and $0\farcs86 \times 0\farcs77$, respectively. For both sources, Younger et al. (2010) derived an angular major axis FWHM of $0\farcs6\pm0\farcs2$ when a Gaussian intensity distribution was assumed. 

The SMGs AzTEC/C5, C17, and C42 were observed with the ALMA Band~7 during the second early science campaign to search for [\ion{C}{II}] or [\ion{N}{II}] line emission (Cycle~1 ALMA project 2012.1.00978.S; PI: A.~Karim). The Common Astronomy Software Applications (CASA; \cite{mcmullin2007}) package\footnote{{\tt https://casa.nrao.edu}.} was used to construct the continuum images from the line-free channels at $\lambda_{\rm obs}=870$~$\mu$m, 857~$\mu$m, and 994~$\mu$m, respectively. The corresponding angular resolutions were $0\farcs30 \times 0\farcs29$, $0\farcs35 \times 0\farcs35$, and $0\farcs52 \times 0\farcs30$, respectively. As described in M15, the deconvolved angular major axis FWHM of AzTEC/C5 was found to be $0\farcs39\pm0\farcs01$ using the {\tt JMFIT} task in AIPS. Using a similar AIPS/{\tt JMFIT} analysis, we derived a rest-frame FIR size of $(0\farcs35\pm0\farcs01) \times (0\farcs19\pm0\farcs01)$ for AzTEC/C17. As we also described in more detail in M15, AzTEC/C42 was resolved into two components with these ALMA observations. We used a two-component Gaussian fit option of {\tt JMFIT} to derive a FIR size of $0\farcs33^{+0.05}_{-0.03}\times 0\farcs24^{+0.06}_{-0.09}$ for the northern component, and $0\farcs38^{+0.10}_{-0.07}\times 0\farcs36^{+0.07}_{-0.08}$ for the southern component. The northern component is well coincident with the 3~GHz radio emission peak position (see also Fig.~\ref{figure:taffy} herein). As a consistency check, we also used CASA (release 4.6.1) to determine the ALMA dust-emitting sizes of the aforementioned three sources (using the {\tt imfit} task), and obtained identical results with AIPS/{\tt JMFIT}. 

\begin{figure*}[!htb]
\begin{center}
\includegraphics[width=0.45\textwidth]{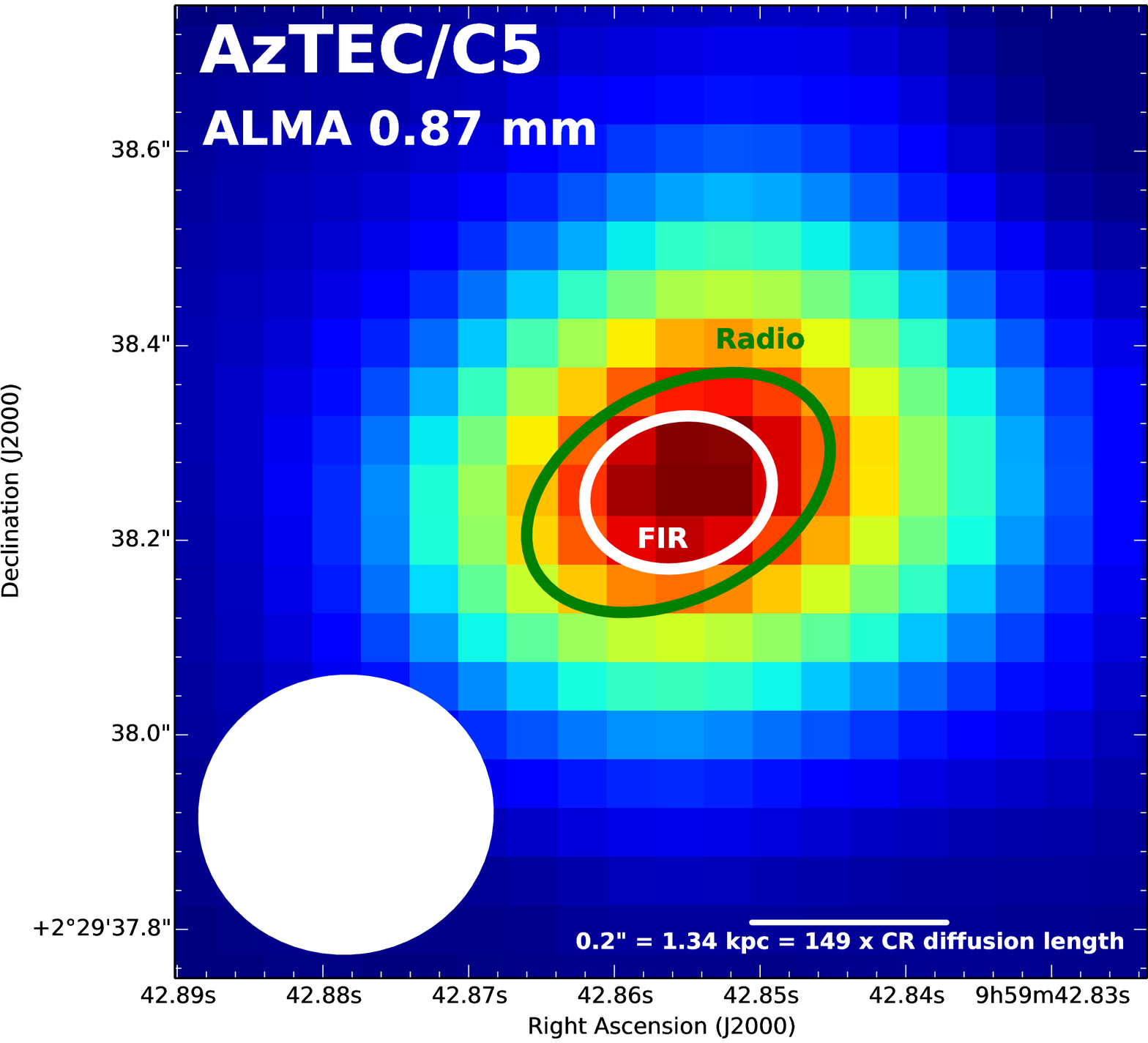}
\includegraphics[width=0.46\textwidth]{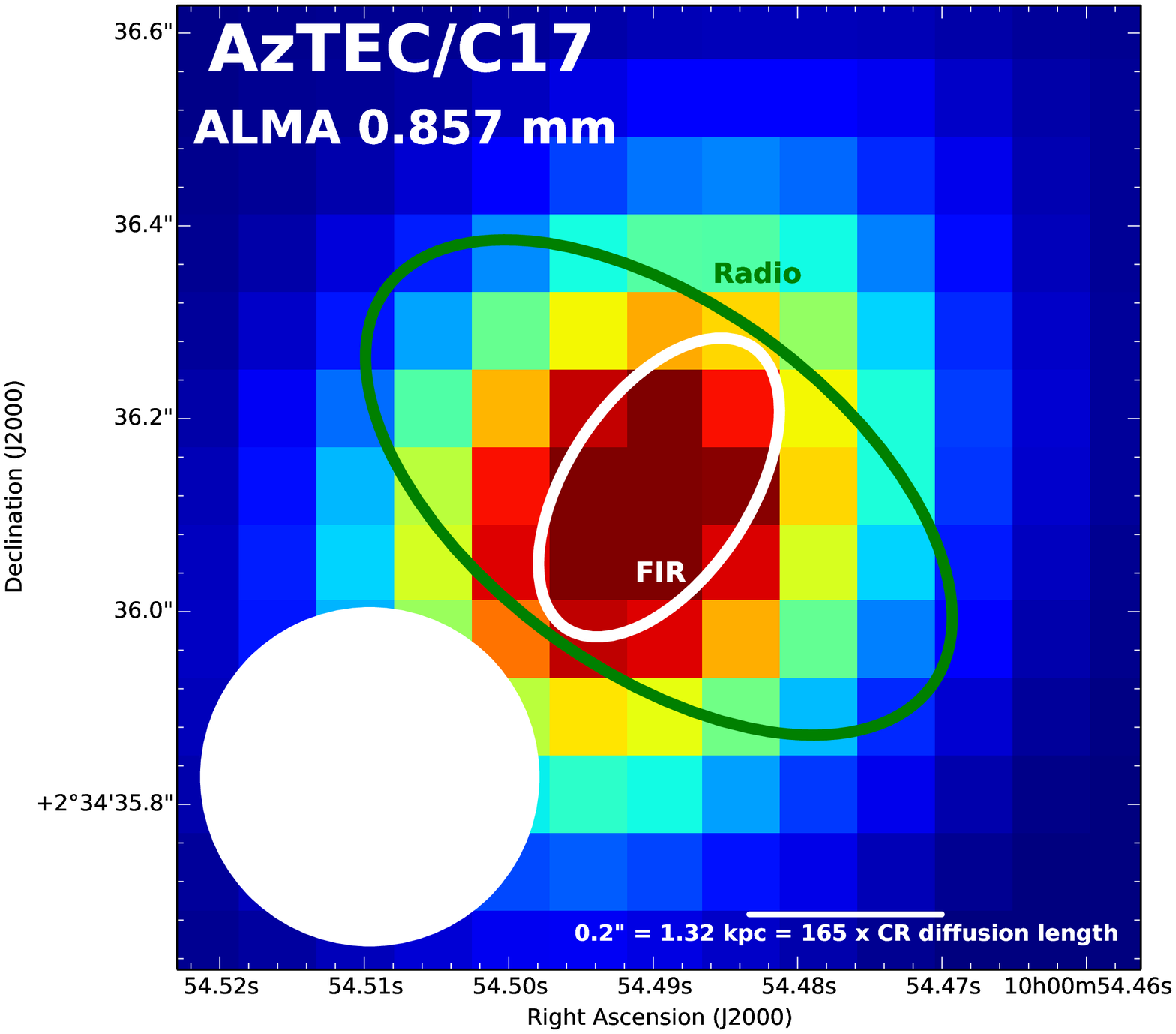}
\caption{\textbf{Left:} Observed-frame 870~$\mu$m ($\lambda_{\rm rest}=163$~$\mu$m) ALMA image of AzTEC/C5. \textbf{Right:} Observed-frame 857~$\mu$m ($\lambda_{\rm rest}=155$~$\mu$m) ALMA image of AzTEC/C17. These ALMA data were obtained as part of our Cycle~1 project (PI: A.~Karim). In both panels, the white and green ellipses, which are centred on the ALMA peak position, show the intrinsic rest-frame FIR and $\nu_{\rm obs}=3$~GHz radio FWHM sizes of the source (AzTEC/C5: $0\farcs39 \times 0\farcs31$ and $0\farcs67 \times 0\farcs43$, respectively; AzTEC/C17: $0\farcs35 \times 0\farcs19$ and $0\farcs70 \times < 0\farcs38$, respectively). A scale bar of $0\farcs2$ ($=1.34$~proper kpc for AzTEC/C5, and $1.32$~proper kpc for AzTEC/C17) projected length is shown in the bottom right corner. The length of the bar is about 2.2~dex longer than the CR electron diffusion length of 9~pc for AzTEC/C5 and 8~pc for AzTEC/C17 (Appendix~D). The ALMA synthesised beam FWHM ($0\farcs30 \times 0\farcs29$, P.A. $98\fdg1$ for AzTEC/C5; $0\farcs35 \times 0\farcs35$ for AzTEC/C17) is shown in the bottom left.}
\label{figure:dust}
\end{center}
\end{figure*}

The rest-frame FIR sizes of the five SMGs under study are about 0.7 (AzTEC/C2a) to 2.9 (C4 and C42) times 
that of their radio-emitting region (see Table~\ref{table:cooling}, the fifth row); on average, the radio size is two times larger than the dust-emitting region, in good agreement with the size comparisons described in Sect.~4.2.2. The cases of AzTEC/C5 and C17 are illustrated in Fig.~\ref{figure:dust}.

%using casa v 4.6.152: C5: $0\farcs39\pm0\farcs02 \times 0\farcs30\pm0\farcs02$ 
%C42: $0\farcs36\pm0\farcs05 \times 0\farcs21\pm0\farcs15$
%C17: $0\farcs35\pm0\farcs01 \times 0\farcs19\pm0\farcs01$ (aips gave identical size) 

\subsection{The cooling timescale and diffusion length scale of cosmic-ray electrons}

In an accompanying paper (O.~Miettinen et al., in prep.; see also \cite{miettinen2017}), 
we derived the total-IR ($8-1\,000$~$\mu$m) luminosities ($L_{\rm IR}$) of AzTEC/C2a, C4, C5, C17, and C42 through fitting their panchromatic SEDs, and we also estimated their gas masses ($M_{\rm gas}$) using the Scoville et al. (2016) dust continuum method. Here, we employ those quantities to improve the estimates of the CR electron cooling timescales and diffusion length scales from M15. 

More precisely, $L_{\rm IR}$ is needed to calculate the radiation field energy density of the galaxy ($u_{\rm rad}$; see \cite{murphy2008}, 2012a for an equation where also the non-absorbed UV radiation is taken into account), to which the cooling time due to the inverse Compton (IC) process is inversely proportional ($\tau_{\rm IC}\propto u_{\rm rad}^{-1}$). The galaxy area needed in the calculation of $u_{\rm rad}$ was defined as 

\begin{equation}
\label{eq:area}
A_{\rm IR}=\pi \times \frac{{\rm FWHM_{maj}}} {2}\times \frac{{\rm FWHM_{min}}} {2}\,,
\end{equation}
where the major and minor axes refer to the ALMA dust emission extents described in Appendix~D.1. 

On the other hand, the mass of the gas component ($M_{\rm gas}\simeq M_{\rm H_2}$) is needed to estimate the hydrogen number density of the galaxy that is being sampled by the electrons; the bremsstrahlung and ionisation cooling times are both inversely proportional to the gas density ($\tau_{\rm brem}\propto n_{\rm H}^{-1}$ and $\tau_{\rm ion}\propto n_{\rm H}^{-1}$). To calculate the volumetric average hydrogen number 
density, we used the gas mass estimates from O.~Miettinen et al. (in prep.). The CO-based gas masses derived for AzTEC/C5 by Yun et al. (2015) and for AzTEC/C17 by Schinnerer et al. (2008) agree with the dust-based values within a factor of two when correcting for the different assumptions used in the calculation. The volume-averaged gas (H$_2$) mass density was calculated as $\rho_{\rm gas}=M_{\rm gas}/V$, where the volume was defined as $V=2\pi R^2 h$ with $R\equiv R_{\rm FIR}$ the radius of the dust-emitting region, and $h$ the disk scale height (half-thickness). For the purpose of our calculation, we adopted a value of $h=1$~kpc, which is characteristic of high-$z$ puffy starbursts (\cite{lacki2010}, and references therein); a lower scale height would make the density estimates higher (e.g. by 1~dex for a thin disk with $h\sim100$~pc). The volume-averaged hydrogen number density was then derived using the standard formula 

\begin{equation} 
n_{\rm H}=n({\rm H})+2n({\rm H}_2)\simeq 2n({\rm H}_2)=2\times \frac{\rho_{\rm gas}}{\mu_{\rm H_2} m_{\rm H}}\,,
\end{equation}
where $\mu_{\rm H_2}=2.82$ is the mean molecular weight per H$_2$ molecule (assuming a He/H abundance ratio of 0.1; \cite{kauffman2008}), and $m_{\rm H}$ is the mass of a hydrogen atom. 

The magnetic field strength ($B$), which (for a given electron energy) determines the critical frequency at which a CR electron emits most of its energy ($\nu_{\rm crit} \propto B$), is needed in the calculation of the synchrotron cooling time ($\tau_{\rm synch}\propto B^{-3/2}$), IC cooling time ($\tau_{\rm IC}\propto B^{1/2}$), and ionisation cooling time ($\tau_{\rm ion}\propto B^{-1/2}/ \ln(\nu_{\rm crit}/B)$). We estimated the $B$-field strength using the assumption of magnetic flux freezing, that is $B=B_0\times \sqrt{n_{\rm H}\,[{\rm cm}^{-3}]}$, where $B_0$, which is the field strength at $n_{\rm H}=1$~cm$^{-3}$, was taken to be 10~$\mu$G (e.g. \cite{crutcher1999}; \cite{beck2001}, and references therein).

The cooling times due to the energy losses from the aforementioned processes, the effective total cooling lifetime for CR electrons ($\tau_{\rm cool}^{\rm tot}$), and the corresponding electron escape scale-length ($l_{\rm esc} \propto (\tau_{\rm cool}^{\rm tot})^{1/2}\nu_{\rm crit}^{1/2}B^{-1/2}$ for a random walk diffusion) in the five SMGs under study are listed in Table~\ref{table:cooling}. Owing to the strong radiation fields in these SMGs, the most rapid cooling occurs via the IC losses ($\tau_{\rm IC}\sim 1.2\times10^4$~yr on average), and the average diffusion (escape) length is only 11~pc. 

\begin{table}
\renewcommand{\footnoterule}{}
\caption{Physical properties of AzTEC/C2a, C4, C5, C17, and C42, and the CR electron cooling times and diffusion length scales.}
{\scriptsize
\begin{minipage}{1\columnwidth}
\centering
\label{table:cooling}
\begin{tabular}{c c c c c c}
\hline\hline 
Parameter & C2a & C4 & C5 & C17 & C42 \\
\hline
$z$\tablefootmark{a} & 3.179 & $5.30^{+0.70}_{-1.10}$ & 4.3415 & 4.542 & $3.63^{+0.37}_{-0.56}$ \\[1ex]
$M_{\rm gas}$\tablefootmark{b} [$10^{11}$~M$_{\sun}$] & 5.6 & 4.7 & 5.5 & 3.8 & 3.1\\[1ex]
$D_{\rm FIR}^{\rm maj}$\tablefootmark{c} [kpc] & 4.5 & 3.7 & 2.6 & 2.3 & 2.4\\[1ex]
$D_{\rm FIR}^{\rm min}$\tablefootmark{c} [kpc] & 3.8 & 2.4 & 2.1 & 1.2 & 1.7\\[1ex]
$D_{\rm radio}^{\rm maj}/D_{\rm FIR}^{\rm maj}$\tablefootmark{d} & $0.7^{+0.6}_{-0.4}$ & $2.9^{+2.3}_{-1.2}$ & $1.7^{+0.5}_{-0.5}$ & $2.0^{+0.6}_{-0.7}$ & $2.9^{+0.5}_{-0.5}$ \\[1ex]
$h$\tablefootmark{e} [kpc] & 1.0 & 1.0 & 1.0 & 1.0 & 1.0\\[1ex]
$n_{\rm H}$\tablefootmark{f} [cm$^{-3}$] & 598 & 967 & 1\,840 & 2\,515 & 1\,388 \\[1ex]
$B$\tablefootmark{g} [$\mu$G] & 245 & 311 & 429 & 502 & 373 \\[1ex]
$L_{\rm IR}$\tablefootmark{h} [$10^{13}$~L$_{\sun}$] & 2.6 & 2.0 & 1.8 & 0.8 & 3.8\\[1ex]
$u_{\rm rad}$\tablefootmark{i} [keV~cm$^{-3}$] & 8.2 & 12.1 & 17.7 & 15.6 & 49.9\\[1ex]
$u_{\rm CMB}$\tablefootmark{i} [eV~cm$^{-3}$] & 80.3 & 414.9 & 214.4 & 248.5 & 121.0\\[1ex]
$\tau_{\rm synch}$ [$10^5$~yr] & 1.0 & 0.6 & 0.4 & 0.3 & 0.5\\[1ex]
$\tau_{\rm IC}$ [$10^3$~yr] & 19.3 & 11.9 & 10.4 & 12.5 & 3.7\\[1ex]
$\tau_{\rm IC,\,CMB}$ [$10^5$~yr] & 19.6 & 3.5 & 8.6 & 7.9 & 15.2\\[1ex]
$\tau_{\rm brem}$ [$10^5$~yr] & 1.4 & 0.9 & 0.5 & 0.3 & 0.6 \\[1ex]
$\tau_{\rm ion}$ [$10^4$~yr] & 4.6 & 3.1 & 1.3 & 0.9 & 1.7 \\[1ex]
$\tau_{\rm cool}^{\rm tot}$\tablefootmark{j} [$10^3$~yr] & 11.0 & 6.8 & 4.5 & 3.9 & 2.7\\[1ex]
$l_{\rm esc}$\tablefootmark{j} [pc] & 17 & 14 & 9 & 8 & 7\\[1ex]
\hline 
\end{tabular} 
\tablefoot{\tablefoottext{a}{See Table~\ref{table:results} for the redshift references.}\tablefoottext{b}{The gas mass estimated by using the Scoville et al.  (2016) dust continuum method (O.~Miettinen et al., in prep.). The CO-based gas masses for AzTEC/C5 (\cite{yun2015}) and AzTEC/C17 (\cite{schinnerer2008}) agree with the quoted values within a factor of two when the values are scaled according to the assumptions used in the calculation.}\tablefoottext{c}{The parameters $D_{\rm FIR}^{\rm maj}$ and $D_{\rm FIR}^{\rm min}$ refer to the major and minor axis FWHM sizes, respectively. The FWHM of the rest-frame FIR-emitting region of AzTEC/C2a and C4 was derived through subarcsec resolution SMA 870~$\mu$m observations by Younger et al. (2010), while that of AzTEC/C5, C17, and C42 was derived from the subarcsec resolution ALMA submm continuum data (PI: A.~Karim). The AzTEC/C42 size refers to the northern ALMA component, which is coincident with the 3~GHz emission peak (M15).}\tablefoottext{d}{The ratio between the observed-frame 3~GHz radio size and the rest-frame FIR size.}\tablefoottext{e}{A disk scale height is assumed to be 1~kpc, i.e. that of high-$z$ puffy starbursts considered by Lacki \& Thompson (2010).}\tablefoottext{f}{The average hydrogen number density was calculated from the $M_{\rm gas}$, FIR-emitting size, and $h$ values (see text for details).}\tablefoottext{g}{The magnetic field strength was estimated using the formula $B=10~{\rm \mu G}\times \sqrt{n_{\rm H}/{\rm cm}^{-3}}$.}\tablefoottext{h}{The total IR (8--1\,000~$\mu$m) luminosity adopted from O.~Miettinen et al. (in prep.).}\tablefoottext{i}{The radiation field and CMB energy densities ($u_{\rm rad}$ and $u_{\rm CMB}$, respectively).}\tablefoottext{j}{For the formulas used to calculate the electron cooling times and escape scale lengths, we refer to Appendix~E in M15, and references therein.}    }
\end{minipage} 
}
\end{table}

\section{Estimating the radial dust temperature profile of AzTEC/C5} 

As we discussed in Sect.~4.5.2, one possible cause of the very compact rest-frame FIR sizes of SMGs is a negative radial dust temperature 
gradient in the galactic disk. To explore this possibility, we employed the aforementioned high-resolution ($0\farcs30 \times 0\farcs29$) ALMA 870~$\mu$m data for AzTEC/C5 to estimate the average temperature of the cold dust component within the rest-frame FIR-emitting region  (see Fig.~\ref{figure:dust}, left panel). Besides the high-resolution ALMA data we possess, AzTEC/C5 is well suited for this analysis because it has a reliable CO and [\ion{C}{ii}]-based spectroscopic redshift (\cite{yun2015}), it exhibits a fairly symmetric morphology in the rest-frame FIR, it is likely seen nearly face-on (\cite{yun2015}; \cite{miettinen2017}), and its SED-based fundamental physical properties are available (O.~Miettinen et al., in prep.; \cite{miettinen2017}, and references therein). 

We modelled the dust emission as a modified blackbody, in which case the surface brightness at rest-frame frequency $\nu$ is given by

\begin{equation} 
\label{eq:rad}
I_{\nu}=B_{\nu}(T_{\rm dust})\left(1-e^{-\tau_{\nu}}\right)+B_{\nu}(T_{\rm CMB})e^{-\tau_{\nu}}\,, 
\end{equation} 
where $B_{\nu}(T_{\rm dust})$ is the Planck blackbody function at the dust temperature $T_{\rm dust}$, $\tau_{\nu}$ is the optical 
thickness, and $T_{\rm CMB}$ is temperature of the CMB, which evolves with redshift as $T_{\rm CMB}(z)=2.725\times (1+z)$~K. At the redshift of AzTEC/C5, 
the value of $T_{\rm CMB}$ is about 14.6~K. 

The dust surface brightness map was converted into a $T_{\rm dust}$ map by solving Eq.~(\ref{eq:rad}) for $T_{\rm dust}$, and 
assuming that the optical thickness obeys a power-law functional form as a function of frequency, 
namely $\tau_{\nu}=\tau_0 \times (\nu/\nu_0)^{\beta}$ (e.g. \cite{hildebrand1983}), where the reference frequency, at which $\tau_{\nu}=\tau_0=1$, 
was set to $\nu_0=3$~THz ($\lambda_0=100$~$\mu$m). The dust emissivity index was fixed at $\beta=1.5$. At $\nu_{\rm rest}=1.84$~THz probed by 
our ALMA 870~$\mu$m observations, $\tau_{\nu}=0.48$, that is the dust has a moderate optical thickness, and the assumption of optically thin ($\tau \ll 1$) dust emission would not be valid. As an alternative dust optical thickness estimate, we used a dust mass of $\log (M_{\rm dust}/{\rm M}_{\sun})=9.17$ derived for AzTEC/C5 by O.~Miettinen et al. (in prep.), and converted that to $\tau_{\nu}$ using the formula  

\begin{equation}
\label{eq:tau}
\tau_{\nu}=\kappa_{\nu}\times \frac{M_{\rm dust}}{A_{\rm IR}}\,,
\end{equation} 
where $\kappa_{\nu}$ is the dust opacity, and $A_{\rm IR}$ is defined in Eq.~(\ref{eq:area}). To be consistent with the SED analysis by O.~Miettinen et al. (in prep.), we assumed that $\kappa_{\nu}$ has a power-law $\nu$-dependence of $\kappa_{\nu}=\kappa_0 \times (\nu/\nu_0)^{\beta}$ with $\kappa_0=0.77$~cm$^2$~g$^{-1}$ at $\nu_0=352.7$~GHz ($\lambda_0=850$~$\mu$m). For a dust-emission region radius of 1.3~kpc (Table~\ref{table:cooling}), Eq.~(\ref{eq:tau}) yields $\tau_{\nu}=0.66$, which is comparable to the aforementioned, simple frequency-scaled value of $\tau_{\nu}=0.48$. 

\begin{figure*}[!htb]
\begin{center}
\includegraphics[width=0.45\textwidth]{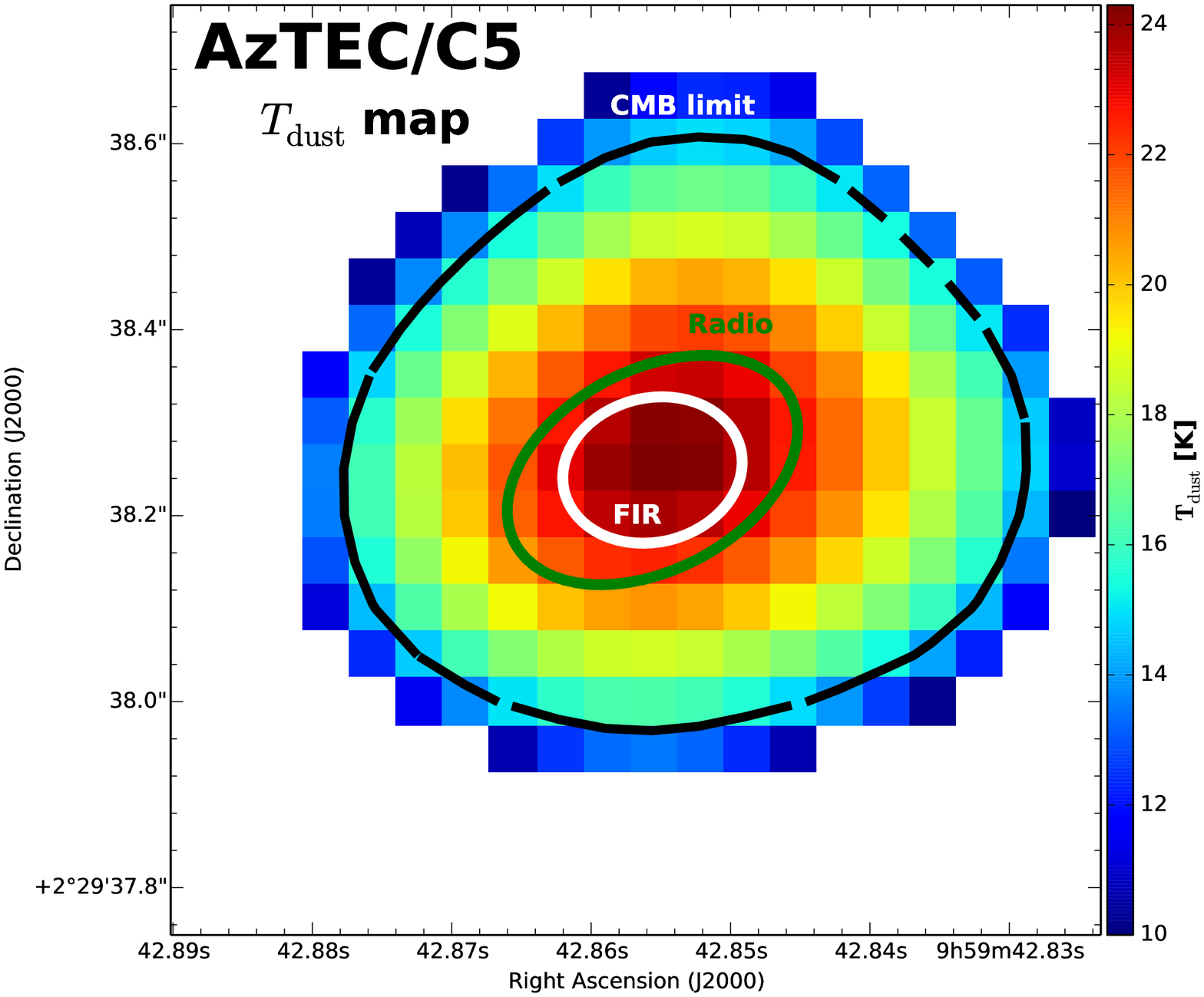}
\includegraphics[width=0.46\textwidth]{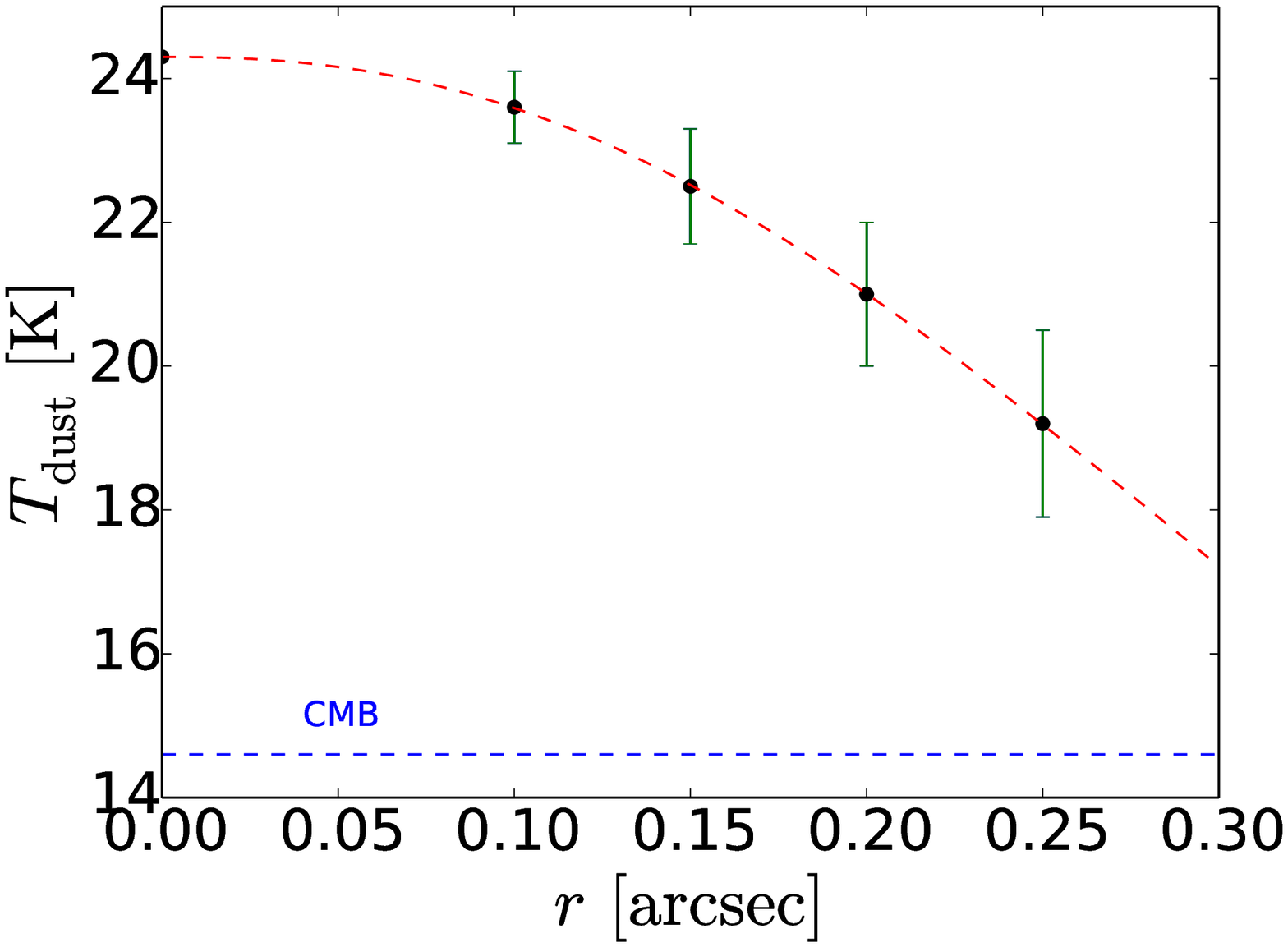}
\caption{\textbf{Left:} Dust temperature map of AzTEC/C5. The dust emissivity index, $\beta$, was fixed at 1.5. The colourbar
on the right indicates the units in kelvin. The two ellipses show the dust and radio-emitting FWHM extents as in Fig.~\ref{figure:dust}. 
The thick black line indicates the border where the dust temperature drops to the CMB temperature at the source redshift ($T_{\rm CMB}(z=4.3415)=14.6$~K). 
\textbf{Right:} Radial profile of the cold dust temperature of AzTEC/C5. The data points represent azimuthally
averaged values inside $0\farcs05$ wide circular annuli, and the vertical error bars represent the standard deviations of these averages. The red dashed line shows the mean radial $T_{\rm dust}(r)$ profile fitted using Eq.~(\ref{eq:plummer}). The CMB temperature floor is shown by a horizontal blue dashed line, and is met with the $T_{\rm dust}(r)$ profile at $r=0\farcs367$.}
\label{figure:Tdust}
\end{center}
\end{figure*}

The $T_{\rm dust}$ map based on the assumption that $(\beta,\, \tau_{\nu})=(1.5,\,0.48)$ is shown in the left panel in Fig.~\ref{figure:Tdust}. To derive 
the $T_{\rm dust}(r)$ profile shown in the right panel in Fig.~\ref{figure:Tdust}, we used one pixel-wide ($0\farcs05$) annuli (concentric circles) centred on the $T_{\rm dust}$ peak position, and derived the azimuthally averaged $T_{\rm dust}$ values. The data points were 
fitted by a Plummer-like function (\cite{plummer1911}) of the form

\begin{equation} 
\label{eq:plummer}
T_{\rm dust}(r)=\frac{T_{\rm dust}^{\rm peak}}{1+\left(\frac{r}{r_0}\right)^m}\,,
\end{equation} 
where $T_{\rm dust}^{\rm peak}$ is the central peak dust temperature, $r$ is radial distance from the centre, $r_0$ is the radius of the flat inner region, and 
$m$ is the the power-law exponent at large radii ($r \gg r_0$). For the case shown in Fig.~\ref{figure:Tdust}, we derived $T_{\rm dust}^{\rm peak}=24.3$~K, 
$r_0=0\farcs436\pm0\farcs002$ and $m=2.378\pm0.015$. The $T_{\rm dust}^{\rm peak}$ value is in good agreement with the temperature of the cold dust component of 25~K derived 
for AzTEC/C5 by O.~Miettinen et al. (in prep.) through their panchromatic SED analysis using the {\tt MAGPHYS} code (\cite{dacunha2015}). 
The CMB temperature floor is reached at $r=0\farcs367$, that is at a radius shorter than $r_0$, and hence the compact size of the rest-frame FIR-emitting region appears to be physical, rather than an observational bias owing to a $T_{\rm dust}(r)$ gradient. The fitting parameters obtained through different assumptions about 
$\tau_{\nu}$ are fairly similar to each other; see Table~\ref{table:t}. The implications of this analysis are discussed further in Sect.~4.5.2.

\begin{table}
\renewcommand{\footnoterule}{}
\caption{Parameters of the $T_{\rm dust}$ profiles derived using different assumptions.}
{\tiny
\begin{minipage}{1\columnwidth}
\centering
\label{table:t}
\begin{tabular}{c c c c c}
\hline\hline 
$\tau_{\nu}$\tablefootmark{a} & $T_{\rm dust}^{\rm peak}$\tablefootmark{b} & $r_0$\tablefootmark{b} & $m$\tablefootmark{b} & $r=r(T_{\rm CMB})$\tablefootmark{c} \\
                             & [K] & [\arcsec] & & [\arcsec] \\
\hline
0.48 ($\beta=1.5$) & 24.3 & $0.436\pm0.002$ & $2.378\pm0.015$ & 0.367 \\
0.38 ($\beta=2$) & 25.6 & $0.397\pm0.018$ & $2.718\pm0.158$ & 0.358 \\
0.66 ($M_{\rm dust}$; $\beta=1.5$) & 22.9 & $0.449\pm0.012$ & $2.462\pm0.080$ & 0.357 \\
1.51 ($M_{\rm dust}$; $\beta=2$) & 20.7 & $0.497\pm0.021$ & $2.377\pm0.104$ & 0.344 \\
\hline 
\end{tabular} 
\tablefoot{\tablefoottext{a}{The optical thickness at $\lambda_{\rm rest}=163$~$\mu$m based on different assumptions of $\beta$ and using either a simple frequency scaling or dust mass method (Eq.~(\ref{eq:tau})).}\tablefoottext{b}{The Plummer profile parameters of Eq.~(\ref{eq:plummer}).}\tablefoottext{c}{The radius at which the CMB temperature is reached.}}
\end{minipage} 
}
\end{table}


\begin{thebibliography}{}

\bibitem[Achterberg et al. 2001]{achterberg2001} Achterberg, A., 
Gallant, Y.~A., Kirk, J.~G., \& Guthmann, A.~W.\ 2001, \mnras, 328, 393

\bibitem[Aretxaga et al. 2011]{aretxaga2011} Aretxaga, I., Wilson, 
G.~W., Aguilar, E., et al.\ 2011, \mnras, 415, 3831 

\bibitem[Astropy Collaboration et al. 2013]{astropy2013} Astropy Collaboration, Robitaille, T.~P., 
Tollerud, E.~J., et al.\ 2013, \aap, 558, A33 

\bibitem[Barcos-Mu{\~n}oz et al. 2015]{barcos2015} Barcos-Mu{\~n}oz, L., Leroy, A.~K., Evans, A.~S., 
et al.\ 2015, \apj, 799, 10 

%\bibitem[Barnes \& Hernquist 1996]{barnes1996} Barnes, J.~E., \& Hernquist, L.\ 1996, \apj, 471, 115

\bibitem[Basu et al. 2015]{basu2015} Basu, A., Wadadekar, Y., Beelen, A., et al.\ 2015, \apj, 803, 51 

\bibitem[Bate 2014]{bate2014} Bate, M.~R.\ 2014, \mnras, 442, 285

\bibitem[Beck 2001]{beck2001} Beck, R.\ 2001, \ssr, Vol.~99, p.~243

\bibitem[Bell 2003]{bell2003} Bell, E.~F.\ 2003, \apj, 586, 794 

\bibitem[Bertoldi et al. 2007]{bertoldi2007} Bertoldi, F., Carilli, C., Aravena, M., et al.\ 2007, \apjs, 172, 132 

\bibitem[Bezanson et al. 2009]{bezanson2009} Bezanson, R., van Dokkum, P.~G., Tal, T., et al.\ 2009, \apj, 697, 1290 

\bibitem[Biggs \& Ivison 2008]{biggs2008} Biggs, A.~D., \& Ivison, R.~J.\ 2008, \mnras, 385, 893 

\bibitem[Blandford \& K{\"o}nigl 1979]{blandford1979} Blandford, R.~D., K{\"o}nigl, A.\ 1979, \apj, 232, 34

\bibitem[Blandford \& Ostriker 1980]{blandford1980} Blandford, R.~D., \& Ostriker, J.~P.\ 1980, \apj, 237, 793

\bibitem[Bogdan \& V{\"o}lk 1983]{bogdan1983} Bogdan, T.~J., \& V{\"o}lk, H.~J.\ 1983, \aap, 122, 129

\bibitem[Bolatto et al. 2013]{bolatto2013} Bolatto, A.~D., Wolfire, M., \& Leroy, A.~K.\ 2013, \araa, 51, 207

\bibitem[Bondi et al. 2008]{bondi2008} Bondi, M., Ciliegi, P., Schinnerer, E., et al.\ 2008, \apj, 681, 1129 

\bibitem[Bothwell et al. 2013]{bothwell2013} Bothwell, M.~S., Smail, I., Chapman, S.~C., et al.\ 2013, \mnras, 429, 3047

\bibitem[Braine et al. 2003]{braine2003} Braine, J., Davoust, E., Zhu, M., et al.\ 2003, \aap, 408, L13

\bibitem[Brisbin et al. 2017]{brisbin2017} Brisbin, D., et al.\ 2017, \aap, \textit{submitted}

\bibitem[Bussmann et al. 2015]{bussmann2015} Bussmann, R.~S., Riechers, D., Fialkov, A., et al.\ 2015, \apj, 812, 43

\bibitem[Carilli \& Yun 1999]{carilli1999} Carilli, C.~L., \& Yun, M.~S.\ 1999, \apjl, 513, L13

\bibitem[Carilli \& Yun 2000]{carilli2000} Carilli, C.~L., \& Yun, M.~S.\ 2000, \apj, 530, 618 

\bibitem[Capak et al. 2007]{capak2007} Capak, P., Aussel, H., Ajiki, M., et al.\ 2007, \apjs, 172, 99

\bibitem[Casey et al. 2009]{casey2009} Casey, C.~M., Chapman, S.~C., Muxlow, T.~W.~B., et al.\ 2009, \mnras, 395, 1249

\bibitem[Cassata et al. 2007]{cassata2007} Cassata, P., Guzzo, L., Franceschini, A., et al.\ 2007, \apjs, 172, 270 

\bibitem[Ceverino et al. 2016]{ceverino2016} Ceverino, D., Almeida, J.~S., Tu{\~n}{\'o}n, C.~M., 
et al.\ 2016, \mnras, 457, 2605 

\bibitem[Chapman et al. 2004]{chapman2004} Chapman, S.~C., Smail, I., Windhorst, R., et al.\ 2004, \apj, 611, 732 

\bibitem[Chen et al. 2015]{chen2015} Chen, C.-C., Smail, I., Swinbank, A.~M., et al.\ 2015, \apj, 799, 194 

\bibitem[Civano et al. 2012]{civano2012} Civano, F., Elvis, M., Brusa, M., et al.\ 2012, \apjs, 201, 30 

\bibitem[Civano et al. 2016]{civano2016} Civano, F., Marchesi, S., Comastri, A., et al.\ 2016, \apj, 819, 62  

\bibitem[Clemens et al. 2010]{clemens2010} Clemens, M.~S., Scaife, 
A., Vega, O., \& Bressan, A.\ 2010, \mnras, 405, 887

\bibitem[Condon 1992]{condon1992} Condon, J.~J.\ 1992, \araa, 30, 575

\bibitem[Condon et al. 1991]{condon1991} Condon, J.~J., Anderson, M.~L., \& Helou, G.\ 1991, \apj, 376, 95

\bibitem[Condon et al. 1993]{condon1993} Condon, J.~J., Helou, 
G., Sanders, D.~B., \& Soifer, B.~T.\ 1993, \aj, 105, 1730 

\bibitem[Condon et al. 2002]{condon2002} Condon, J.~J., Helou, G., \& Jarrett, T.~H.\ 2002, \aj, 123, 1881 

\bibitem[Conselice 2003]{conselice2003} Conselice, C.~J.\ 2003, \apjs, 147, 1 

\bibitem[Coppin et al. 2015]{coppin2015} Coppin, K.~E.~K., Geach, J.~E., Almaini, O., et al.\ 2015, \mnras, 446, 1293 

\bibitem[Crutcher 1999]{crutcher1999} Crutcher, R.~M.\ 1999, \apj, 520, 706 

\bibitem[da Cunha et al. 2013]{dacunha2013} da Cunha, E., Groves, B., Walter, F., et al.\ 2013, \apj, 766, 13

\bibitem[da Cunha et al. 2015]{dacunha2015} da Cunha, E., Walter, F., Smail, I., et al.\ 2015, \apj, 806, 110 

\bibitem[Deeg et al. 1993]{deeg1993} Deeg, H.-J., Brinks, E., Duric, N., et al.\ 1993, \apj, 410, 626

\bibitem[de Jong et al. 1985]{dejong1985} de Jong, T., Klein, U., Wielebinski, R., \& Wunderlich, E.\ 1985, \aap, 147, L6 

\bibitem[Dekel et al. 2009]{dekel2009} Dekel, A., Sari, R., \& Ceverino, D.\ 2009, \apj, 703, 785 

\bibitem[Delhaize et al. 2017]{delhaize2017} Delhaize, J., Smol{\v c}i{\'c}, V., Delvecchio, I., et al. 2017, \aap, \textit{in press}

\bibitem[Devereux \& Young 1990]{devereux1990} Devereux, N.~A., \& Young, J.~S.\ 1990, \apjl, 350, L25

\bibitem[Donevski \& Prodanovi{\'c} 2015]{donevski2015} Donevski, D., \& Prodanovi{\'c}, T.\ 2015, \mnras, 453, 638

\bibitem[Draine et al. 2007]{draine2007} Draine, B.~T., Dale, D.~A., Bendo, G., et al.\ 2007, \apj, 663, 866 

\bibitem[Drzazga et al. 2011]{drzazga2011} Drzazga, R.~T., Chy{\.z}y, K.~T., Jurusik, W., 
\& Wi{\'o}rkiewicz, K.\ 2011, \aap, 533, A22

\bibitem[Elvis et al. 2009]{elvis2009} Elvis, M., Civano, F., Vignali, C., et al.\ 2009, \apjs, 184, 158

\bibitem[Engel et al. 2010]{engel2010} Engel, H., Tacconi, L.~J., Davies, R.~I., et al.\ 2010, \apj, 724, 233 

%\bibitem[Fixsen 2009]{fixen2009} Fixsen, D.~J.\ 2009, \apj, 707, 916

\bibitem[Ezawa et al. 2004]{ezawa2004} Ezawa, H., Kawabe, R., Kohno, K., \& Yamamoto, S.\ 2004, \procspie, 5489, 763 

\bibitem[Falcke 1996]{falcke1996} Falcke, H.\ 1996, \apjl, 464, L67 

\bibitem[Forbes et al. 2011]{forbes2011} Forbes, D.~A., Spitler, L.~R., Strader, J., et al.\ 2011, \mnras, 413, 2943

\bibitem[Franco \& Cox 1986]{franco1986} Franco, J., \& Cox, D.~P.\ 1986, \pasp, 98, 1076 

\bibitem[Franx \& Illingworth 1990]{franx1990} Franx, M., \& Illingworth, G.\ 1990, \apjl, 359, L41 

\bibitem[Fu et al. 2013]{fu2013} Fu, H., Cooray, A., Feruglio, C., et al.\ 2013, \nat, 498, 338

\bibitem[Gao et al. 2003]{gao2003} Gao, Y., Zhu, M., \& Seaquist, E.~R.\ 2003, \aj, 126, 2171 

\bibitem[Gardner et al. 2006]{gardner2006} Gardner, J.~P., Mather, J.~C., Clampin, M., et al.\ 2006, \ssr, 123, 485

\bibitem[Goldsmith \& Langer 1978]{goldsmith1978} Goldsmith, P.~F., \& Langer, W.~D.\ 1978, \apj, 222, 881 

\bibitem[Gonz{\'a}lez-L{\'o}pez et al. 2017]{gonzalez2017} Gonz{\'a}lez-L{\'o}pez, J., Bauer, F.~E., Romero-Ca{\~n}izales, C., et al.\ 2017, \aap, 597, A41

\bibitem[Hales et al. 2012]{hales2012} Hales, C.~A., Murphy, T., Curran, J.~R., et al.\ 2012, \mnras, 425, 979

\bibitem[Helou et al. 1985]{helou1985} Helou, G., Soifer, B.~T., 
\& Rowan-Robinson, M.\ 1985, \apjl, 298, L7

\bibitem[Helsel 2005]{helsel2005} Helsel, D.~R.\ 2005, \textit{Nondetects And Data Analysis: 
Statistics for Censored Environmental Data}, John Wiley and Sons, New York

\bibitem[Hildebrand 1983]{hildebrand1983} Hildebrand, R.~H.\ 1983, \qjras, 24, 267

\bibitem[Hodge et al. 2013]{hodge2013} Hodge, J.~A., Karim, A., Smail, I., et al.\ 2013, \apj, 768, 91 

\bibitem[Hodge et al. 2016]{hodge2016} Hodge, J.~A., Swinbank, A.~M., Simpson, J.~M., et al.\ 2016, \apj, 833, 103

%\bibitem[Hunt et al. 2015]{hunt2015} Hunt, L.~K., Draine, B.~T., Bianchi, S., et al.\ 2015, \aap, 576, A33 

\bibitem[Ikarashi et al. 2015]{ikarashi2015} Ikarashi, S., Ivison, 
R.~J., Caputi, K.~I., et al.\ 2015, \apj, 810, 133 

\bibitem[Ikarashi et al. 2017]{ikarashi2017} Ikarashi, S., Ivison, R.~J., Caputi, K.~I., et al.\ 2017, \apj, 835, 286

\bibitem[Immeli et al. 2004]{immeli2004} Immeli, A., Samland, M., Westera, P., \& Gerhard, O.\ 2004, \apj, 611, 20 

\bibitem[Iono et al. 2016]{iono2016} Iono, D., Yun, M.~S., Aretxaga, I., et al.\ 2016, \apjl, 829, L10

\bibitem[Ivison et al. 2010]{ivison2010} Ivison, R.~J., Alexander, D.~M., Biggs, A.~D., et al.\ 2010, \mnras, 402, 245

\bibitem[Ivison et al. 2011]{ivison2011} Ivison, R.~J., 
Papadopoulos, P.~P., Smail, I., et al.\ 2011, \mnras, 412, 1913

\bibitem[Johnson et al. 2013]{johnson2013} Johnson, S.~P., Wilson, 
G.~W., Wang, Q.~D., et al.\ 2013, \mnras, 431, 662 

\bibitem[Karim et al. 2013]{karim2013} Karim, A., Swinbank, 
A.~M., Hodge, J.~A., et al.\ 2013, \mnras, 432, 2 

\bibitem[Kauffmann et al. 2008]{kauffman2008} Kauffmann, J., Bertoldi, F., Bourke, T.~L., et al.\ 2008, \aap, 487, 993

\bibitem[Lacki \& Thompson 2010]{lacki2010} Lacki, B.~C., \& Thompson, T.~A.\ 2010, \apj, 717, 196

%\bibitem[Lisenfeld \& V{\"o}lk 2000]{lisenfeld2000} Lisenfeld, U., V{\"o}lk, H.~J.\ 2000, \aap, 354, 423 

\bibitem[Laigle et al. 2016]{laigle2016} Laigle, C., McCracken, H.~J., Ilbert, O., et al. 2016, \apjs, 224, 24

\bibitem[Leroy et al. 2011]{leroy2011} Leroy, A.~K., Evans, A.~S., 
Momjian, E., et al.\ 2011, \apjl, 739, L25 

%\bibitem[Longair 1992]{longair1992} Longair, M.~S.\ 1992, \textit{High 
%Energy Astrophysics, Vol.~1: Particles, photons and their detection}, 
%Cambridge, UK: Cambridge University Press

\bibitem[Lindroos et al. 2016]{lindroos2016} Lindroos, L., Knudsen, K.~K., Fan, L., et al.\ 2016, \mnras, 462, 1192 

\bibitem[Magnelli et al. 2015]{magnelli2015} Magnelli, B., Ivison, R.~J., Lutz, D., et al.\ 2015, \aap, 573, A45

\bibitem[Marvil et al. 2015]{marvil2015} Marvil, J., Owen, F., \& Eilek, J.\ 2015, \aj, 149, 32

\bibitem[Mathis et al. 1983]{mathis1983} Mathis, J.~S., Mezger, P.~G., \& Panagia, N.\ 1983, \aap, 128, 212

\bibitem[McMullin et al. 2007]{mcmullin2007} McMullin, J.~P., Waters, B., Schiebel, D., et al.\ 2007, 
\textit{Astronomical Data Analysis Software and Systems XVI}, ASP Conference Series, Vol.~376, p.~127

\bibitem[Miettinen et al. 2015]{miettinen2015} Miettinen, O., Novak, M., 
Smol{\v c}i{\'c}, V., et al.\ 2015, \aap, 584, A32 (M15)

\bibitem[Miettinen et al. 2017]{miettinen2017} Miettinen, O., Delvecchio, I., 
Smol{\v c}i{\'c}, V., et al.\ 2017, \aap, 597, A5 

\bibitem[Murphy 2013]{murphy2013} Murphy, E.~J.\ 2013, \apj, 777, 58

\bibitem[Murphy et al. 2008]{murphy2008} Murphy, E.~J., Helou, G., Kenney, J.~D.~P., et al.\ 2008, \apj, 678, 828

\bibitem[Murphy et al. 2012a]{murphy2012a} Murphy, E.~J., Porter, T.~A., Moskalenko, I.~V., et al.\ 2012a, \apj, 750, 126

\bibitem[Murphy et al. 2012b]{murphy2012b} Murphy, E.~J., Bremseth, J., Mason, B.~S., et al.\ 2012b, \apj, 761, 97

\bibitem[Murphy et al. 2013]{murphyetal2013} Murphy, E.~J., Stierwalt, S., Armus, L., et al.\ 2013, \apj, 768, 2

\bibitem[Murphy et al. 2017]{murphy2017} Murphy, E.~J., Momjian, E., Condon, J.~J., et al.\ 2017, \apj, \textit{in press}, 
{\tt arXiv:1702.06963} 

\bibitem[Nagar et al. 2000]{nagar2000} Nagar, N.~M., Falcke, H., Wilson, A.~S., \& Ho, L.~C.\ 2000, \apj, 542, 186 

\bibitem[Narayanan et al. 2012]{narayanan2012} Narayanan, D., Krumholz, M.~R., Ostriker, E.~C., \& Hernquist, L.\ 2012, 
\mnras, 421, 3127

%\bibitem[Niklas et al. 1997]{niklas1997} Niklas, S., Klein, U., \& Wielebinski, R.\ 1997, \aap, 322, 19

\bibitem[Oteo et al. 2016]{oteo2016} Oteo, I., Ivison, R.~J., Dunne, L., et al.\ 2016, \apj, 827, 34 

\bibitem[Peletier et al. 1990]{peletier1990} Peletier, R.~F., Davies, R.~L., Illingworth, G.~D., et al.\ 1990, \aj, 100, 1091 

\bibitem[Perley et al. 2011]{perley2011} Perley, R.~A., Chandler, C.~J., Butler, B.~J., \& Wrobel, J.~M.\ 2011, \apjl, 739, L1

\bibitem[Plummer 1911]{plummer1911} Plummer, H.~C.\ 1911, \mnras, 71, 460

\bibitem[Reynolds \& Ellison 1992]{reynolds1992} Reynolds, S.~P., \& Ellison, D.~C.\ 1992, \apjl, 399, L75 

\bibitem[Riechers et al. 2011a]{riechers2011a} Riechers, D.~A., 
Carilli, L.~C., Walter, F., et al.\ 2011a, \apjl, 733, L11 

\bibitem[Riechers et al. 2011b]{riechers2011b} Riechers, D.~A., 
Hodge, J., Walter, F., et al.\ 2011b, \apjl, 739, L31 

\bibitem[Riechers et al. 2013]{riechers2013} Riechers, D.~A., Bradford, C.~M., Clements, D.~L., et al.\ 2013, \nat, 496, 329 

\bibitem[Riechers et al. 2014]{riechers2014} Riechers, D.~A., Carilli, C.~L., Capak, P.~L., et al.\ 2014, \apj, 796, 84

\bibitem[Rujopakarn et al. 2016]{rujopakarn2016} Rujopakarn, W., Dunlop, J.~S., Rieke, G.~H., et al.\ 2016, \apj, 833, 12

\bibitem[Salvato et al. 2011]{salvato2011} Salvato, M., Ilbert, 
O., Hasinger, G., et al.\ 2011, \apj, 742, 61 

\bibitem[Sargent et al. 2007]{sargent2007} Sargent, M.~T., Carollo, C.~M., 
Lilly, S.~J., et al.\ 2007, \apjs, 172, 434 

\bibitem[Sargent et al. 2010]{sargent2010} Sargent, M.~T., 
Schinnerer, E., Murphy, E., et al.\ 2010, \apjl, 714, L190

\bibitem[Scarlata et al. 2007]{scarlata2007} Scarlata, C., Carollo, C.~M., Lilly, S., et al.\ 2007, \apjs, 172, 406 

\bibitem[Schinnerer et al. 2007]{schinnerer2007} Schinnerer, E., 
Smol{\v c}i{\'c}, V., Carilli, C.~L., et al.\ 2007, \apjs, 172, 46 

\bibitem[Schinnerer et al. 2008]{schinnerer2008} Schinnerer, E., 
Carilli, C.~L., Capak, P., et al.\ 2008, \apjl, 689, L5

\bibitem[Schinnerer et al. 2010]{schinnerer2010} Schinnerer, E., 
Sargent, M.~T., Bondi, M., et al.\ 2010, \apjs, 188, 384

\bibitem[Scott et al. 2008]{scott2008} Scott, K.~S., Austermann, 
J.~E., Perera, T.~A., et al.\ 2008, \mnras, 385, 2225 

\bibitem[Scoville et al. 2007]{scoville2007} Scoville, N., Aussel, H., Brusa, M., et al.\ 2007, \apjs, 172, 1 

%\bibitem[Scoville et al. 2014]{scoville2014} Scoville, N., Aussel, H., Sheth, K., et al.\ 2014, \apj, 783, 84 

\bibitem[Scoville et al. 2016]{scoville2016} Scoville, N., Sheth, K., Aussel, H., et al.\ 2016, \apj, 820, 83

\bibitem[Searle et al. 1973]{searle1973} Searle, L., Sargent, W.~L.~W., \& Bagnuolo, W.~G.\ 1973, \apj, 179, 427 

\bibitem[Sharon et al. 2015]{sharon2015} Sharon, C.~E., Baker, A.~J., Harris, A.~I., et al.\ 2015, \apj, 798, 133 

\bibitem[Shetty et al. 2009]{shetty2009} Shetty, R., Kauffmann, J., Schnee, S., et al.\ 2009, \apj, 696, 2234 

\bibitem[Shimizu et al. 2012]{shimizu2012} Shimizu, I., Yoshida, N., \& Okamoto, T.\ 2012, \mnras, 427, 2866 

\bibitem[Simpson et al. 2014]{simpson2014} Simpson, J.~M., 
Swinbank, A.~M., Smail, I., et al.\ 2014, \apj, 788, 125

\bibitem[Simpson et al. 2015]{simpson2015} Simpson, J.~M., Smail, I., 
Swinbank, A.~M., et al.\ 2015, \apj, 799, 81 (S15)

\bibitem[Simpson et al. 2016]{simpson2016} Simpson, J.~M., Smail, I., Swinbank, A.~M., et al.\ 2016, \apj, \textit{submitted}, {\tt arXiv:1611.03084} 

\bibitem[Smol{\v c}i{\'c} et al. 2012]{smolcic2012} Smol{\v c}i{\'c}, V., Aravena, M., 
Navarrete, F., et al.\ 2012, \aap, 548, A4 

\bibitem[Smol{\v c}i{\'c} et al. 2017]{smolcic2017} Smol{\v c}i{\'c}, V., Novak, M., Bondi, M., et al.\ 2017, \aap, \textit{in press}

\bibitem[Spilker et al. 2015]{spilker2015} Spilker, J.~S., Aravena, M., Marrone, D.~P., et al.\ 2015, \apj, 811, 124

\bibitem[Swinbank et al. 2006]{swinbank2006} Swinbank, A.~M., 
Chapman, S.~C., Smail, I., et al.\ 2006, \mnras, 371, 465

\bibitem[Swinbank et al. 2010]{swinbank2010} Swinbank, A.~M., Smail, I., Chapman, S.~C., et al.\ 2010, \mnras, 405, 234

\bibitem[Swinbank et al. 2011]{swinbank2011} Swinbank, A.~M., Papadopoulos, P.~P., Cox, P., et al.\ 2011, \apj, 742, 11

\bibitem[Tacconi et al. 2006]{tacconi2006} Tacconi, L.~J., Neri, 
R., Chapman, S.~C., et al.\ 2006, \apj, 640, 228

\bibitem[Tacconi et al. 2008]{tacconi2008} Tacconi, L.~J., Genzel, R., 
Smail, I., et al.\ 2008, \apj, 680, 246

\bibitem[Tamura et al. 2000]{tamura2000} Tamura, N., Kobayashi, 
C., Arimoto, N., et al.\ 2000, \aj, 119, 2134 

\bibitem[Tasca et al. 2009]{tasca2009} Tasca, L.~A.~M., Kneib, J.-P., Iovino, A., et al.\ 2009, \aap, 503, 379 

\bibitem[Thompson et al. 1980]{thompson1980} Thompson, A.~R., Clark, B.~G., Wade, C.~M., \& Napier, P.~J.\ 1980, \apjs, 44, 151

\bibitem[Thomson et al. 2014]{thomson2014} Thomson, A.~P., Ivison, R.~J., Simpson, J.~M., et al.\ 2014, \mnras, 442, 577

\bibitem[Toft et al. 2014]{toft2014} Toft, S., Smol{\v c}i{\'c}, V., Magnelli, B., et al.\ 2014, \apj, 782, 68

\bibitem[Troncoso et al. 2014]{troncoso2014} Troncoso, P., Maiolino, R., Sommariva, V., et al.\ 2014, \aap, 563, A58 

\bibitem[van der Kruit 1971]{vanderkruit1971} van der Kruit, P.~C.\ 1971, \aap, 15, 110

\bibitem[Varenius et al. 2016]{varenius2016} Varenius, E., Conway, J.~E., Mart{\'{\i}}-Vidal, I., et al.\ 2016, \aap, 593, A86 

\bibitem[Voigt \& Bridle 2010]{voigt2010} Voigt, L.~M., \& Bridle, S.~L.\ 2010, \mnras, 404, 458

\bibitem[Wang et al. 2013]{wang2013} Wang, S.~X., Brandt, W.~N., Luo, B., et al.\ 2013, \apj, 778, 179 

\bibitem[Wang et al. 2016]{wang2016} Wang, T., Elbaz, D., Daddi, E., et al.\ 2016, \apj, 828, 56

\bibitem[Weiler et al. 1986]{weiler1986} Weiler, K.~W., Sramek, R.~A., 
Panagia, N., et al.\ 1986, \apj, 301, 790 

\bibitem[Wei{\ss} et al. 2009]{weiss2009} Wei{\ss}, A., 
Kov{\'a}cs, A., Coppin, K., et al.\ 2009, \apj, 707, 1201

\bibitem[Wellons et al. 2015]{wellons2015} Wellons, S., Torrey, P., Ma, C.-P., et al.\ 2015, \mnras, 449, 361 

\bibitem[Williams et al. 2011]{williams2011} Williams, C.~C., Giavalisco, M., Porciani, C., et al.\ 2011, \apj, 733, 92 

\bibitem[Wuyts et al. 2010]{wuyts2010} Wuyts, S., Cox, T.~J., Hayward, C.~C., et al.\ 2010, \apj, 722, 1666

\bibitem[Yan \& Ma 2016]{yan2016} Yan, H., \& Ma, Z.\ 2016, \apjl, 820, L16 

\bibitem[Yoast-Hull et al. 2013]{yoasthull2013} Yoast-Hull, T.~M., Everett, J.~E., Gallagher, J.~S., III, \& Zweibel, E.~G.\ 2013, \apj, 768, 53 

\bibitem[Younger et al. 2010]{younger2010} Younger, J.~D., Fazio, 
G.~G., Ashby, M.~L.~N., et al.\ 2010, \mnras, 407, 1268 

\bibitem[Yun et al. 2001]{yun2001} Yun, M.~S., Reddy, N.~A., 
\& Condon, J.~J.\ 2001, \apj, 554, 803 

\bibitem[Yun et al. 2015]{yun2015} Yun, M.~S., Aretxaga, I., Gurwell, M.~A., 
et al.\ 2015, \mnras, 454, 3485

\bibitem[Zahid et al. 2015]{zahid2015} Zahid, H.~J., Damjanov, I., Geller, M.~J., \& Chilingarian, I.\ 2015, \apj, 806, 122 

\bibitem[Zamojski et al. 2007]{zamojski2007} Zamojski, M.~A., Schiminovich, D., Rich, R.~M., et al.\ 2007, \apjs, 172, 468

\bibitem[Zhang et al. 2016]{zhang2016} Zhang, Z.-Y., Papadopoulos, P.~P., Ivison, R.~J., et al.\ 2016, Royal Society Open Science, Vol.~3, Issue~6, id.~160025 

\bibitem[Zhu et al. 2007]{zhu2007} Zhu, M., Gao, Y., Seaquist, 
E.~R., \& Dunne, L.\ 2007, \aj, 134, 118

\bibitem[Zirakashvili \& V{\"o}lk 2006]{zirakashvili2006} Zirakashvili, V.~N., \& V{\"o}lk, H.~J.\ 2006, \apj, 636, 140

\end{thebibliography}
\end{document}